\documentclass[a4paper,aps,pra,reprint,notitlepage,superscriptaddress,twocolumn,notitlepage,floatfix]{revtex4-2}

\usepackage{graphicx}
\usepackage{dcolumn}
\usepackage{bm}
\usepackage[utf8]{inputenc}
\usepackage[english]{babel}
\usepackage{amsfonts}
\usepackage{hyperref}
\usepackage{physics}
\usepackage{epstopdf}
\usepackage{mathtools}
\usepackage[capitalise]{cleveref}
\usepackage{float}
\usepackage[space]{grffile}
\usepackage{color}
\usepackage[normalem]{ulem}
\usepackage{subfigure}
\usepackage{natbib}
\def \hrho{\hat{\rho}}
\def \ha{\hat{a}}
\def \hb{\hat{b}}
\def \hc{\hat{c}}
\def \hH{\hat{H}}

\def\ket #1{\left| #1 \right>}
\def\bra #1{\left< #1 \right|}
\def\avg #1{\left< #1 \right>}

\def\diamondcomma{\ \raise.3ex\hbox{$\diamond$}\kern-0.4em\lower.7ex\hbox{$,$}\ }
\def\lesssim{\ \raise.3ex\hbox{$<$}\kern-0.8em\lower.7ex\hbox{$\sim$}\ }
\def\gesim{\ \raise.3ex\hbox{$>$}\kern-0.8em\lower.7ex\hbox{$\sim$}\ }

\begin{document}

    %\title{How do cavity photon change the $T_1$-decay of superconducting circuit QED?  \ryocom{Needs better title}}
    \title{Intrinsic mechanisms for drive-dependent Purcell decay in superconducting quantum circuits}
    \author{Ryo Hanai}
	\affiliation{Asia Pacific Center for Theoretical Physics, Pohang 37673, Korea}
	\affiliation{Pritzker School of Molecular Engineering, University of Chicago, Chicago, IL 60637, USA}
	
	\author{Alexander McDonald}
	\affiliation{Pritzker School of Molecular Engineering, University of Chicago, Chicago, IL 60637, USA}
	\affiliation{Department of Physics, University of Chicago, Chicago, IL 60637, USA}	
	
	\author{Aashish Clerk}
	\affiliation{Pritzker School of Molecular Engineering, University of Chicago, Chicago, IL 60637, USA}

	\date{\today}

	%%%%%%%%%%%%%%%%%%%%%%%%%%%%%%%%%%%%%%%%%%%%%%%%%%%%%%%%%

	\begin{abstract}
    We develop a new approach to understanding intrinsic mechanisms that cause the $T_1$-decay rate of a multi-level superconducting qubit to depend on the photonic population of a coupled, detuned cavity.
    Our method yields simple analytic expressions 
    for both the coherently driven or thermally excited cases which are in good agreement with full master equation numerics, and also facilitates direct physical intuition.  It also predicts several new phenomena.  In particular, we find that in a wide range of settings, the cavity-qubit detuning controls whether a non-zero photonic population increases or decreases qubit Purcell decay.  Our method combines insights from a Keldysh treatment of the system, and Lindblad perturbation theory.   
    % We  investigate  the  dissipation  properties  of  superconducting  qubits  coupled  to  a  cavity beyond  the  conventional  transmon  theory.  In  the  standard  approach,  the  weakly  anharmonic superconducting  qubits  are  described  by  the  so-called  “blackbox  quantization”  scheme, where  the  number  non-conserving  nonlinear  terms  are  dropped  [1]. This  simplification, however, misses the strong cavity excitation dependence to the qubit lifetimes observed in experiments.  In  this  work,  we  develop  methods  that  systematically  include  higher  order dissipative   effects   that   offers   advantages   over   Schrieffer-Wolff   methods.   These   include approaches  based  on  the  Keldysh  technique  as  well  as  on  Lindblad  perturbation  theory.    We find  novel  correlated  dissipative  processes  that  modifies  the  standard  cavity-induced  qubit dissipation  rates.  Our  work  provides  new  theoretical  tools  to  improve  understandings  of superconducting qubits that is essential for designing high-fidelity devices. 
	\end{abstract}

\maketitle

%\ryocom{Where should I put the explanation of how we determined numerically qubit $T_1$-decay rate?}

% \ryocom{Should we sell this work as (1) a practically useful theory for superconducting circuits, or (2) a quantum optics theory that discovered an interesting mechanism that modify the dissipation processes? I think (1) would make more sense (and I will start writing this way) but have a concern that, in the coherent driving case, our $T_1$-decay rate (where we drop the correlated noise term) always gives us a decreasing $T_1$-decay rate, in disagreement with experiments.}

%%%%%%%%%%%%%%%%%%
%%%%%%%%%%%%%%%%%%%%%%%%%%%%%%%%%%%%%
\section{Introduction}
Circuit quantum electrodynamics (cQED) systems based on superconducting circuits \cite{Blais2004,BlaisReview2020} are
a leading platform for quantum information processing \cite{BlaisNature2020}, and for explorations of basic quantum-optical and many-body phenomena
\cite{Houck2012,CarusottoNature2020}. 
The study of quantum dissipation in these systems is also of crucial interest 
(see e.g.~\cite{Boissonneault2009,Wilson2010,Beaudoin2011,Sete2014,Houck2017,Malekakhlagh2020,Petrescu2020}).
%, both for fundamental reasons, but also as a necessary step towards the enhanced coherence and gate fidelities required for improved 
%quantum processors.  
In many respects, the physics of cQED systems parallel that of atomic cavity QED systems.  cQED systems incorporate nonlinear Josephson junction circuits that mimic artificial atoms, and linear microwave cavities that mimic photonic cavities. 
A paradigmatic dissipative effect in cavity QED is Purcell decay~\cite{Purcell1946}, the modification of atomic decay by a cavity.  cQED systems motivate studying a  modified version of this effect:  what happens to Purcell decay when the cavity is now populated with photons (either by coherent driving or thermal noise)?  This is of crucial relevance to understanding the experimentally-observed excess qubit decay during dispersive 
measurement~\cite{Sank2016,Minev2019}
%~\cite{Johnson2012,Slichter2012,Sank2016,Minev2019}, 
as well as the effect of background thermal radiation on qubit coherence.

Surprisingly, a full understanding of how a photonic population impacts Purcell decay (in a form relevant to cQED) is currently lacking.  Ref.~\cite{Sete2014} analyzed a driven Jaynes-Cumming (JC) model (i.e.~a two-level qubit), finding that populating a cavity increases the qubit $T_1$-decay time (see also \cite{Boissonneault2009}).  A similar trend was found in \cite{Slichter2016}, which used a closely related
Golden-Rule calculation to study a multi-level transmon qubit.  
% These results do not directly apply to cQED systems, which typically use weakly-anharmonic, multi-level systems. 
However, in a more recent work in Refs.~\cite{Malekakhlagh2020,Petrescu2020} that extended blackbox quantization theory \cite{Nigg2012} to describe transmon-cavity systems, an opposite-signed effect, i.e.~decreasing $T_1$ with increasing cavity photon number, was numerically suggested.
% More recent work in Refs.~\cite{Malekakhlagh2020,Petrescu2020} extended blackbox quantization theory \cite{Nigg2012} to describe transmon-cavity systems. Their numerics suggested an opposite-signed effect, i.e.~$T_1$ appeared to decrease with increasing cavity photon number.
Unfortunately, the complexity of the method did not lend itself to simple analytic expressions nor to an intuitive picture of the underlying physics.

%%%%%%%%%%%%%%%%%%%%%%%%%%%%%%%%%%%%%
    %%%%%%%%%%%
	% FIG 1 Coherent drive 
	%%%%%%%%%%%
	\begin{figure}[t]
	\centering
    \includegraphics[width=0.8\linewidth,keepaspectratio]{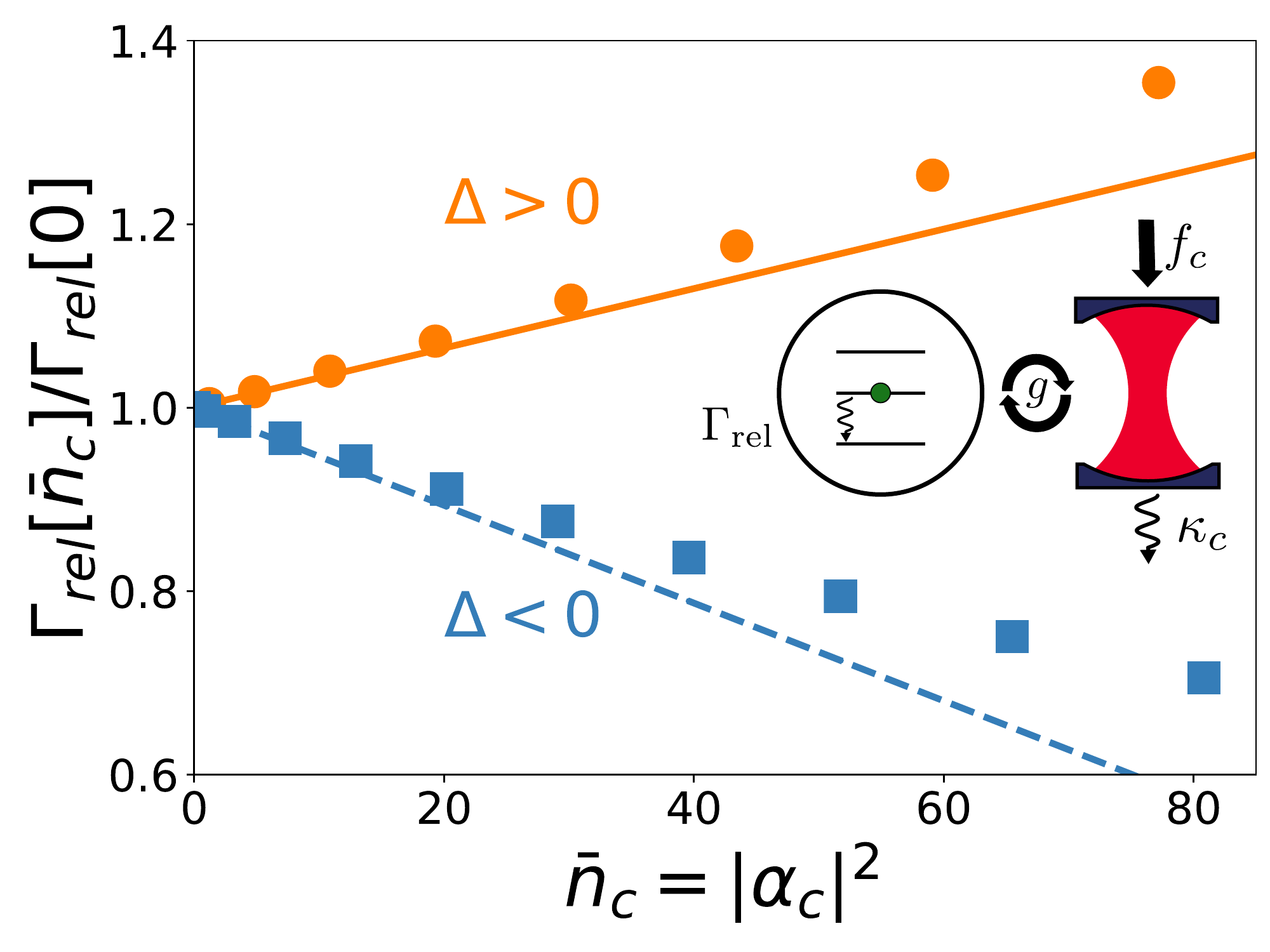}
	\caption{
	Inset: Schematic of a cavity coupled to a weakly anharmonic qubit.  Main:  Qubit $T_1$-decay rate $\Gamma_{\rm rel}$ as a function of drive-induced cavity photon number $\bar{n}_c = |\alpha_c|^2$.
	Orange indicates results for a positive qubit-cavity detuning $\Delta = \omega_a - \omega_c > 0$, blue for a negative detuning.  Solid symbols are master equation numerics, solid lines correspond to our analytic result (Eq.~\eqref{eq:Gamma_rel coherent drive}).  One sees a striking dependence on the sign of $\Delta$.    Parameters correspond to a qubit cavity coupling $g = 0.1 |\Delta|$, qubit nonlinearity $U = 0.1 |\Delta|$, cavity damping rate $\kappa_c = 0.01 |\Delta|$ and drive frequency $\omega_D = \omega_c - 0.1 |\Delta|$.  
	Setting $|\Delta|/(2\pi)=1{\rm GHz}$, the above parameters correspond to $U/(2\pi)=g/(2\pi)=100{\rm MHz}$ and $\kappa_c/(2\pi)=10{\rm MHz}$.
	All baths are at zero temperature, and we assume that qubit decay is only due to Purcell effects (i.e. $\kappa_a = 0$). 
    % 	under coherent cavity drive.  
    % 	Taking $\Delta=\omega_a-\omega_c$, we set $\omega_c-\omega_{\rm D} = 0.1|\Delta|, g=0.1|\Delta|,U=0.1|\Delta|,\kappa_c=0.01|\Delta|,\kappa_a=0$ at zero temperature $\bar n_c=\bar n_a=0$.
    % 	The lines are our analytical result (Eq.~\eqref{eq:Gamma_rel coherent drive}) and the points are our numerical result.
    }
	\label{fig:coherent drive}
	\end{figure}
    %%%%%%%%%%%
%%%%%%%%%%%%%%%%%%%%%%%%%%%%%%%%%%%%%

In this paper, we describe a new theoretical approach to understanding Purcell decay in transmon-cavity systems in the presence of driving that complements and extends previous studies. Our approach combines  insights from Keldysh theory with Lindblad 
perturbation theory (see e.g.~\cite{Li2014,LiPRX2016}). 
It provides compact {\it analytic} expressions that could be easily compared against experiment, and also facilitates simple intuitive explanations.  It also reveals several new surprising effects not previously discussed.  In particular, we show that whether or not $T_1$ increases or decreases with cavity drive is crucially dependent on the {\it sign} of the cavity-qubit frequency detuning (Figs.~\ref{fig:coherent drive} and \ref{fig:thermal drive}).  We also analyze the impact of thermal cavity photons, and show that the basic physics in this case is strikingly different from the coherent-drive case.  In the thermal case, the unexpected interplay between a non-resonant dissipative process and a non-resonant Hamiltonian process yields the dominant contribution.  We discuss how this process would be completely missed if one resorted to standard secular approximations or considered a JC model instead of the transmon-cavity model analyzed here. 
Our work reveals new understanding into the basic quantum dissipative mechanisms of driven circuit QED systems.  It also outlines a new kind of analytic approach that could be useful in studying a host of driven-dissipative systems.  

\section{Transmon model and theoretical framework}
We consider a standard setup where a multi-level transmon-style superconducting qubit is coupled to a linear microwave resonator, with each system subject to dissipation. 
We first consider the case with no coherent driving; this will be analyzed later.
% We model the superconducting circuit as a weakly nonlinear oscillator coupled to a linear cavity. 
% We first consider the case where there is only incoherent thermal pumping and add a coherent drive to our system later. 
% It turns out that thermal occupancy gives rise to a larger impact to Purcell decay rate than the coherent cavity drive case, when comparing the same photon number occupancy in the cavity. (Compare Figs.~\ref{fig:coherent drive} and \ref{fig:numerics vs analytics}(a) that uses the same parameters.)
The total Hamiltonian is $\hH = \hH_{\rm s} + \hH_{\rm diss}$, with $\hH_{\rm s}$ describing the isolated qubit and cavity, and $\hH_{\rm diss}$ the dissipative environment and its coupling to the system. We will focus on regimes where the qubit can be treated as an anharmonic (Kerr) oscillator, hence
% (which is the same as Refs.\cite{Malekakhlagh2020,Petrescu2020}), 
% \alexcom{removed essentially, since it is the same to leader order and taking into account the RWA}, 
\begin{equation}\label{eq:Bare_Hamiltonian}
    \hH_{\rm s} = 
    \omega_a\hat a_0^\dagger\hat a_0
    + g(\hat a_0^\dagger \hat c_0 + {\rm h.c.})
    + \omega_c\hat c_0^\dagger \hat c_0
    - \frac{U}{2}\hat a_0^\dagger\hat a_0^\dagger \hat a_0 \hat a_0.
\end{equation}
%describes the superconducting circuit. 
Here, $\hat a_0$ and $\hat c_0$ are bosonic annihilation operators describing (bare) qubit and cavity excitations, with $\omega_a$ and $\omega_c$ their resonant frequencies. The qubit-cavity coupling is denoted by $g$, while $U > 0$ is the Kerr nonlinearity of the qubit (obtained by expanding the full Josephson junction cosine potential \cite{Koch2007}).
We will be interested throughout in the typical regime where the cavity-qubit detuning $\Delta = \omega_a-\omega_c$ may be comparable in magnitude to $U$, but where $U \ll \omega_a,\omega_c$. 
We also focus on modest drives and temperatures; together, this implies that additional nonlinear terms play no significant role \footnote{In the regime of interest, there are many localized-in-phase energy levels associated with the transmon, and the instability physics studied in Ref.~\cite{Verney2019} plays no role.}.

We will further focus on the standard dispersive regime of cQED, where  $|g/\Delta|$ is small, but not so small that leading $(g/\Delta)^2$ corrections can be ignored. 
%Given this (and the fact that $U\lesssim|\Delta|$), 
We will work in the so-called ``blackbox" basis \cite{Nigg2012}, and thus first
diagonalize the quadratic part of $\hat{H}_{\rm s}$: 
%performing a transformation that diagonalizes the quadratic part of the Hamiltonian 
$\hH_0=\omega_c\hat c_0^\dagger \hat c_0 + g(\hat a_0^\dagger \hat c_0 + {\rm h.c.})+\omega_a\hat a_0^\dagger\hat a_0
= \tilde\omega_c\hat c^\dagger\hat c
     +\tilde\omega_a\hat a^\dagger\hat a$. 
% $\hH_0=\omega_c\hat c_0^\dagger \hat c_0 + g(\hat a_0^\dagger \hat c_0 + {\rm h.c.})+\omega_a\hat a_0^\dagger\hat a_0$ as 
% \begin{eqnarray}
%     \hH_0 
%     = \tilde\omega_c\hat c^\dagger\hat c
%     +\tilde\omega_a\hat a^\dagger\hat a,
% \end{eqnarray}
Here, dressed cavity and qubit ``polariton'' operators are given by $\hat c \simeq [1-g^2/(2\Delta^2)]\hat c_0 - (g/\Delta) \hat a_0 %+{\mathcal O}(g^3/\Delta^3)
$ and $\hat a \simeq [1-g^2/(2\Delta^2)]\hat a_0 + (g/\Delta) \hat c_0
%+{\mathcal O}(g^3/\Delta^3)
$, respectively. The corresponding renormalized frequencies are
$\tilde\omega_a \simeq \omega_a + g^2/\Delta 
%+ {\mathcal O}(g^4/\Delta^3)
$ and $\tilde\omega_c \simeq \omega_c -g^2/\Delta %+ {\mathcal O}(g^4/\Delta^3)
$. 
%The baths operators trasforms accordingly as $(\hat b_c, \hat b_a)^{\mathsf T} 
%= \bm V (\hat B_c, \hat B_a)^{\mathsf T}$.
%To order $g^2/\Delta^2$ 
The Kerr nonlinearity takes the form
$\hH_{\rm int}=-(U/2)\hat a_0^\dagger \hat a_0^\dagger\hat a_0 \hat a_0
\simeq \hH_{\rm int}^{\rm slf}+\hH_{\rm int}^{\rm crs}
+\hH_{\rm int}^{\rm nc} + \hat{V}_{\rm int}$. 
% $\hH_{\rm int}=-(U/2)\hat a_0^\dagger \hat a_0^\dagger\hat a_0 \hat a_0
% \simeq \hH_{\rm int}^{\rm slf}+\hH_{\rm int}^{\rm crs}
% +\hH_{\rm int}^{\rm nc}$. 
Here, the first two terms are  
usual self and cross-Kerr nonlinearities
%a sum of secular terms (direct and cross-Kerr-nonlinearity)
\begin{eqnarray}
\label{eq:secular nonlinearity}
    \hH_{\rm int}^{\rm slf}
    &=& 
    %-\frac{U}{2}
    %\big(1-\frac{g^2}{2\Delta^2}\big)
    \chi_{aa}
    \hat a^\dagger \hat a^\dagger \hat a \hat a,
    \qquad
    %+O\big(\frac{g^4U}{\Delta^4}\big), 
    %\\
    \hH_{\rm int}^{\rm crs}
    =
    %-\frac{2g^2U}{\Delta^2}
    \chi_{ca}
    \hat c^\dagger\hat c 
    \hat a^\dagger\hat a
    %+O\big(\frac{g^4U}{\Delta^4}\big)
    % \hH_{\rm int}^{\rm sec}
    % &=& -\frac{U}{2}
    % \big(1-\frac{g^2}{2\Delta^2}\big)
    % \hat a^\dagger \hat a^\dagger \hat a \hat a
    % -\frac{2g^2U}{\Delta^2}
    % \hat c^\dagger\hat c 
    % \hat a^\dagger\hat a
    % +O\big(\frac{g^4U}{\Delta^4}\big)
    % \nonumber\\
    % &=& -\chi_{aa}
    % \hat a^\dagger \hat a^\dagger \hat a \hat a
    % -\chi_{ca}
    % \hat c^\dagger\hat c 
    % \hat a^\dagger\hat a
    % +O\big(\frac{g^4U}{\Delta^4}\big)
\end{eqnarray}
with 
$\chi_{aa}=-(U/2)
    [1-g^2/(2\Delta^2)]$
and
$\chi_{ca}=-2g^2U/\Delta^2$.
%that conserves the number of excitations in respective species 
In contrast
\begin{equation}
\label{eq:nonlinear conversion}
    \hH_{\rm int}^{\rm nc}
    = \tilde \chi
    (\hat a^\dagger\hat a^\dagger\hat a\hat c
    +\hat c^\dagger\hat a^\dagger\hat a\hat a)
    %+O\big(\frac{g^2U}{\Delta^2}\big)
\end{equation}
(with $\tilde \chi = g U / \Delta$) describes a nonlinear 
process in which a cavity photon is converted into a qubit excitation (or vice-versa) with an amplitude that depends on qubit excitation number.
While this process is non-resonant, it will play a crucial role in mediating photon-number dependent dissipative effects.
Finally, the last interaction term $\hat{V}_{\rm int}$ contains non-resonant terms of order $(g/\Delta)^3$ or higher, and will play no role in what follows; we thus set it to zero.  
%Although often ignored, the latter (Eq.~\eqref{eq:nonlinear conversion}) is one of the key players in our analysis \cite{Malekakhlagh2020}. 
 We will often refer to the cavity/qubit polariton modes $\hc$ and $\ha$ as simply the `cavity/qubit modes', while $\hc_0$ and $\ha_0$ will be called the `bare modes'. 

We now turn to the modelling of dissipation. As is standard, we take the bare qubit and cavity to each be coupled linearly to independent, Markovian bosonic reservoirs \cite{Blais2004,BlaisReview2020} (though certain extensions to non-Markovian cavity baths are discussed below).
Using a Keldysh approach, one can integrate out these reservoirs and derive a formally exact dissipative action describing the system.  As shown in the Appendix \ref{SM:Keldysh_Section}, in the small dissipation limit of interest, this action is equivalent to the following Lindblad master equation:  
\begin{eqnarray}
\label{eq:Lindblad master equation bare}
&&\partial_t\hat\rho 
=
-i[\hH_{\rm s},\hat\rho]
\nonumber\\
&&+\sum_{\mu=a,c}
\kappa_\mu \Big[
(1+\bar n_\mu^0)
{\mathcal D}[\hat d_\mu^0]\hat\rho
+ \bar n_\mu^0
{\mathcal D}[\hat  d_\mu^{0\dagger}]\hat\rho
\Big] 
\equiv {\mathcal L}\hat\rho,
\end{eqnarray}
where $\hat d_c^0=\hat c_0$, $\hat d_a^0=\hat a_0$, and $\kappa_\mu$ ($\bar n_\mu^0$) are the decay rates (thermal occupancies) of the \emph{bare} cavity and qubit environments.
We also take ${\mathcal D}[\hat L]\rho = \hat L\rho\hat L^\dagger - \frac{1}{2}\{\hat L^\dagger \hat L, \hrho\} $ as the usual Lindblad dissipator.
% \ACnew{We stress that despite the nonlinearity in $\hat{H}_{\rm s}$, this master equation is exact in the case where the intrinsic qubit and cavity baths are each Markovian.} 
% It also provides an excellent description in the physically-relevant case where the cavity (qubit) bath density of states is relatively flat in the vicinity of $\omega_c$ ($\omega_a$).
In what follows, we will focus attention on the experimentally relevant regime where the cavity damping rate is much larger than the intrinsic qubit decay rate, $\kappa_a\ll \kappa_c$.
As our focus is on describing Purcell decay, we do not include an intrinsic qubit dephasing dissipator \footnote{Our focus here is on Purcell decay: understanding how cavity dissipation gives rise to qubit relaxation.  We note that high-frequency qubit dephasing noise can also lead to photon-number dependent qubit population decay.  This dressed-dephasing mechanism \cite{Boissonneault2008,Slichter2012} is expected to be suppressed in systems where dephasing is weak and primarily due to low-frequency noise.}.

Note that Eq.~\eqref{eq:Lindblad master equation bare}
describes a {\it dissipative} coupling between polariton modes.  To see this explicitly, we transform it to the blackbox basis, where it takes the form
\begin{eqnarray}
    \label{eq:Linblad master equation}
    \partial_t\hat\rho 
    = {\mathcal L}\hat\rho
    = {\mathcal L}_{\rm ind}\hat\rho+{\mathcal L}_{\rm cd}\hat\rho.
\end{eqnarray}
The Liouvillian $\mathcal{L}_{\rm ind}$ describes a model where each polariton is coupled to independent effective reservoirs; letting $\hat d_c=\hat c$ and $\hat d_a=\hat a$, we have
\begin{eqnarray}
\label{eq:Lindblad blackbox}
    && {\mathcal L}_{\rm ind}\hat\rho  
    % \partial_t\hat\rho 
    = -i[\hH_{\rm s},\hat\rho]
    \nonumber\\
    &&+\sum_{\mu=a,c}
    \tilde\kappa_\mu \Big[
    (1+\tilde n_\mu)
    {\mathcal D}[\hat d_\mu]\hat\rho
    + \tilde n_\mu
    {\mathcal D}[\hat d_\mu^\dagger]\hat\rho
    \Big].
  % \equiv{\mathcal L}_{\rm ind}\hat\rho,
\end{eqnarray}
The damping rates and thermal occupancies corresponding to the effective qubit-polariton bath are given by:
% where for each bath, $\tilde\kappa_\mu$ is the decay rate and $\tilde n_\mu$ the thermal occupancy.  We also define 
% by relating the bare and renormalized parameters as
\begin{eqnarray}
    \label{eq:kappa_tilde}
    \tilde\kappa_a
    &=&\kappa_a 
    +\frac{g^2}{\Delta^2}
    (\kappa_c - \kappa_a)
    \equiv 
    \kappa_a + \kappa_{\rm P},
    \\
    \label{eq:n_tilde}
    \tilde n_a 
    &=&\frac{1}{\tilde\kappa_a}
    \left[\kappa_a\bar n_a^0
    +\frac{g^2}{\Delta^2}
    (\kappa_c
    \bar n_c^0
    -\kappa_a\bar n_a^0) \right].
\end{eqnarray}
The cavity-polariton bath parameters $\tilde\kappa_c$ and $\tilde n_c$ are given by analogous expressions (one simply exchanges $c$ and $a$).
%in Eqs.~\eqref{eq:kappa_tilde} and \eqref{eq:n_tilde}).  
Note that the
qubit-polariton decay rate in Eq.~\eqref{eq:kappa_tilde} is simply the sum of the intrinsic qubit decay rate $\kappa_a$ and standard (zero temperature) Purcell decay rate $\kappa_{\rm P}$.  
%Note that, when $\kappa_a\ll (g^2/\Delta^2)(\kappa_c-\kappa_a)$ (as in most experiments), $\tilde n_a \approx \tilde n_c \approx \bar n_c^0$, i.e., the qubit is occupied to the same level as the cavity  when the bare cavity is thermally driven. 

Eq.~\eqref{eq:Lindblad blackbox} also has a term  $\mathcal{L}_{\rm cd}$ describing \emph{correlated dissipation} that provides a dissipative coupling of qubit and cavity polaritons:
% , we find that the full treatment of the baths (Eq.~\eqref{eq:Linblad master equation}) generates additional, \emph{correlated dissipation} between the two modes,
\begin{eqnarray}
    \label{eq:correlated dissipation}
    {\mathcal L}_{\rm cd}\hat\rho
    &=&-\frac{1}{2}(\tilde\gamma_{\uparrow}
    +\tilde\gamma_{\downarrow})
    \{\hat a^\dagger \hat c+\hat c^\dagger\hat a,\hat\rho\}
    \nonumber\\
    &+&\tilde\gamma_\downarrow
    (\hat a\hat\rho\hat c^\dagger + \hat c\hat\rho\hat a^\dagger)
    +\tilde\gamma_\uparrow
    (\hat a^\dagger\hat\rho\hat c + \hat c^\dagger\hat\rho\hat a),
\end{eqnarray}
where 
\begin{align}
    \tilde\gamma_\downarrow &= 
        (g/\Delta) [\kappa_c (1+\bar n_c^0) - \kappa_a(1+\bar n_a^0)] \\
    \tilde\gamma_\uparrow &=
    (g/\Delta)[\kappa_c\bar n_c^0 - \kappa_a\bar n_a^0].  
\end{align}
% $\tilde\gamma_\downarrow = (g/\Delta) [\kappa_c (1+\bar n_c^0) - \kappa_a(1+\bar n_a^0)]$ and $\tilde\gamma_\uparrow =(g/\Delta)[\kappa_c\bar n_c^0 - \kappa_a\bar n_a^0]$.  
The first line in Eq.~(\ref{eq:correlated dissipation}) describes an effective non-Hermitian beam-splitter coupling between polaritons, whereas the last line
describes correlated noise.
% can be interpreted as describing correlated noise driving both modes.  

We stress again that the correlated polariton dissipation described by Eq.~(\ref{eq:correlated dissipation}) follows from our exact treatment.  Nonetheless, it is common at this point to simply omit ${\mathcal L}_{\rm cd}$.  This corresponds to a standard secular approximation:  as $\mathcal{L}_{\rm cd}$ describes non-resonant processes (detuning $\sim \Delta$), and as $|\Delta| \gg \tilde{\gamma}_\uparrow,\tilde{\gamma}_\uparrow$, 
${\mathcal L}_{\rm cd}$ is expected to have a marginal effect.  We will not make this approximation in what follows \footnote{Note that Refs.~\cite{Malekakhlagh2020,Petrescu2020} 
 use an alternate approach that also avoids making unjustified secular approximations.  Their approach captures some but not all aspects of our correlated-dissipation master equation (whose form follows from an exact derivation).}.
  % partially takes the correlated dissipation into account, but has vanishing effect as $U\rightarrow 0$. This is in contrast to our correlated dissipation (Eq.~\eqref{eq:correlated dissipation}) derived exactly, where they are present irrespective of the magnitude of $U$.}}
Surprisingly, we show that in the case of a thermal cavity population, the correlated dissipation described by $\mathcal{L}_{\rm cd}$ provides the dominant temperature-dependent correction to the qubit Purcell decay rate.  

\section{Qubit decay rate in the presence of temperature}
We can now examine how qubit dissipation is modified by populating the cavity.  The general picture is that non-resonant coupling processes (both 
Hamiltonian and dissipative) will alter the Purcell contribution to the qubit $T_1$ population decay rate $\Gamma_{\rm rel}$.  Our approach will be to treat these processes systematically using Lindblad perturbation theory (see Appendix \ref{SM:Lindblad perturbation theory} for a short review).
To this end, we write our full Liouvillian as 
$\mathcal{L} = \mathcal{L}_0 + \mathcal{L}_1$, 
where $\mathcal{L}_0$ describes 
{\it all processes which do not couple qubit and cavity polaritons}.
%(i.e.~set $\chi_{ca}=\tilde\chi=\tilde\gamma_\uparrow=\tilde\gamma_\downarrow = 0$).
In contrast, $\mathcal{L}_1 = \mathcal{L} - \mathcal{L}_0$ describes both nonlinear and dissipative polariton-polariton coupling terms:
\begin{align}
	\mathcal{L}_1 \hrho
	=
	-i[\epsilon_{\rm crs} \hat{H}^{\rm crs}_{\rm int} 
	+ 
	\epsilon_{\rm ns} \hat{H}^{\rm nc}_{\rm int}, \hrho]
	+ 
	\epsilon_{\rm nc} \mathcal{L}_{\rm cd }\hrho 
\end{align}
$\mathcal{L}_1$ will be treated perturbatively, as it scales as the small parameter $g/\Delta$. 
To make the physical origin of different contributions clear in what follows, we have introduced book-keeping constants  $\epsilon_{\rm crs}=\epsilon_{\rm ns}=\epsilon_{\rm cd}=1$.  
We stress that the qubit self-Kerr interaction $\hat{H}^{\rm slf}_{\rm int}$ in Eq.~\eqref{eq:secular nonlinearity} is included in $\mathcal{L}_0$. 
% Writing $\mathcal{L} = \mathcal{L}_0 + \mathcal{L}_1$, our starting unperturbed Liouvillian consists of the uncoupled qubit and cavity $\mathcal{L}_0 = \mathcal{L}_0^a + \mathcal{L}_0^c$. 

The first step is to identify the qubit $T_1$-decay mode to zeroth order in perturbation theory.  This can be done unambiguously, as $\mathcal{L}_0$ has a set of eigenmodes which {\it only} describe qubit population decay (in the Fock basis).  We identify the eigenvalue of the {\it slowest} of these eigenmodes as the qubit $T_1$-decay rate $\Gamma_{\rm rel}$; it dominates the relaxation of an initial qubit excited state.  We find \cite{Chaturvedi1991,Dykman1984,Honda2010} (see Appendix \ref{SM:Section_Thermal_Occupation}.)  
\begin{equation}
	\label{eq:T1 decay rate non-perturbative}
	\Gamma_{\rm rel}^{(0)}
	%    = -\lambda_{\rm rel}^{(0)}
	= \tilde\kappa_a
	=\kappa_a + \frac{g^2}{\Delta^2}
	(\kappa_c-\kappa_a).
\end{equation}
Note that this leading-order decay rate is independent of both the temperature $\bar{n}_\mu^0$ and the self-Kerr nonlinearity $\chi_{aa}$, as has been noted in other contexts (see e.g.~\cite{Scarlatella2019}). 

We next calculate the leading-order correction to $\Gamma_{\rm rel}$ arising from $\mathcal{L}_1$.  This amounts to perturbatively calculating the eigenvalue shift of the relevant Liouvilian eigenmode, which emerges at second order.
% , and again provides an unambiguous method for identifying the relaxation rate.  
% $\mathcal{L}_1$ describes both Hamiltonian and dissipative couplings between polaritons:
% \ACnew{As all of the above terms are non-resonant, the perturbation expansion is ultimately controlled by the small parameters $g / \Delta$ and $\tilde{\kappa}_j / \Delta$.} 
% \AC{Hmm, is this right?  What about cross-Kerr?}
% A straightforward but tedious calculation yields the desired correction.  
Focusing on the experimentally relevant regime of weak intrinsic qubit loss ($|\chi_{aa}|\sim U\gg\tilde\kappa_{a}$) and low temperature ($\bar n_c^0,\bar n_a^0 \ll 1$), 
a straightforward but tedious calculation yields~(See Appendix \ref{SM:Section_Thermal_Occupation} for derivation),
% we obtain~\cite{SM}
%the compact expression
\begin{eqnarray}
	\label{eq:Gamma_rel thermal}
	\Gamma_{\rm rel} 
	% 	\simeq \tilde\kappa_a 
	%     &+&\frac{\tilde\chi}{\Delta-U}[8\tilde\gamma_\downarrow\tilde n_a
	%     -4\tilde\gamma_\uparrow]
	%     \nonumber\\
	%     &+&\frac{\tilde \chi^2}{(\Delta-U)^2}
	% 	(\tilde\kappa_a + \tilde\kappa_c)
	% 	(4\tilde n_a-2\tilde n_c )
	% 	\nonumber\\
	\simeq \tilde\kappa_a 
	&+&
	\frac{g^2}{\Delta^2}\frac{\epsilon_{\rm cd}\epsilon_{\rm nc} U}{\Delta-U}
	\left[8(\kappa_c - \kappa_a) \tilde n_a
	-4(\kappa_c\bar n_c^0 - \kappa_a\bar n_a^0) \right]
	\nonumber\\
	&+&
	\frac{g^2}{\Delta^2}
	\frac{\epsilon_{\rm nc}^2U^2}{(\Delta-U)^2}
	(\tilde\kappa_a + \tilde\kappa_c)
	(4\tilde n_a - 2\tilde n_c )
% 	+\mathcal{O}\left[ (g/\Delta)^4 \right].
\end{eqnarray}
where all neglected terms are $\mathcal{O}[(g/\Delta)^4]$ or higher.
This is the first main result of this paper.
The second and third terms here describe temperature-dependent contributions to Purcell decay.  We find a surprising dependence both on bath temperatures, and on {\it the sign} of the qubit cavity detuning $\Delta$.
The third term in Eq.~(\ref{eq:Gamma_rel thermal})
$\propto \epsilon_{\rm nc}^2$ is solely due to the nonlinear conversion process in Eq.~(\ref{eq:nonlinear conversion}), and can be linked to Fermi's Golden rule rates involving the qubit $n=2$ state (See Appendix \ref{SM:Section_Thermal_Occupation}).  More interesting is the second term ($\propto \epsilon_{\rm cd}
\epsilon_{\rm nc}$), which dominates the third term in the usual limit where $\Delta \gesim U$.  This process results from a subtle interplay between the Hamiltonian nonlinear conversion interaction, and the dissipative polariton coupling described by $\mathcal{L}_{\rm cd}$.  
Note that both correction terms vanish at zero temperature.

The surprising interplay of coherent and dissipative conversion processes in determining qubit relaxation can be understood intuitively.  Consider a model of two linear classical oscillators whose amplitudes $\beta_a,\beta_c$ obey:
\begin{eqnarray}
\label{eq:coupled linear oscillator}
    i\partial_t
    \begin{pmatrix}
        \beta_c \\
        \beta_a
    \end{pmatrix}
    =
    \begin{pmatrix}
        \tilde\omega_c - i\tilde\kappa_c/2 & r - i \gamma
        \\
        r - i \gamma
        & \tilde\omega_a - i\tilde\kappa_a/2
    \end{pmatrix}
    \begin{pmatrix}
        \beta_c \\
        \beta_a
    \end{pmatrix}.
    \end{eqnarray}
This describes two modes with resonant frequencies $\tilde\omega_c,\tilde\omega_a$, decay rates $\tilde\kappa_c,\tilde\kappa_a$ that are coupled both coherently (rate $r$) and dissipatively (rate $\gamma$)
which roughly mimics the nonlinear conversion and correlated dissipation, respectively. 
%\gamma$ here mimics the correlated polariton dissipation in Eq.~\eqref{eq:correlated dissipation}, while $r$ mimics the coherent conversion process in Eq.~\eqref{eq:nonlinear conversion}.
% The same notation for the frequency and the decay rates are used to emphasize the parallel we are making here.
For weak couplings, the eigenmodes of the above dynamical matrix remain localized.  A simple diagonalization shows that the decay rate for the $a$-like mode is modified by the couplings as 
$\tilde\kappa_a \rightarrow 
    \tilde\kappa_a
    +2\gamma r/(\tilde{\omega}_a - \tilde{\omega}_c )
    + \mathcal{O}(r^2)$.
% $\tilde\kappa_a\rightarrow\tilde\kappa_a
%     +2\gamma(r/\Delta)
%     -(\tilde\kappa_c-\tilde\kappa_a)(r/|\Delta|)^2/2$.
We see that the dominant shift in the lifetime involves the {\it product} of dissipative and coherent couplings, in direct analogy to Eq.~(\ref{eq:Gamma_rel thermal}).  Of course, in our system nonlinearity modifies the form of the correction.  Still, the basic mechanism involving coherent and dissipative couplings working in consort is the same (as is the striking dependence on the sign of the detuning, reflecting an avoided crossing).

% Note how the term composed of product of the coherent and dissipative coupling $r\gamma$ provides the dominant correction when $r/|\Delta|\ll 1$, similarly to Eq.~\eqref{eq:Gamma_rel thermal}.
% As a result, this would increase or decrease Purcell decay depending on the sign of $\Delta$.
% This is analogous to what is seen in Eq.~\eqref{eq:Gamma_rel thermal}, except for the absence of extra energy shift from the self-Kerr nonlinearity $\chi_{aa}\simeq U$.
% This arises due to fact that the coupling from the nonlinear conversion depends on the state of the qubit, and in particular involves exciting to the $n=2$ Fock states of the qubit (unlike the simplified problem of Eq.~\eqref{eq:coupled linear oscillator} which involves excitations to $n=1$ Fock state). 
% This arises due to fact that the coupling from the nonlinear conversion is actually nonlinear, which involves the excitation from the $n=1$ to $n=2$ Fock state of the qubit that has the energy difference $E_{12}^a=\tilde\omega_a-U$ instead of $E_{01}^a=\tilde\omega_a$ between $n=0$ and $n=1$ involved in the linear problem.
% We note that, in our problem, the coupling is nonlinear, where the nonlinear conversion only turns on when the qubit and the cavity are thermally excited. Therefore, $r$ would effectively be proportional to the thermal occupation $\bar n_{c,a}^0$, giving correction to Purcell decay only at finite temperature.  

Another striking prediction of 
Eq.~(\ref{eq:Gamma_rel thermal}) is that the 
sign of the dominant temperature-dependent term is sensitive to the relative importance of Purcell decay to intrinsic qubit decay.  For $\kappa_P \gg \kappa_a$,  Eq.~\eqref{eq:n_tilde} tells us that $\tilde n_a\approx\tilde n_c\approx\bar n_c^0$, whereas for 
$\kappa_P \ll \kappa_a$ we have $\tilde n_a\ll \bar n_c^0$.  It follows that in these two limiting cases, 
% The sign of the correction to $\Gamma_{\rm rel}$ also depends on whether the qubit loss is dominated by the Purcell decay $\kappa_P \gg \kappa_a$ or by the intrinsic bare qubit decay $\kappa_a \ll \kappa_P$ even when the bare qubit bath is at zero temperature ($\bar n_c^0>0,\bar n_a^0=0$).
% In the former regime, Eq.~\eqref{eq:n_tilde} tells us that $\tilde n_a\approx\tilde n_c\approx\bar n_c^0$, while in the latter, it gives $\tilde n_a\ll \bar n_c^0$.
% Using these, 
Eq.~\eqref{eq:Gamma_rel thermal} can be approximated as
\begin{eqnarray}
    \Gamma_{\rm rel}
	\simeq \tilde\kappa_a 
    &\pm &\frac{4g^2}{\Delta^2}\frac{U}{\Delta-U}
    \kappa_c
    \bar n_c^0,
\end{eqnarray}
where $+$ ($-$) corresponds to $\kappa_{\rm P} \gg \kappa_a$ ($\kappa_{\rm P} \ll \kappa_a$).  We see that the impact of a cavity thermal population is opposite in these two regimes.
	\begin{figure}[t]
    \centering
    \includegraphics[width=1\linewidth,keepaspectratio]{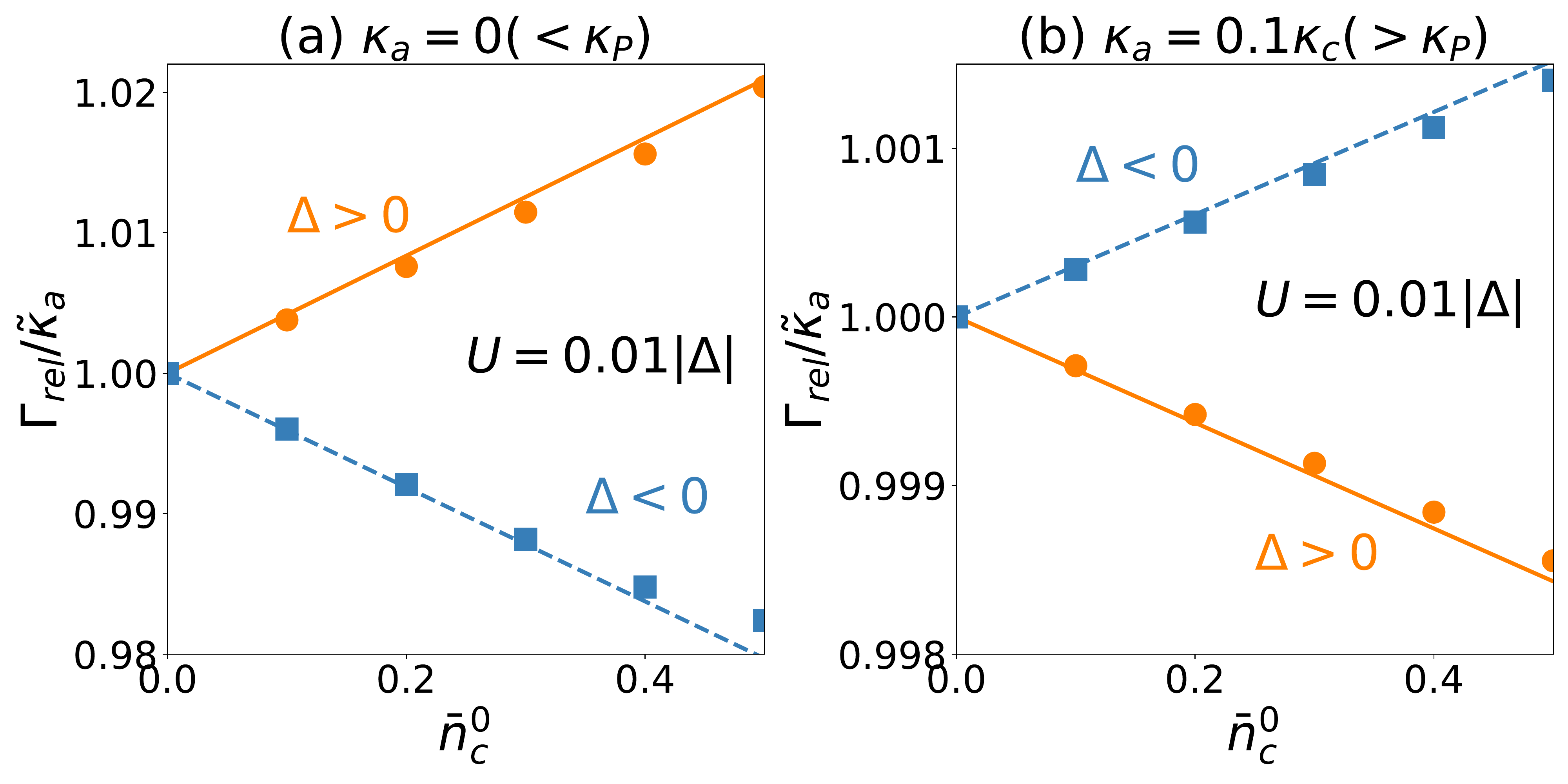}
	\caption{
	Qubit $T_1$-decay rate $\Gamma_{\rm rel}$ as a function of cavity thermal population $\bar n_c^0$.  
	The parameters are  $g=0.1|\Delta|,U=0.01|\Delta|,\kappa_c=0.01|\Delta|, \bar n_a = 0$, with the bare qubit decay rate 
	(a) $\kappa_a=0(<\kappa_{\rm P}\equiv(g^2/\Delta^2)(\kappa_c-\kappa_a))$ and (b) $\kappa_a=0.1\kappa_c  (>\kappa_{\rm P})$.
	The lines and points are our analytical (Eq.~\eqref{eq:Gamma_rel thermal}) and numerical results, respectively.  One sees that the sign of the temperature dependence depends both on detuning $\Delta$ and on the ratio of the intrinsic qubit decay rate and the Purcell decay rate $\kappa_a / \kappa_{\rm P}$.   
% 	This applies also to Fig.~\ref{fig:numerics vs analytics}(a).
	}
	\label{fig:thermal drive}
	\end{figure}
    %%%%%%%%%%%

% \alexcom{Should we mention here or in the SM how we get the T1 modes numerically?}

\begin{figure}[t]
	\centering
    \includegraphics[width=1\linewidth,keepaspectratio]{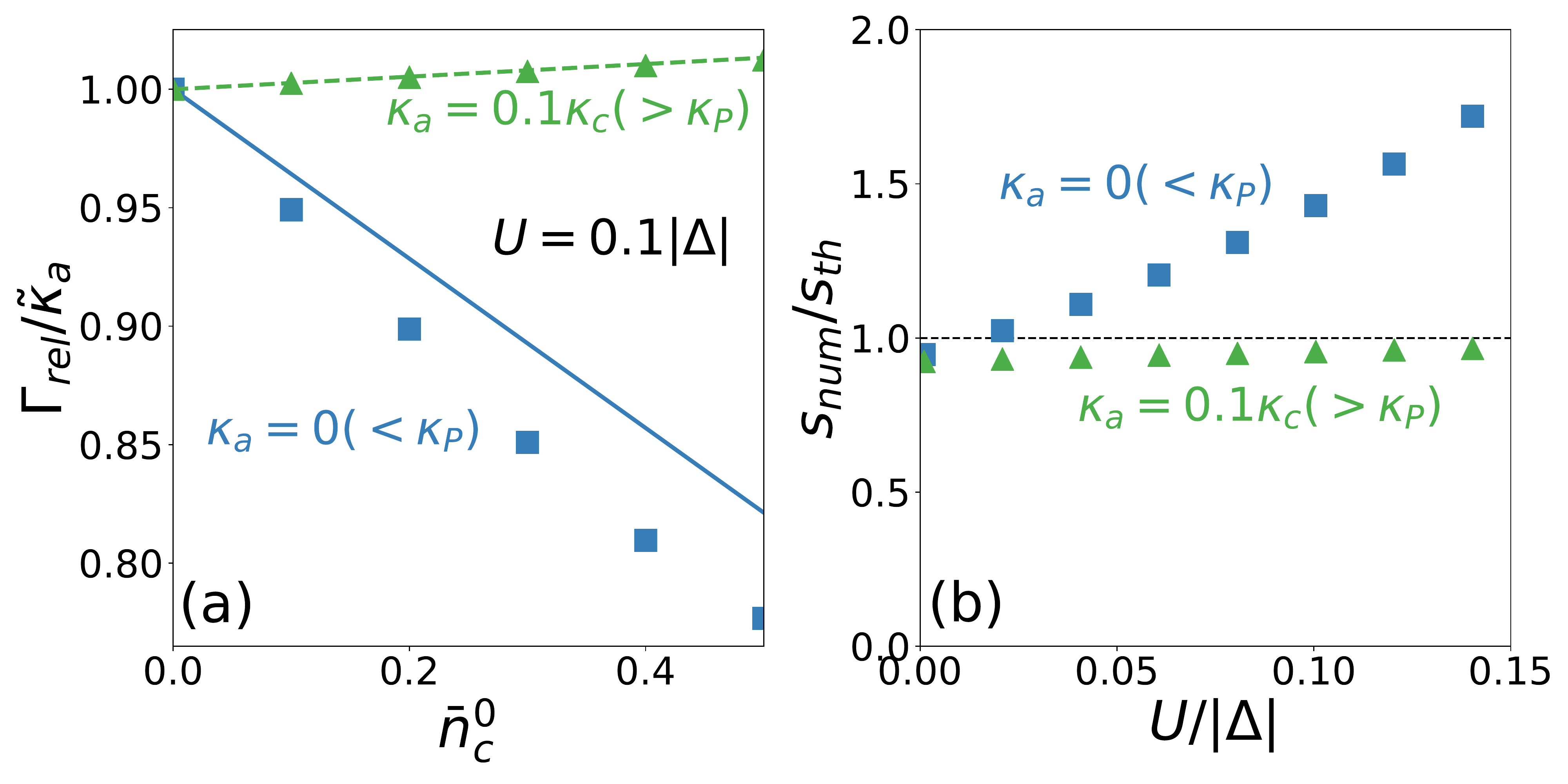}
	\caption{
	(a) Thermal cavity drive dependence on the qubit $T_1$-decay rate $\Gamma_{\rm rel}$ for a larger nonlinearity $U=0.1|\Delta|$.
	Lines (points) correspond to analytical (numerical) results.
	Higher order corrections are more important here, but the analytic results still give good qualitative agreement.
	(b) The ratio of the numerically ($s_{\rm num}$) to analytically ($s_{\rm th}$) evaluated slope parameter $s = d \Gamma_{\rm rel} / d \bar{n}_c^0$ (evaluated at zero temperature). 
	For both the panels, we set $g=0.1|\Delta|,\kappa_c = 0.01|\Delta|$ and $\Delta<0$.
    For large $U$ and $\kappa_{\rm P} > \kappa_a$, higher order corrections become important (as discussed in Appendix \ref{SM:Section_Thermal_Occupation}), but our analytical expression still provides a reasonable qualitative agreement. 
	}
	\label{fig:numerics vs analytics}
	\end{figure}

The results of Eq.~(\ref{eq:Gamma_rel thermal}) are numerically confirmed in Fig.~\ref{fig:thermal drive}, where we compare against a direct master equation simulation of Eq.~\eqref{eq:Linblad master equation}~\cite{QUTIP,QUTIP2} using experimentally-relevant parameters
\cite{BlaisNature2020}.
The numerical $T_1$-decay rate corresponds to the time-dependent decay of an initial state where an excitation is added to the qubit (See Appendix \ref{sec:numerical simulation} for details).  Our analytic, perturbative expressions quantitatively agree the numerical results at low temperatures and small-to-modest nonlinearity.  
%Key qualitative features are also evident, e.g. the crucial dependence on the sign of detuning $\Delta$.    
% As seen in the figure, we obtain a quantitative agreement at low temperature $\bar n_c^0,\bar n_a^0\ll 1$ at small nonlinearity $U\ll |\Delta|$, where the sign change of the slope when changing the detuning is reproduced.
The qualitative agreement at larger nonlinearity $U= 0.1|\Delta|$ is also reasonable (see Fig.~\ref{fig:numerics vs analytics}), though here, higher order contributions become important, especially if $\kappa_a<\kappa_{\rm P}$. (See Appendix \ref{SM:Section_Thermal_Occupation} for a detailed discussion on this point.)
% Even at a decently large nonlinearity $U= 0.1|\Delta|$, the agreement remains good
% except at $\kappa_a<\kappa_{\rm P}$ (See Fig.~\ref{fig:numerics vs analytics}(a)).
% This is quantified in Fig.~\ref{fig:numerics vs analytics}(b), where the slope $s=\partial\Gamma_{\rm rel}/\partial \bar n_c^0|_{\bar n_c,\bar n_a=0}$ obtained from Eq.~\eqref{eq:Gamma_rel thermal} ($s=s_{(11)}$) is compared against the numerical result ($s=s_{\rm num}$).
% Clearly, while the case $\kappa_a>\kappa_{\rm P}$ gives a good agreement, the case $\kappa_a<\kappa_{\rm P}$ deviates from the prediction of Eq.~\eqref{eq:Gamma_rel thermal}.
% We argue \textcolor{red}{in the SM} that this deviation is due to the missing higher-order correction that contains resonant processes that can become comparable to the third term when $\kappa_a\ll \kappa_{\rm P}$.
% We emphasize that even there, the order of magnitude and the sign of the slope is well captured by Eq.~\eqref{eq:Gamma_rel thermal}, where the most dominant contribution (i.e. the second term) is  incorporated. 

    %%%%%%%%%%%
	% FIG 3 Numeric  vs.Analytic 
	%%%%%%%%%%%
    %%%%%%%%%%%

% To understand this in a intuitive way, let us consider the following toy model. 
% ...
% This explains how the dissipation rate is sensitive to the sign of the detuning $\Delta$.

% The sign difference at different regimes of $¥kappa_a$ is attributed to the property that the effective pumping and damping gives rise to opposite effects.

% \ryocom{Add heuristic arguments here.}
% \begin{eqnarray}
%     \Gamma_{\rm rel}
%     \simeq
%     \tilde\kappa_a
%     +W_{21}
%     +W_{12}
% \end{eqnarray}

%\subsection{Limitation of Eq.~\eqref{eq:Gamma_rel thermal}}

% So far, we have analytically derived the second order corrections to the qubit $T_1$-decay rate $\Gamma_{\rm rel}$ in terms of the nonlinear conversion and correlated dissipation ${\mathcal L}_1$.
% As seen in Figs.~2 and 3 in the main text, the obtained formula (Eq.~\eqref{eq:Gamma_rel thermal} in the main text) gives an excellent agreement with our numerical simulation (which we provide details in Sec.~\ref{sec:numerical simulation}) in most regimes.
% However, we see a slight deviation when the nonlinearity is relatively large $U=0.1|\Delta|$ and is in the regime where the qubit decay $\tilde\kappa_a$ is dominated by Purcell decay contribution  $\kappa_a\ll\kappa_{\rm P}$.
% Notably, the formula recovers its predictive power in the opposite regime $\kappa_a>\kappa_{\rm P}$, even with large nonlinearity $U=0.1|\Delta|$, see Fig.~3.

%%%%%%%%%%%%%%%%%%%%%%%%%%%%%%%%%%%%%%%%%%%%%%%%%%
%%%%%%%%%%%%%%%%%%%%%%%%%%%%%%%%%%%%%%%%%%%%%%%%%%
%%%%%%%%%%%%%%%%%%%%%%%%%%%%%%%%%%%%%%%%%%%%%%%%%%
\section{Qubit decay rate in the presence of coherent drive} 
Consider now a coherent linear driving of the cavity, as described by the additional system Hamiltonian term 
$\hat H_{\rm D} =
    f_c e^{-i\omega_{\rm D}t}\hat c_0 + {\rm h.c.}$
We move to a rotating frame at the drive frequency $\omega_{\rm D}$, which effectively shifts $\omega_\mu\rightarrow\omega_\mu - \omega_{\rm D}$.  We further make a standard displacement transformation of both modes:
${\mathcal L}'
=\hat D^\dagger[\alpha_c,\alpha_a]
{\mathcal L}
\hat D[\alpha_c,\alpha_a]$, where 
$\hat D[\alpha_c,\alpha_a]$ is a displacement operator that displaces the bare cavity (qubit) operator $\hc_0(\ha_0)$ by the time-independent amplitude $\alpha_c(\alpha_a)$.
%with $\hat D^\dagger[\alpha_c,\alpha_a]\hat c_0\hat D[\alpha_c,\alpha_a]=\hat c_0 + \alpha_c$ and $\hat D^\dagger[\alpha_c,\alpha_a]\hat a_0 \hat D[\alpha_c,\alpha_a]=\hat a_0 + \alpha_a$. 
By choosing displacements to cancel linear terms, the displaced Lindbladian ${\mathcal L}'$ has the same dissipative terms as ${\mathcal L}$ in  Eq.~\eqref{eq:Lindblad master equation bare}, but a modified system Hamiltonian 
$\hH_{\rm s}'=\hH_0+\hH_{\rm int}'=\hH_0+\hat D^\dagger[\alpha_c,\alpha_a]\hH_{\rm int}\hat D[\alpha_c,\alpha_a]$.  For weak drives with small induced amplitudes
$|\alpha_a|^2\ll 1$, it is sufficient to only keep terms in $\hH_{\rm int}'$ that are at most $\mathcal{O}[\alpha_a^2]$:
% has an form identical to ${\mathcal L}$ in  Eq.~\eqref{eq:Lindblad master equation bare} {\it except} for the replacement of the Hamiltonian 
% $\hH_{\rm s}\rightarrow \hH_{\rm s}'=\hH_0+\hH_{\rm int}'=\hH_0+\hat D^\dagger[\alpha_c,\alpha_a]\hH_{\rm int}\hat D[\alpha_c,\alpha_a]$ with
\begin{align}
    \label{eq:HintDisplaced}
    \hH_{\rm int}'
    & \simeq \hH_{\rm int}
    -U(\alpha_a \hat a_0^\dagger \hat a_0^\dagger \hat a_0 + {\rm h.c.})
    + \hH_{\rm quad} 
\end{align}    
The second term here describes an effective nonlinear single photon drive, whereas
\begin{align}
    \hH_{\rm quad} & = 
    -2U|\alpha_a|^2 \hat a_0^\dagger \hat a_0
    - \frac{U}{2} \left( \alpha_a^2 \hat{a}_0^\dagger \hat{a}_0^\dagger
    + {\rm h.c.} \right)
\end{align}
% \begin{eqnarray}
%     \label{eq:Hint'}
%     \hH_{\rm int}'
%     \simeq \hH_{\rm int}
%     -U(\alpha_a \hat a_0^\dagger \hat a_0^\dagger \hat a_0 + {\rm h.c.})
%     -2U|\alpha_a|^2 \hat a_0^\dagger \hat a_0.
% \end{eqnarray}
% for weak drive strength $|\alpha_a|^2\ll 1$. 
describes a mean-field frequency shift and weak squeezing drive. As these terms are quadratic, they can be accounted for exactly by defining our blackbox polaritons to be the eigenmodes of  
$\hH_0 + \hH_{\rm quad}$.  The squeezing terms will play no role in what follows, so we drop them.  The remaining frequency shift terms then lead to a modification of the qubit-cavity detuning $\Delta$:  $\Delta \rightarrow \Delta - 2U|\alpha_a|^2 \equiv
\Delta'[\alpha_a]$.  
% We then transform to a \emph{modified} blackbox basis that diagonalizes $\hat H_0'$ instead of $\hat H_0$.
% As this effectively gives rise to an energy shift of the bare qubit frequency $\omega_a'(\alpha_a)=\omega_a - 2U|\alpha_a|^2$ and therefore the detuning $\Delta'(\alpha_a)=\Delta - 2U|\alpha_a|^2$,
% this changes the nature of the hybridization of the bare modes to be either weaker or stronger.

The modification of the qubit polariton by the drive directly leads to a modification of $\tilde{\kappa}_a$, the intrinsic (linear-theory) qubit polariton damping rate:
%the detuning is effectively modified to and the one-body loss rate of the qubit to 
%As a result, the mixing of the two bare modes are increased (decreases), which makes the renormalized 
\begin{eqnarray}
    \label{eq:kappatilde Hartree}
    \tilde\kappa'_a[\alpha_a]
    & = &
    \kappa_a +\frac{g^2}{(\Delta'[\alpha_a])^2}
    (\kappa_c - \kappa_a) 
    \nonumber\\
    &\simeq &
    \kappa_a + \kappa_{\rm P}
    +\frac{g^2}{\Delta^2}\frac{4U}{\Delta}(\kappa_c-\kappa_a)
    |\alpha_a|^2,
\end{eqnarray}
where in the last line we have expanded to leading order in $|\alpha_a|^2$.  We see that the simple mean-field shift of the cavity frequency directly yields a change in the linear-theory polariton decay rate, one that is odd in $\Delta$.  This term simply reflects the modified qubit-cavity hybridization resulting from the drive-induced cavity frequency shift. 
The fact that driving a nonlinear system gives rise to a notion of drive dependent polaritons has been discussed in many different contexts 
(see e.g.~\cite{Lemonde2015}).

%Purcell effect is effectively increased (decreases),
% For $\Delta>0(<0)$, the magnitude of the effective detuning $|\Delta'(\alpha_a)|$ is decreased (increased) by the drive, making the mixing of the two bare modes stronger (weaker).
% As a result, Purcell decay of the qubit is effectively increased (decreased) by the drive when $\kappa_c>\kappa_a$, reflected in the second term of Eq.~\eqref{eq:kappatilde Hartree}.

To calculate the full modification of the qubit $T_1$-decay, we must also include the perturbative contribution of the $\mathcal{O}[\alpha_a]$ nonlinear drive term in Eq.~(\ref{eq:HintDisplaced}).  Again using Lindblad perturbation theory, as derived in Appendix \ref{SM:Section_Coherent_Drive}, we finally obtain (for zero temperature, and to order $\mathcal{O}[|\alpha_a|^2]$):
\begin{eqnarray}
    \label{eq:Gamma_rel coherent drive}
    \Gamma_{\rm rel} 
    % \simeq \tilde\kappa_a'(\alpha_a)
    % +\frac{U^2}{(\omega_a-\omega_{\rm D}-2U|\alpha_a|^2)^2}
    % (-2|\alpha_a|^2)
    % \nonumber\\
    &\simeq &
    %\tilde\kappa_a 
    %+\frac{g^2}{\Delta^2}\frac{4U}{\Delta}(\kappa_c-\kappa_a)
    %|\alpha_a|^2
    \tilde\kappa_a'[\alpha_a]
    -\frac{2U^2}{(\tilde\omega_a-\omega_{\rm D}-U)^2}\tilde\kappa_a
    |\alpha_a|^2.
\end{eqnarray}
This is the second main result of this paper.  Note this result is contingent on a perturbative treatment of 
Eq.~(\ref{eq:HintDisplaced}) being valid, which requires drive detuning 
$|\tilde\omega_a-\omega_{\rm D}-U| \gg \tilde\kappa_a$.
% $|\tilde\omega_a-\omega_{\rm D}-U| \ll U |\alpha_a|^2$.
% Here, the driving frequency is assumed to be at least slightly off resonance from the qubit and cavity modes such that $|\tilde\omega_\mu-\omega_{\rm D}|\gg \tilde\kappa_\mu$ is satisfied (as is needed for a perturbative treatment to be valid.
%The second term is the contribution from the Hartree shift and the third term is from the second-order perturbation of the second term in Eq.~\eqref{eq:Hint'}.
%The second term that originates from the Hartree shift 
For the typical experimental scenario where the bare cavity decay rate dominates that of the qubit (i.e.~$\kappa_a\ll\kappa_c$), the drive-dependence of the qubit $T_1$-decay rate is dominated by that of $\tilde{\kappa}_a'[\alpha_a]$.  The sign of the drive dependence thus exhibits a striking dependence on the sign of $\Delta$.  Our results are in excellent agreement with full master equation numerics, see Fig. \ref{fig:coherent drive}.  
The fact that driving the cavity can either increase or decrease qubit $T_1$ depending on $\Delta$ was not noted in previous work.
% The coherent photon number $|\alpha_a|^2$-dependence in $\tilde\kappa_a'(\alpha_a)$
% gives the dominant contribution to $\Gamma_{\rm rel}$ in the experimentally relevant regime $\kappa_a\ll\kappa_c$ unless the drive is too close to resonance with the qubit.
% The comparison to numerics is done in Fig. \ref{fig:coherent drive}, giving an excellent agreement.
Note that this result is easily extended to the case where the intrinsic cavity bath has a different density of states at $\omega = \omega_c$ and $\omega = \omega_a$, see 
Eq.~\eqref{app:Coherent_Drive_Replacement} in Appendix~\ref{SM:Keldysh_Section}.

%%%%%%%%%%%%%%%%%%%%%%%%%%%%%%%%%%%%%%%%%%%%%%%%%%%
%%%%%  MOVE TO APPENDIX
%%%%%%%%%%%%%%%%%%%%%%%%%%%%%%%%%%%%%%%%%%%%%%%%%%%
\section{Comparison to Jaynes-Cummings model}

% \AC{I just cut and paste this section.  Need to fix the text here, also add some text at the start saying we don't expect an agreement with the transmon case....}

In this section, we compare our main results for the transmon model against the more commonly used JC model. 
As we have shown in the previous sections, in the transmon model, the presence of the qubit $n=2$ Fock state played a crucial role in determining the dissipative properties (see the discussions below Eq.~\eqref{eq:Gamma_rel thermal}). 
Therefore, we expect the JC model, which treats the qubit as a two-level system, to give very different results from our transmon model results. 
Here, we show that the dissipative properties of the JC model is indeed very different from the transmon model for both the thermal and coherent drive cases. 

The equation of motion of the JC model is given by,
\begin{eqnarray}
    \partial_t\hat\rho
    =-i[\hat H_{\rm JC},\hat\rho]
    +\kappa_c\big[
    (1+\bar n_c^0){\mathcal D}[\hat c_0]\hat\rho
    +\bar n_c^0{\mathcal D}[\hat c_0^\dagger]\hat\rho
    \big]
\end{eqnarray}
where $\hat H_{\rm JC}=\hat H_0^{\rm JC}+\hat V_{\rm JC}+\hat H_{\rm D}$ with
\begin{eqnarray}
	\hat H_0^{\rm JC} &=& \omega_a \hat\sigma_z 
	+ \omega_c \hat c_0^\dagger\hat c_0, 
	\\
	\hat V_{\rm JC} &=& g(\hat\sigma_+\hat c_0+\hat\sigma_-\hat c_0^\dagger).
\end{eqnarray}
Here, $\hat\sigma_z$ and $\hat\sigma_\pm=(\hat\sigma_x\pm i\hat\sigma_y)/2$ are Pauli matrices and have assumed a vanishing bare qubit dissipation rate  $\kappa_a=0$.

We first consider the case where the thermal bare cavity occupancy is present but has no coherent drive.
Treating the Rabi coupling term ${\mathcal L}_1\bullet=-i[\hat V_{\rm JC},\bullet]$ as the perturbation within the Lindblad perturbation theory, 
the qubit $T_1$-decay rate at low temperature $\bar n_c^0\ll 1$  in the dispersive limit $g/|\Delta|\ll1$ can be computed within the second order perturbation as,
\begin{eqnarray}
    \label{app:qubit T1 JC}
    \Gamma_{\rm rel}^{\rm JC}\approx\frac{g^2}{\Delta^2}\kappa_c(1+2\bar n_c^0).
\end{eqnarray}
In stark contrast to the rich behavior seen in our Figs.~2 and 3(a), we find that the JC model \emph{always} gives an increase of the qubit $T_1$ decay rate. 
This result can be understood by regarding the dissipative cavity as a bath for the qubit that has a Lorentzian spectrum 
$S_c(\omega)=(\kappa_c/(2\pi))/[(\omega- \omega_c)^2+\kappa_c^2/4]$.
The second-order process of the qubit-cavity coupling $g$ gives rise to an effective dissipation rate to the qubit given by $\gamma = 2\pi g^2 S_c(\omega=\omega_a)\approx g^2\kappa_c/\Delta^2$ \cite{RMP_Clerk}.
Since the cavity at a finite temperature gives both the absorption and emission, the qubit $T_1$-decay rate can be estimated as $\Gamma_{\rm rel}^{\rm JC}\approx \gamma(1+2\bar n_c^0)$, which coincides with Eq.~\eqref{app:qubit T1 JC}.

Although our scheme can be applied to the coherently driven JC model, we do not provide it here since the coherent drive case is analyzed in detail in Ref.~\cite{Sete2014}. 
It is found that the drive \emph{always} decreases the qubit $T_1$-decay rate.
This is, again, in stark contrast to our result (Eqs.~\eqref{eq:kappatilde Hartree} and \eqref{eq:Gamma_rel coherent drive}) for the weakly-nonlinear oscillator that can give positive or negative contribution dependent on the sign of the detuning $\Delta$. This is not surprising, as the mean-field shift to the frequency, which was responsible for the sign change in the transmon model, is absent in the JC model.

%%%%%%%%%%%%%%%%%%%%%%%
%%%%%%%%%%%%%%%%%%%%%%%

\section{Conclusion}
We have presented a systematic formalism for analyzing dissipation in driven cQED systems, deriving simple expressions that describe the modification of qubit Purcell decay due to thermal or coherent photons.  Our results highlight the importance of the sign of the cavity-qubit detuning, and the interplay between non-resonant  coherent and dissipative processes.

We note that, in most experiments, it is likely that there are other extrinsic effects (e.g., drive-induced heating) that can also affect the photon-number dependence of qubit $T_1$-decay rate, causing them to vary experiment to experiment. 
Our contribution here is to set the ``fundamental limit'' to such photon-number dependent dissipation; the discovered \emph{intrinsic} mechanisms are unavoidable, even if the extrinsic dissipation channels are terminated.
% Our results set the ``fundamental limit'' of this photon-number dependent Purcell decay rate, which are unavoidable \emph{intrinsic} mechanisms that lead to photon-number dependent qubit $T_1$-decay rate, even if those extrinsic factors are reduced.
% The intrinsic mechanism of photon dependent Purcell decay found in this work is unavoidable, even if other extrinsic mechanism (e.g. heating effects) are present (which are indeed present in many  experiments). 
% Since the intrinsic mechanism found in this work is unavoidable even if extrinsic effects are absent, our study may, in a sense, be regarded as the fundamental limit to the dissipation rate in the presence of photon population.}
%In that sense, our study provides the \emph{fundamental limit} to the dissipation rate in the presence of photon population.
More generally, our work provides a new set of tools that can also be applied to other relevant problems, e.g.~dissipation in multi-cavity systems with transmon-mediated interactions as outlined in Appendix \ref{SM:extensions}.

%%%%%%%%%%%%%%%%%%%%%%%%%%%%%%%%%%%%%
%%%%%%%%%%%%%%%%%%%%%%%%%%%%%%%%%%%%%
\appendix
%%%%%%%%%%%%%%%%%%%%%%%%%%%%%%%%%%%%%
%%%%%%%%%%%%%%%%%%%%%%%%%%%%%%%%%%%%%
\section{Keldysh formalism for multi-level superconducing qubit}\label{SM:Keldysh_Section}

\subsection{Derivation of master equation~(4) via the Keldysh formalism}

We derive here the master equation Eq.~(4) by using the Keldysh formalism \cite{Kamenev_2011} which treats the effects of the dissipative baths exactly. Our starting point is a Hamiltonian $\hat H$ which describes the qubit and cavity coupled to two independent Markovian baths. By integrating out these baths, we obtain a Keldysh action that is identical to the one corresponding to the master equation Eq.~(4) and thus the two theories are equivalent.  

As in the main text, the Hamiltonian which describes the qubit, cavity, and their environments takes the form
\begin{equation}
\label{app:H_everything}
    \hat{H} 
    = 
    \hat{H}_s
    +\hat H_{\rm diss}.
\end{equation}  
Here, $\hat{H}_s$ is the system Hamiltonian (Eq.~(1) in the main text)
%\begin{align}
%\hat{H}_s
%=
%\omega_a \ha_0^\dagger \ha_0 + \omega_c \hc_0^\dagger \hc_0
%+g (\ha_0^\dagger \hc_0 +{\rm{h.c.}})
%-
%\frac{U}{2}\ha_0^\dagger \ha_0^\dagger \ha_0 \ha_0
%\end{align}
and 
\begin{eqnarray}
    \hat H_{\rm diss}=
    \hat{H}_{a, A}
	+
	\hat{H}_{c, C}
	+
	\hat{H}_{A}
	+
	\hat{H}_{C}
\end{eqnarray}
describes the environment and its coupling to the system, with $A$ and $C$ labeling the independent baths coupled to the qubit and cavity respectively. The coupling between the environments and the system of interest take the standard form 
\begin{align}
		\hat{H}_{a, A}
		&
		=
		- i \sqrt{\kappa_a}
		\left(
		\ha^\dagger \hat{\xi}_A
		-
		\ha \hat{\xi}_A^\dagger
		\right)
		,
		\\
		\hat{H}_{c, C}
		&=
		- i \sqrt{\kappa_c}
		\left(
		\hc^\dagger \hat{\xi}_C
		-
		\hc \hat{\xi}_C^\dagger
		\right).
\end{align}
We assume that the baths are a collection of independent harmonic oscillators in a Gaussian state, which is captured by the terms $\hH_A$ and $\hH_C$. The operators $\hat{\xi}_A$ and $\hat{\xi}_C$ are linear combination of bath annihilation operators. 
%For example, if we take the Markovian limit and assume that both baths density of states are flat, then one can recover the Heisenberg-Langevin equations for the system operators $\ha$ and $\hc$. \cite{RMP_Clerk}. By moving to the Heisenberg picture one can then derive, in a standard manner, the starting master equation  Eq.~(4) \cite{Gardiner_Zoller}.
%\ryoedit{We note, however, that the advantage of the Keldysh approach is that we do not need to assume that the baths are Markovian. This will be briefly addressed in the next subsection.}

%\begin{align}
%	&\hat{H}_A
%	=
%	\sum_{i}
%	\omega_i \hat{A}^\dagger_i \hat{A}_i,
%	\\
%	& \hat{H}_C
%	=
%	\sum_{j}
%	\omega_j \hat{C}^\dagger_j \hat{C}_j.
%\end{align}

Due to the Gaussian nature of the baths and the linear coupling, all the information on how they affect the system is captured by the relevant two-point correlation functions (which can be frequency-dependent). As with any theory where the path integral plays the central role, we must first identify the action. The Keldysh action corresponding to the Hamiltonian of Eq.~(\ref{app:H_everything}) can be written as \cite{Kamenev_2011}
\begin{align}\label{app:S_everything}
    S
    =
    S_{\rm s}
    +
    S_{a, A} + S_{c, C}
    +
    S_{A}+ S_{C}.
\end{align}
The first term describes the coherent dynamics between the qubit and the cavity,
while the last two terms $S_A$ and $S_c$ describe the dynamics of a set of independent harmonic oscillators.
The terms $S_{a,A}$ and $S_{c,C}$ describe the system-environment coupling.
% The first term describes the coherent dynamics between the qubit and the cavity
% Since these terms take a standard form and play no role when we integrate out the baths, we will write its explicit form later in this section. 
% the terms $S_A$ and $S_c$ describe the dynamics of a set of independent harmonic oscillators.
% and thus, in a similar vein, we will not write them out explicitly. 
By defining the complex vectors
\begin{align}
&
\bm a^\dagger(t) = 
\begin{pmatrix}
	a^*_{\rm{cl}}(t) & a^*_{\rm{q}}(t)
\end{pmatrix},
\hspace{0.5cm}
\bm c^\dagger(t) = 
\begin{pmatrix}
	c^*_{\rm{cl}}(t) & c^*_{\rm{q}}(t)
\end{pmatrix},
% \hspace{0.5cm}
\\
&
\boldsymbol{\xi}_A^\dagger(t) = 
\begin{pmatrix}
	\xi^*_{A,\rm{cl}}(t) & \xi^*_{A,\rm{q}}(t)
\end{pmatrix},
\hspace{0.5cm}
\boldsymbol{\xi}_C^\dagger(t) = 
\begin{pmatrix}
	\xi^*_{C,\rm{cl}}(t) & \xi^*_{C,\rm{q}}(t)
\end{pmatrix},
\hspace{0.5cm}
\end{align}
where $\rm{cl}$ and $\rm{q}$ label the classical and quantum fields, respectively, we can write the system-environment coupling terms in the action as 
\begin{align}\label{app:Coupling_a_A}
	&S_{a,A}
	=
	i\sqrt{\kappa_a}
	\int_{-\infty}^{\infty}
	dt
	\left(
	\bm a^\dagger(t)
	\boldsymbol{\sigma}_x
	\boldsymbol{\xi}_A(t)
	-
	\boldsymbol{\xi}_A^\dagger(t)
	\boldsymbol{\sigma}_x
	\bm a(t)
	\right)
	\\ \label{app:Coupling_c_C}
	&
	S_{c,C}
	=
	i\sqrt{\kappa_c}
	\int_{-\infty}^{\infty}
	dt
	\left(
	\bm c^\dagger(t)
	\boldsymbol{\sigma}_x
	\boldsymbol{\xi}_C(t)
	-
	\boldsymbol{\xi}_C^\dagger(t)
	\boldsymbol{\sigma}_x
	\bm c(t)
	\right)
\end{align} 
where $\boldsymbol{\sigma}_x$ is a Pauli matrix. 
Similarly, the system term $S_{\rm s}$ can be written as a function of complex vectors $\bm a(t)$ and $\bm c(t)$, with its form reflecting the system Hamiltonian $\hat H_{\rm s}$. 
The environment terms $S_A,S_C$ are quadratic functions of $\boldsymbol{\xi}_A$ and $\boldsymbol{\xi}_C$.
%Here, $\bar{n}_{a/c}^0[\omega_i] = (\exp(\hbar \omega_i/(k_B T_{a/c}))-1)^{-1}$ is the average occupation of each bath mode $i/j$, whree $T_{a/c}$ is the temperature of each bath. Implicit in this expression is the assumption that both baths are in thermal equilibrium, which is a standard assumption. The term $0^+$ is a positive infinitesimal and is best thought of as a regularization factor. Strictly speaking, in this continuum representation of the action, it is the only term which keeps track of the occupation far in the infinite past \cite{Kamenev_2011}. It will vanish in what follows.

We now want a description of our system which only involves the qubit and cavity modes. When working directly with the density matrix, this means tracing over the bath degrees of freedom. In the context of the path integral, the analogous step is to integrate over all bath fields. To do so, we first make a linear transformation to each bath field
\begin{align}\label{app:Linear_Trans_Qubit}
	\boldsymbol{\xi_A(t)}
	\to
	\boldsymbol{\xi_A(t)}
	+
	i \sqrt{\kappa_a}
	\int_{-\infty}^\infty
	dt' 
	\boldsymbol{G}_{A}(t-t')
	\boldsymbol{\sigma}_x
	\boldsymbol{a}(t')
	\\ \label{app:Linear_Trans_Cavity}
	\boldsymbol{\xi}_C(t)
	\to
	\boldsymbol{\xi}_C(t)
	+
	i\sqrt{\kappa_c}
	\int_{-\infty}^\infty
	dt' 
	\boldsymbol{G}_{C}(t-t')
		\boldsymbol{\sigma}_x
	\boldsymbol{c}(t')
\end{align}
where $\boldsymbol{G}_{A}(t)$ and $\boldsymbol{G}_{C}(t)$ are the matrix Green's function of the baths
\begin{align}
	\boldsymbol{G}_{A}(t)
	=
	\begin{pmatrix}
	-i \langle \{\hat{\xi}_A(t), \hat{\xi}_A^\dagger(0)\} \rangle & -i \Theta(t) \langle [\hat{\xi}_A(t), \hat{\xi}_A^\dagger(0)] \rangle  \\
	i \Theta(-t) \langle [\hat{\xi}_A(0), \hat{\xi}_A^\dagger(t)] \rangle & 0	
	\end{pmatrix}
	\\
    \boldsymbol{G}_{C}(t)
	=
	\begin{pmatrix}
	-i \langle \{\hat{\xi}_C(t), \hat{\xi}_C^\dagger(0)\} \rangle & -i \Theta(t) \langle [\hat{\xi}_C(t), \hat{\xi}_C^\dagger(0)] \rangle  \\
	i \Theta(-t) \langle [\hat{\xi}_C(0), \hat{\xi}_C^\dagger(t)] \rangle & 0	
	\end{pmatrix}
\end{align} 
with $\Theta(t)$ the Heaviside step function, $\{\cdot, \cdot \}$ the anticommutator and $[\cdot, \cdot]$ the commutator respectively. Here, the bath operators are in the Heisenberg picture generated by their free evolution, and the expectation values are taken with respect a stationary-state of each bath, which is what allowed us to assume that $\boldsymbol{G}_A$ and $\boldsymbol{G}_C$ only depend on the difference between $t$ and $t'$. This transformation does not change the functional measure of the baths fields and, more importantly, leads to an action in which the baths are uncoupled from the system of interest. The oscillator degrees of freedom can then be integrated out exactly, leaving us with an action that describes only the qubit and cavity.
Due to the linear transformations Eqs.~(\ref{app:Linear_Trans_Qubit})-(\ref{app:Linear_Trans_Cavity}) and the coupling term Eqs.~(\ref{app:Coupling_a_A})-(\ref{app:Coupling_c_C}), however, the system action acquires an additional term that is non-local in time and take the form
\begin{align}\label{eq:Self_Energies}
    -\int_{-\infty}^\infty
    dt
    \int_{-\infty}^\infty
    dt'
    \left(
    \boldsymbol{a}^\dagger(t)
    \boldsymbol{\Sigma}_a(t-t')
    \boldsymbol{a}(t)
    +
    \boldsymbol{c}^\dagger(t)
    \boldsymbol{\Sigma}_c(t-t')
    \boldsymbol{c}(t)
    \right)
\end{align}
where the qubit and cavity self energy are directly related to the bath's Green's functions:  
\begin{align}
	&
	\boldsymbol{\Sigma}_a(t)
	=
	\kappa_a
	\boldsymbol{\sigma}_x
	\boldsymbol{G}_A(t)
	\boldsymbol{\sigma}_x
	\equiv
	\begin{pmatrix}
	0 & \Sigma^A_a(t)  \\
	\Sigma^R_a(t) & \Sigma^K_a(t)
	\end{pmatrix}
	\\ 
	\label{eq:Self_Energy_c}
	&
    \boldsymbol{\Sigma}_c(t)
	=
    \kappa_c
	\boldsymbol{\sigma}_x
	\boldsymbol{G}_C(t)
	\boldsymbol{\sigma}_x
	\equiv
	\begin{pmatrix}
	0 & \Sigma^A_c(t)  \\
	\Sigma^R_c(t) & \Sigma^K_c(t)
	\end{pmatrix}
\end{align}
where $\Sigma_{a/c}^A(t)$, $\Sigma_{a/c}^R(t)$ and $\Sigma_{a/c}^K(t)$ are the advanced, retarded and Keldysh component of the self-energy respectively. The first two capture the response properties of the baths. For the linear, Gaussian baths under consideration, these quantities are independent of the state of the baths. Only the Keldysh component of the self-energy carries this information.

To obtain a Markovian description of the dynamics, we assume that the bath density of states of the qubit and cavity are flat.  Within this approximation, both $\hat{\xi}_A(t)$ and $\hat{\xi}_C(t)$ become the operator equivalent of Gaussian white noise. In particular, the commutator between the bath operators at different times is simply a delta function $[\hat{\xi}_A(t), \hat{\xi}_A(t')] = [\hat{\xi}_C(t), \hat{\xi}_C(t')] = \delta(t-t') $. Physically, since the the commutator $[\hat{\xi}_{A/C}(t), \hat{\xi}^\dagger_{A/C}(t')]$ is directly linked to the linear response properties of the bath via the Kubo formula, this implies that the bath auto-correlation time is vanishingly small. We further assume that the baths are in thermal equilibrium. Since dissipation is weak, we would only be probing frequencies near the resonance frequency of the qubit or cavity. In the spirit of the Markov approximation, we may then set $\langle \hat{\xi}^\dagger_{A/C}(t) \hat{\xi}_{A/C}(t')\rangle = \delta(t-t') \bar{n}_{a/c}^0$, where $\bar{n}_{a/c}^0$ is the thermal occupation number evaluated at the qubit/cavity frequency. We stress that this is a standard approximation, and is necessary if we want Markovian dynamics.
 
 %If we assume that the baths are in thermal equilibrium then the anti-commutator of the baths is a delta function times $2\bar{n}_{a/c}^0+1$ where $\bar{n}_{a/c}$ is the thermal equilibrium occupation number evaluated at the frequency of interest (see Ref.~\cite{RMP_Clerk} for further discussion).
 %\AC{This isn't correct... the anti-commutator of these operators is not a number.  Don't be so brief as to cause confusion.}
  Using both of these results, the self-energies can then be written as
  \begin{align}
	&
	\boldsymbol{\Sigma}_a(t)
	=
	-i\kappa_a\delta(t)
	\begin{pmatrix}
	0 & - \Theta(-t)  \\
	\Theta(t) & (2 \bar{n}_a^0+1)
	\end{pmatrix}
	\\ 
	\label{eq:Self_Energy_c2}
	&
    \boldsymbol{\Sigma}_c(t)
    -i\kappa_c\delta(t)
	\begin{pmatrix}
	0 & - \Theta(-t)  \\
	\Theta(t) & (2 \bar{n}_c^0+1).
	\end{pmatrix}
\end{align}
Using the identity
\begin{align}
	\int_{-\infty}^{t}
	dt'
	\delta(t-t')
	=
	\frac{1}{2}
\end{align}
we thus arrive at the final system action
\begin{widetext}
\begin{align} \nonumber
    S
    =
    \int_{-\infty}^\infty
    dt
    &
    \Big(
    a_{\rm q}^*
    \left(
    i\partial_t -\omega_a+i \frac{\kappa_a}{2})
    \right)
     a_{\rm cl}
     +
     a_{\rm cl}^*
    \left(
    i\partial_t -\omega_a-i \frac{\kappa_a}{2}
    \right)
     a_{\rm q}
    +i \kappa_a(2 \bar{n}_a^0
                    +1) 
    a_{\rm q}^* a_{\rm q}
    \\ \nonumber
    &
    +
    \frac{U}{2}
    \left(
    a_{\rm q}^* a_{\rm cl}
    +a_{\rm cl}^* a_{\rm q}
    \right)
    \left(
    a_{\rm q}^* a_{\rm q}
    +a_{\rm cl}^* a_{\rm cl}
    \right)
    +
    g
    \left(
    a_{\rm q}^* c_{\rm cl}
    +
    c_{\rm q}^* c_{\rm cl}
    +
    a_{\rm cl}^* c_{\rm q}
    +
    c_{\rm cl}^* c_{\rm q}
    \right)
    \\ \label{app:Full_Action}
    &
    +
    c_{\rm q}^*
    \left(
    i\partial_t -\omega_c+i \frac{\kappa_c}{2}
    \right)
     c_{\rm cl}
     +
     c_{\rm cl}^*
    \left(
    i\partial_t -\omega_c-i \frac{\kappa_c}{2}
    \right)
     c_{\rm q}
    +i \kappa_c( 2 \bar{n}_c^0
+1) 
    c_{\rm q}^* c_{\rm c}
    \Big)
\end{align}
\end{widetext}
where, for notational compactness, we have suppressed the temporal arguments of the fields.
%\AC{A bit awkward to go back to noise operators, then suddenly jump back to the action.  It would be much much cleaner to clearly state what the self energies are.}

We now wish to compare this action to the one we would obtain if we started with the master equation Eq.~(4) in the main text, which we rewrite here for convenience
\begin{align} \nonumber
    \partial_t \hrho
    & =
    -i[\hat{H}_s, \hrho]
    +
    \kappa_a(\bar{n}_a^0+1) \mathcal{D}[\ha_0]\hrho 
    + \kappa_a \bar{n}_a^0 \mathcal{D}[\ha_0^\dagger]\hrho
    \\ \label{app:Bare_Master_Equation}
    &
    +
    \kappa_c(\bar{n}_c^0+1) \mathcal{D}[\hc_0]\hrho 
    + \kappa_c \bar{n}_c^0 \mathcal{D}[\hc_0^\dagger]\hrho.
\end{align}
One can readily obtain a Keldysh action from a master equation using a standard procedure (see Ref.~\cite{Sieberer2016} for a pedagogical review). In short, assuming the operators are normal-ordered, creation or annihilation operators acting on the left or right of the density matrix are associated with a field on forward or backward branch of the contour. After rotating to the classical and quantum basis, the contribution to the action from the dissipation is
\begin{align}
    S_{a, \rm diss}
    =
    \int_{-\infty}^\infty
    dt
    \boldsymbol{a}^\dagger(t)
    \begin{pmatrix}
    0 & -i \frac{\kappa_a}{2} \\
    i \frac{\kappa_a}{2} & i \kappa_a(2 \bar{n}_a^0+1)
    \end{pmatrix}
    & \boldsymbol{a}(t)
    \\
    S_{c, \rm diss}
    =
    \int_{-\infty}^\infty
    dt
    \boldsymbol{c}^\dagger(t)
    \begin{pmatrix}
    0 & -i \frac{\kappa_c}{2} \\
    i \frac{\kappa_c}{2} & i \kappa_c(2 \bar{n}_c^0+1)
    \end{pmatrix}
    & \boldsymbol{c}(t).
\end{align}
In addition to the contribution to the action from the coherent Hamiltonian, the total Keldysh action is in fact Eq.~(\ref{app:S_everything}). The upshot is then that the two theories are equivalent, as promised. The only approximations we have made are standard ones, namely that the cavity and qubit baths are independent and Markovian.  

We briefly note that the same equation~(4) can be reproduced from an alternative approach, namely, by
constructing Heisenberg-Langevin equations for the system operators $\ha$ and $\hc$ \cite{RMP_Clerk} by writing down the equation of motion of those operators. By moving to the Heisenberg picture, one can then derive, in a standard manner, the starting master equation  Eq.~(4) \cite{Gardiner_Zoller}.
We note, however, that the advantage of the above Keldysh approach is that we can readily extend our theory to systems which do not have Markovian baths. This will be briefly addressed in the next subsection.

%\ryoedit{We note, however, that the advantage of the Keldysh approach is that we do not need to assume that the baths are Markovian. This will be briefly addressed in the next subsection.}

% \alexnew{Although it is clear that our approach to characterizing the effects of dissipation differs from the one used in Refs.~\cite{Malekakhlagh2020}-\cite{Petrescu2020}, it is worth stressing that the effective starting point (within the rotating wave approximation) is equivalent. In both theories, it is the bare qubit and cavity which are coupled to independent Markovian reservoirs. Before integrating out the baths, Refs~\cite{Malekakhlagh2020}-\cite{Petrescu2020} first approximately diagonalizes the system Hamiltonian using the Schrieffer-Wolf approach, which leads to the (renormalized) qubit and cavity modes being coupled to the same bath. Consequently, after the baths have been eliminated and the standard secular approximation has been made, both modes appear in the same jump operator, see Eqs.~(41a)-(41b) of Ref.~\cite{Malekakhlagh2020}. The upshot is that both theories capture the relevant resonant and dissipative processes.}

\subsection{Beyond the Markovian approximation - coherent drive case} 

% \ryocom{Change the section title?}
% \alexcom{Changed the title.}

Here, we will extend the result presented in the last section of the main text by relaxing the assumption that the bath density of states of the cavity is completely flat.
This, in turn, implies that the self-energies are no longer frequency independent. For clarity of presentation, we will assume that the qubit is not explicitly coupled to a thermal bath: the only loss it experiences is through its interaction with the cavity. We note that we can easily extend this result to the case where the intrinsic qubit decay rate is large.

Without the Markovian assumption, it is convenient to express the action in frequency space. The quadratic part of the action then takes the form
\begin{align}
\int_{-\infty}^\infty
\frac{d \omega}{2\pi}
\begin{pmatrix}
a^*_{\rm cl} & c^*_{\rm cl} & a_{\rm q}^* & c_{\rm q}^*
\end{pmatrix}
\boldsymbol{G}_0^{-1}[\omega]
\begin{pmatrix}
a_{\rm cl} \\
c_{\rm cl} \\
a_{\rm q} \\
c_{\rm q}
\end{pmatrix}
\end{align}
where the free Green's function is given by,
\begin{align}
    \boldsymbol{G}_0^{-1}[\omega]
    =
    \begin{pmatrix}
    0 & (\boldsymbol{G}_0^{-1}[\omega])^A 
    \\
    (\boldsymbol{G}_0^{-1}[\omega])^R & (\boldsymbol{G}_0^{-1}[\omega])^K
    \end{pmatrix}
\end{align}
with
\begin{align}
    (\boldsymbol{G}_0^{-1}[\omega])^R
    =
    (\boldsymbol{G}_0^{-1}[\omega])^A)^\dagger
    &=
    \begin{pmatrix}
        \omega-\omega_A & -g \\
        -g & \omega-\omega_c - \Sigma^R_c[\omega]
    \end{pmatrix}
    \\
    (\boldsymbol{G}_0^{-1}[\omega])^K
    &=
    \begin{pmatrix}
    0 & 0 \\
    0 & - \Sigma^K_c[\omega]
    \end{pmatrix}.
\end{align}
Here, we have suppressed the frequency dependence of the fields for notational simplicity. The retarded and Keldysh part of the self-energy is
\begin{align}
    &\Sigma^R_c[\omega]
    =
    -\frac{i}{2} \kappa_c[\omega]
    \\
    &\Sigma^K_c[\omega]
    =
    -i \kappa_c[\omega](2 \bar{n}_c[\omega]+1).
\end{align}
Without a flat density of states, the self-energies are frequency-dependent and, consequently, we obtain a theory that is non-local in time. 

%\begin{widetext}
We can however still make progress by assuming that $\kappa_c[\omega]$ is smooth and a slow-varying function of frequency. In this case, it is best to diagonalize the quadratic coherent problem by moving to a basis polaritons. After this transformation, the (inverse) Green's function in this basis take the form
\begin{widetext}
\begin{align}
    (\boldsymbol{G}_0^{-1}[\omega])^R
    =
     \begin{pmatrix}
        \omega-\tilde{\omega}_a- \Sigma^R_c [\omega] \frac{g^2}{\Delta^2} & -\Sigma^R_c [\omega] \frac{g}{\Delta} \\
        -\Sigma^R_c [\omega] \frac{g}{\Delta}& \omega-\tilde{\omega}_c - \Sigma^R_c[\omega](1- \frac{g^2}{\Delta^2})
    \end{pmatrix}
\end{align}
\begin{align}
    (\boldsymbol{G}_0^{-1}[\omega])^K
    =
    -
    \Sigma_c^K[\omega]
     \begin{pmatrix}
    \frac{g^2}{\Delta^2} & \frac{g}{\Delta} 
    \\
    \frac{g}{\Delta} & 1- \frac{g^2}{\Delta^2}
    \end{pmatrix}
\end{align}
where, as in the main text, we have ignored terms of order $g^3/\Delta^3$. The off-diagonal elements of these matrices correspond to dissipation induced coupling between the polaritons (because they are proportional to the self energies).

In the penultimate section of the main text, we considered how the presence of coherent photons modified the $T_1$ decay rate of the qubit. We found that the largest contribution to the change in the decay rate does not come from these off-diagonal terms: we can thus safely ignore them. Within this approximation, the quadratic part of the action is thus diagonal in the polariton basis. We may then apply the Markovian approximation to each polariton separately: since dissipation is weak and $\Sigma^R_c[\omega]$ and $\Sigma^K_c[\omega]$ are slowly varying functions of $\omega$, the largest contribution to the frequency integral will be near $\tilde{\omega}_a$ or $\tilde{\omega}_c$ depending on which polariton we are concerned with. Under this approximation, the Green's function now take the form
\begin{align}
    (\boldsymbol{G}_0^{-1}[\omega])^R
    =
     \begin{pmatrix}
        \omega-\tilde{\omega}_a+i \frac{\kappa_c [\tilde{\omega}_a]}{2} \frac{g^2}{\Delta^2} & 0 \\
        0 & \omega-\tilde{\omega}_c +i \frac{\kappa_c [\tilde{\omega}_c]}{2}(1- \frac{g^2}{\Delta^2})
    \end{pmatrix}
\end{align}
\begin{align}
    (\boldsymbol{G}_0^{-1}[\omega])^K
    =
     \begin{pmatrix}
    i \kappa_c[\tilde{\omega}_a](2 \bar{n}_c[\tilde{\omega}_a])\frac{g^2}{\Delta^2} & 0
    \\
    0 & i \kappa_c[\tilde{\omega}_c](2 \bar{n}_c[\tilde{\omega}_c]+1) (1- \frac{g^2}{\Delta^2})
    \end{pmatrix}
\end{align}
\end{widetext}
Once this replacement has been made, the analysis of 
the coherently driven circuit
%the next-to-last section of the main text 
is nearly identical. The upshot is then that the second main result, Eq.~(20) still holds with the replacement
\begin{align}\label{app:Coherent_Drive_Replacement}
    \tilde{\kappa}_a \to \frac{g^2}{\Delta^2}\kappa_{c}[\tilde{\omega}_a].
\end{align}

We briefly note that, for the thermal case, the non-Markovian bath extension does not seem straightforward, as the off-diagonal term corresponding to the correlated dissipation plays crucial role there. This issue is left as our future work.
% \AC{The discussion above would be much clearer if you explicitly state the definition of $\kappa_c[\omega]$ in terms of self energies.  This won't be obvious to your intended audience.}

\section{Lindblad perturbation theory}
\label{SM:Lindblad perturbation theory}

% In Ref. \cite{Malekakhlagh2020}, Eq.~\eqref{Lindblad blackbox} [with the number non-conserving nonlinear terms included] were analyzed through the standard Schrieffer-Wolfe transformation technique, regarding $\epsilon = U/\omega_{a}$ as a small quantity. 
% However, as pointed out earlier, this method can miss relevant dissipation processes and their expression can soon become complicated, making their physical picture somewhat opaque. 
%However, this canonical transformation based methods has no assurance to capture contributions from all possible processes. 
%In addition, it ...  unnecessary for our purpose and makes their physical picture somewhat opaque.

%\ryocom{One might argue that the "opaqueness" of expression is subjective. Our final expression is simple and have a clear physical meaning, but we can easily make our expression messier by keeping more terms...}

As its use is not widespread, we briefly outline here the basics of the Lindblad perturbation theory used in the main text, following Ref.~\cite{Li2014,LiPRX2016}.
% We give here a brief review of this approach. 
% We briefly note that, in contrast to the original work \cite{Li2014} where a \emph{singular value decomposition} (SVD) was used, we use the \emph{eigenstates} of the Lindbladian as our basis to describe the dynamics. 
% The SVD was used to avoid the potential difficulty that the eigenstates may not form  a complete set, due to the non-Hermiticity of the Lindbladian ${\mathcal L}$.
% However, as long as we restrict ourselves to weak dissipation, it should be safe to assume the absence of such issues. 
% On the other hand, the eigenstates and the corresponding eigenvalues of the Lindbladian has the advantage that they are directly related to the physical decay modes and rates. 
% In what follows, we therefore use eigenstates and eigenvalues in our perturbative treatment of the Lindbladian. 
% We will take such advantage and use eigenstates as the basis, which is safe as long as we restrict our interest to weak dissipation regimes where these issues would be absent. 
Within this framework,
the original Lindbladian ${\mathcal L}$ is split into non-perturbative (${\mathcal L}_0$) and perturbative (${\mathcal L}_1$) parts,
${\mathcal L}={\mathcal L}_0 + \epsilon{\mathcal L}_1$; 
$\epsilon=1$ is introduced as a book-keeping constant.  The eigenvalues $\lambda_\alpha$ and right eigenvectors $\hat{r}_\alpha$ of ${\mathcal L}$ are defined via
\begin{equation}
    {\mathcal L}\hat r_\alpha = \lambda_\alpha\hat r_\alpha.
\end{equation} 
As is done in standard Rayleigh-Schr\"odinger perturbation theory~\cite{Sakurai}, we write these quantities as a formal power series in $\epsilon$:
$\lambda_\alpha=\sum_{j=0}^\infty\epsilon^j\lambda_\alpha^{(j)}$ and 
$\hat r_\alpha=\sum_{j=0}^\infty
\epsilon^j\hat r_\alpha^{(j)}$. 
% Here,  $\lambda_\alpha^{(0)},\hat r_\alpha^{(0)}$ is the non-perturbative part satisfying  ${\mathcal L}_0\hat r_\alpha^{(0)}=\lambda_\alpha^{(0)}\hat r_\alpha^{(0)}$ and $\lambda_\alpha^{(j)},\hat r_\alpha^{(j)}~(j\ge 1)$ are the $j$-th order correction to them with in ${\mathcal L}_1$. 
Comparing order by order, we obtain the recursive relation
\begin{equation}
\label{app:general order perturbation}
    ({\mathcal L}_0-\lambda_\alpha^{(0)})\hat r_\alpha^{(j)}
    =-{\mathcal L}_1\hat r_\alpha^{(j-1)}
    +\sum_{k=1}^j\lambda_\alpha^{(k)}\hat r_\alpha^{(j-k)}.
\end{equation}
From this relation at $j=1$, 
\begin{equation}
    ({\mathcal L}_0-\lambda_\alpha^{(0)})\hat r_\alpha^{(1)}
    =-{\mathcal L}_1\hat r_\alpha^{(0)}
    +\lambda_\alpha^{(1)}\hat r_\alpha^{(0)},
\end{equation}
we get the first-order correction to the eigenvalue,
\begin{equation}
    \label{app:eigenvalue first order correction}
    \lambda_\alpha^{(1)} 
    = \langle \hat l_\alpha^{(0)},{\mathcal L}_1 \hat r_\alpha^{(0)} \rangle.
\end{equation}
Here, we have introduced the left eigenstate of the non-perturbative part ${\mathcal L}_0$ defined as
%${\mathcal L}_0\hat l_\alpha^{(0)\dagger}
%=\lambda_\alpha^{(0)}\hat l_\alpha^{(0)\dagger}
%$ 
${\mathcal L}^\dagger_0 \hat l_\alpha^{(0)}
=\lambda_\alpha^{(0)*}\hat l_\alpha^{(0)}$
where $\mathcal{L}^\dagger_0$ is the adjoint of the Liouvillian superoperator \cite{Li2014}. 
We have also used $\langle \hat l_\alpha^{(0)},\hat r_\beta^{(0)} \rangle=\delta_{\alpha,\beta}$ with $\langle \hat A,\hat B\rangle
={\rm tr}[\hat A^\dagger \hat B]$. 

The first-order correction to the right eigenstate is given by
\begin{equation}
	({\mathcal L}_0 - \lambda_\alpha^{(0)})\hat r_\alpha	^{(1)}
	=-\sum_{\beta\ne\alpha}
	\hat r_{\beta}^{(0)}\avg{\hat l_\beta^{(0)},{\mathcal L}_1 \hat r_\alpha^{(0)} }.
\end{equation}
Projection to the state $\beta\ne\alpha$ gives
\begin{equation}
	(\lambda_\beta^{(0)} - \lambda_\alpha^{(0)})
	\avg{\hat l_\beta^{(0)},\hat r_\alpha^{(1)}}
	=-\avg{\hat l_\beta^{(0)},{\mathcal L}_1 \hat r_\alpha^{(0)} }.
\end{equation}
Assuming that the spectrum of the unperturbed Lindbladian is not degenerate, we  have
\begin{equation}
	\avg{\hat l_\beta^{(0)},
	\hat r_\alpha^{(1)}}
	=
	- \frac{\avg{\hat l_\beta^{(0)},{\mathcal L}_1 \hat r_\alpha^{(0)} }}{\lambda_\beta^{(0)} - \lambda_\alpha^{(0)}}.
	\qquad (\beta\ne\alpha)
\end{equation}
Assuming further that $\{ \hat r_\alpha^{(0)}\}$ gives a complete set, which is equivalent to assuming that the Lindbladian ${\mathcal L}^{(0)}$ is diagonalizable
(which is always true in our problem), we get the first-order correction to the eigenstates,
\begin{equation}
	\hat r_\alpha^{(1)}	
	=-\sum_{\beta\ne\alpha} 
	\hat r_\beta^{(0)}
	\frac{\avg{\hat l_\beta^{(0)},{\mathcal L}_1 \hat r_\alpha^{(0)} }}{\lambda_\beta^{(0)} - \lambda_\alpha^{(0)}}.
\end{equation}
Without loss of generality, we have chosen to set $\avg{\hat l_\alpha^{(0)},\hat r_\alpha^{(1)}}=0$. 

Then, using Eq.~\eqref{app:general order perturbation} 
for $j=2$,
\begin{equation}
    ({\mathcal L}_0-\lambda_\alpha^{(0)})\hat r_\alpha^{(2)}
    =-{\mathcal L}_1\hat r_\alpha^{(1)}
    +\lambda_\alpha^{(1)}\hat r_\alpha^{(1)}
    +\lambda_\alpha^{(2)}\hat r_\alpha^{(0)}
\end{equation}
the second-order correction to the eigenvalue is given by,
\begin{equation}
\label{eq:second order eigenvalue}
	\lambda_\alpha^{(2)} = \avg{\hat l_\alpha^{(0)},{\mathcal L}_1 \hat r_\alpha^{(1)} }
	=-\sum_{\beta\ne\alpha} 
		\frac{\avg{\hat l_\alpha^{(0)},{\mathcal L}_1\hat r_\beta^{(0)}}
\avg{\hat l_\beta^{(0)},{\mathcal L}_1 \hat r_\alpha^{(0)} }}{\lambda_\beta^{(0)} - \lambda_\alpha^{(0)}}
\end{equation}
which is our central relation we will use in the following. We note in passing the similarity to usual second-order perturbation theory with the left and right eigenstates replacing the usual orthogonal eigenvectors of a Hermitian Hamiltonian. 
% In deriving Eq.~\eqref{eq:second order eigenvalue}, we have assumed that $\{ \hat r_\alpha^{(0)}\}$ gives a complete set, omitting the possibility that the Lindbladian ${\mathcal L}_0$ can be non-diagonalizable. 
% % \ryocom{Should we comment on the difference to Ref.\cite{Li2014}, where they do SVD?}
% Since we are comparing order by order, 
% it is assured to capture all the contribution in given order of ${\mathcal L}_1$,
% which poses an advantage over the conventional canonical transformation based methods.
% %this approach has an advantage over the  conventional canonical transformation based methods that it is assured to capture all the processes of given order in ${\mathcal L}_1$.

\section{Thermal  occupation}\label{SM:Section_Thermal_Occupation}

We show here how Lindblad perturbation theory leads to Eq.~\eqref{eq:Gamma_rel thermal} in the main text for the qubit $T_1$-decay in the presence of thermal excitations; this perturbative expression is valid for small thermal occupancy in the dispersive limit $g/|\Delta|\ll 1$. 
We will give more quantitative constraints on the validity of our perturbative expansion in what follows. 

As done in the main text, we regard the decoupled system
\begin{eqnarray}
    &&\label{app:L0}
    {\mathcal L}_0\hat\rho
    \equiv-i[\hat H_{\rm s}^{(0)},\hat\rho]
    +
    \tilde\kappa_c\big(
    (1+\tilde n_c)
    {\mathcal D}[\hat c]\hat\rho
    + \tilde n_c
    {\mathcal D}[\hat c^\dagger]
    \hat\rho
    \big)
    \nonumber\\
    &&
    \ \ \ \ \ \ \ \ \ \ \ \ \ \ \ \ \ \ \ \ 
    +
    \tilde\kappa_a\big(
    (1+\tilde n_a)
    {\mathcal D}[\hat a]\hat\rho
    + \tilde n_a
    {\mathcal D}[\hat a^\dagger]
    \hat\rho
    \big)
    % &+&\sum_{\mu=a,c}
    % \tilde\kappa_\mu\big(
    % (1+\tilde n_\mu)
    % {\mathcal D}[\hat d_\mu]\hat\rho
    % + \tilde n_\mu
    % {\mathcal D}[\hat d_\mu^\dagger]\hat\rho
    % \big),
\end{eqnarray}
as the non-perturbative part, 
where
\begin{eqnarray}
    \hat H_{\rm s}^{(0)}
    = \hat H_0 + \hat H_{\rm int}^{\rm slf}
    = 
    \tilde\omega_c\hat c^\dagger \hat c
    +\tilde\omega_a\hat a^\dagger\hat a
    + \chi_{aa}\hat a^\dagger\hat a^\dagger \hat a \hat a.
    \nonumber\\
\end{eqnarray}
We treat the remaining part,
\begin{equation}
    \label{app:L1}
    {\mathcal L}_1\hat\rho
    \equiv[{\mathcal L}-{\mathcal L}_0]\hat\rho=-i[
    \epsilon_{\rm crs}\hH_{\rm int}^{\rm crs}
    +\epsilon_{\rm ns}\hH_{\rm int}^{\rm nc},\hat\rho]
    +\epsilon_{\rm cd}{\mathcal L}_{\rm cd}\hat\rho,
\end{equation}
as a perturbation that is at most ${\mathcal O}(g/\Delta)$.

\subsection{Characterization of the non-perturbative part ${\mathcal L}_0$}

As is clear from Eq.~\eqref{eq:second order eigenvalue}, the first step in our approach is to characterize the spectral properties of the unperturbed Linbladian ${\mathcal L}_0$ (which includes the qubit self-Kerr interaction).
% In this subsection, we provide such characterization of ${\mathcal L}_0$. 
% \alexcom{Ryo, you keep using the direct sum $\oplus$ to describe the unperturbed Liouvillian, but this is incorrect. If you want a precise notation, you should write $\mathcal{L}_0 = \hat{1} \otimes L_a+ L_c \otimes \hat{1}$ which is not equivalent to ${\mathcal L}_0={\mathcal L}_0^c \oplus{\mathcal L}_0^a$. I think writing ${\mathcal L}_0={\mathcal L}_0^c + {\mathcal L}_0^a$ is fine, since its understood that there is a tensor product there. }
Since the cavity and qubit photons are completely decoupled in ${\mathcal L}_0={\mathcal L}_0^c\otimes \hat 1 + \hat 1\otimes{\mathcal L}_0^a$, the unperturbed eigenstates have a direct product structure: $\hat r_{\alpha_c,\alpha_a}^{(0)}=\hat r_{\alpha_c}^{c(0)}\otimes\hat r_{\alpha_a}^{a(0)}$,
$\hat l_{\alpha_c,\alpha_a}^{(0)}=\hat l_{\alpha_c}^{c(0)}\otimes\hat l_{\alpha_a}^{a(0)}$.
Here, the cavity-photon part of the right eigenstates are right eigenstates of a thermal harmonic oscillator Lindbladian,
\begin{eqnarray}
    \label{app:L0c}
    {\mathcal L}_0^c\hat r_{\alpha_c}^{c(0)}
    &=&-i[\tilde\omega_c\hat c^\dagger \hat c,\hat r_{\alpha_c}^{c(0)}]
    \nonumber\\
    &+&
    \tilde\kappa_c\big(
    (1+\tilde n_c)
    {\mathcal D}[\hat c]
    \hat r_{\alpha_c}^{c(0)}
    + \tilde n_c
    {\mathcal D}[\hat c^\dagger]
    \hat r_{\alpha_c}^{c(0)}
    \big)
    \nonumber\\
    &=&\lambda_{\alpha_c}^{c(0)}
    \hat r_{\alpha_c}^{c(0)},
\end{eqnarray}
and similarly, the right eigenvectors of the qubit satisfy
\begin{eqnarray}
    \label{app:L0a}
    {\mathcal L}_0^a\hat r_{\alpha_a}^{a(0)}
    &=&-i[\tilde\omega_a\hat a^\dagger \hat a    +\chi_{aa}\ha^\dagger\ha^\dagger\ha \ha,\hat r_{\alpha_a}^{a(0)}]
    \nonumber\\
    &+&
    \tilde\kappa_a\big(
    (1+\tilde n_a)
    {\mathcal D}[\hat a]\hat r_a^{a(0)}
    + \tilde n_a
    {\mathcal D}[\hat a^\dagger]
    \hat r_{\alpha_a}^{a(0)}
    \big)
    \nonumber\\
    &=&
    \lambda_{\alpha_a}^{a(0)}
    \hat r_{\alpha_a}^{a(0)}.
\end{eqnarray}
The eigenvalue of ${\mathcal L}_0$ corresponding to $\hat r_{\alpha_c,\alpha_a}^{(0)}$ is given by $\lambda_{\alpha_c,\alpha_a}^{(0)}=\lambda_{\alpha_c}^{c(0)}+\lambda_{\alpha_a}^{a(0)}$.

It is instructive to point out that ${\mathcal L}_0^c$ and ${\mathcal L}_0^a$ commutes with the superoperator 
${\mathcal M}_c\bullet=[\hat c^\dagger\hat c,\bullet]$ and  ${\mathcal M}_a\bullet=[\hat a^\dagger\hat a,\bullet]$, respectively. 
One can readily verify that the spectrum of $\mathcal{M}_a$ and $\mathcal{M}_c$ consist of the integers $m_a, m_c \in \mathbb{Z}$, and each of these eigenvalues are infinitely degenerate: any outer product of Fock states constitutes an eigenvector. The corresponding eigenvalue is simply the photon number in the ket state minus the photon number in the bra state. Using the familiar result from linear algebra that any two commuting operators share a set of eigenvectors, we conclude that the cavity (qubit) part of the Linbladian ${\mathcal L}_0$ takes on a  block-diagonal form ${\mathcal L}_0^{c(a)}=\otimes_{m=- \infty}^{\infty}{\mathcal L}_{0m}^{c(a)}$ \cite{Scarlatella2019},
where ${\mathcal L}_{0m}^{c(a)}$ only acts on the eigensubspace of ${\mathcal M}_{c(a)}$ characterized by the integer eigenvalue $m$. In other words, $m$ is a good quantum number we may use to label our eigenstates. Although this block-diagonal decomposition greatly simplifies our problem, it is worth pointing out that each block is still infinite in size. 

Our task now reduces to diagonalizing each superoperator ${\mathcal L}_{0m}^{\mu}$. We may write down the eigenvalue problem as 
%This property makes it possible for the cavity (qubit) part of the Linbladian ${\mathcal L}_0$ to be expressed in a block-diagonal form ${\mathcal L}_0^{c(a)}=\oplus_{m=0}^{\infty}{\mathcal L}_{0m}^{c(a)}$ [CITE Scarlatella],
%where ${\mathcal L}_{0m}^{c(a)}$ only acts on the eigensubspace of ${\mathcal M}_{c(a)}$ characterized by the integer eigenvalue $m$. \alexcom{Why does m only take values from 0 to infinity? Negative eigenvalues corresponds to states of the form $\ket{n}\bra{n+|m|}$...}
%The $k$-th right eigenstate in the $m$-subspace ($\mu=c,a$)
% (where $i=0,1,2,...$ in the descending order of the real part of the eigenvalue),
\begin{eqnarray}
    % {\mathcal L}_{0m}^{\mu}\hat r_{k,m}^{\mu(0)}
    % =\lambda_{k,m}^{\mu(0)}
    % \hat r_{k,m}^{\mu(0)}
    {\mathcal L}_{0m}^{\mu}\hat r_{k,m}^{\mu(0)}
    =\lambda_{k,m}^{\mu(0)}
    \hat r_{k,m}^{\mu(0)}.
\end{eqnarray}
with the constraint that $\hat r_{k,m}^{\mu(0)}$ must be an eigenvector of $\mathcal{M}_\mu$ with eigenvalue $m$. It must then necessarily take the form 
\begin{equation}
    \label{app:right eigenstates 0}
    \hat r_{k,m}^{\mu(0)}
    =
    \begin{cases}
    \sum_{n=0}^\infty
    r_{k,m,n}^{\mu(0)}
    \ket{(n+m)_\mu}\bra{n_{\mu}}
    &
    m \geq 0
    \\
    \sum_{n=0}^\infty
    r_{k,m,n}^{\mu(0)}
    \ket{n_\mu}\bra{(n-m)_{\mu}}
    &
    m < 0
    \end{cases}
\end{equation}
where $\ket {n_{c(a)}}$ is the Fock state for the cavity (qubit). A similar relation holds for the left eigenstates,
\begin{eqnarray}
    {\mathcal L}_{0m}^{\mu\dagger}\hat l_{k,m}^{\mu(0)}
    =\lambda_{k,m}^{\mu(0)*}
    \hat l_{k,m}^{\mu(0)}
\end{eqnarray}
% \alexcom{Same mistake you made near Eq. C4 left eigenvectors are defined by $\mathcal{L}^\dagger \hat{l} = E \hat{l}$, not $\mathcal{L} \hat{l}^\dagger = E \hat{l}^\dagger$. Otherwise, that would just mean $\hat{l}^\dagger$ is a right eigenvector of $\mathcal{L}...$ }
with
\begin{equation}
    \label{app:left eigenstates 0}
    \hat l_{k,m}^{\mu(0)}
    =
    \begin{cases}
    \sum_{n=0}^\infty
    l_{k,m,n}^{\mu(0)}
    \ket{(n+m)_\mu}\bra{n_{\mu}}
    &
    m \geq 0
    \\
    \sum_{n=0}^\infty
    l_{k,m,n}^{\mu(0)}
    \ket{n_\mu}\bra{(n-m)_{\mu}}
    &
    m < 0
    \end{cases}
\end{equation}
We may now, in a very precise way, identity the unperturbed $T_1$ modes we discussed in the main text: they correspond to eigenmodes labelled by $m = 0$. The right and left eigenvectors of these modes only involve Fock-state projectors, and hence only describe the decay of Fock-state populations.  In contrast, we refer to states labelled by $m \neq 0$ as $T_2$ modes: these necessarily involve decay of off-diagonal elements of the density matrix in the Fock basis.  We also point out that the (unique) unperturbed steady-state is necessarily a $T_1$ mode, i.e.~there are no steady-state Fock-state coherences.

%These properties let us classify the eigenstates into three categories: (a) the steady state $\hat\rho_{ss}^{\mu(0)}\equiv\hat r_{k=0,m=0}^{\mu(0)}$ that has vanishing eigenvalue $\lambda_{k=0,m=0}^{\mu(0)}=0$, (b) the $T_1$-decay modes that has a real eigenvalue $\lambda_{k (\ge 1),m=0}^{\mu(0)}\ne 0$ in the $m=0$ eigensubspace, that involves no coherence, and (c) $T_2$-decay modes, is in the $m\ne 0$ eigensubspace and does involve coherence. 

% The steady state $\hat\rho_{\rm ss}^{c(a)(0)}=\hat r_{i=0,m=0}^{c(a)(0)}$ with $\lambda_{i=0,m=0}^{c(a)(0)}=0$ sits in the $m=0$ eigensubspace. 
% For other eigenstates that has non-zero (real) eigenvalues in this $m=0$ eigensubspace, 
% since they involve no coherence, 
% it is natural to call them (i.e., $\hat r_{l,m=0}^{c(a)(0)}$) the $T_1$-decay modes of the cavity (qubit) photons.
% The eigenstates of the cavity (qubit) in the $m=0$ eigensubspace involves no coherence.
% For eigenstates that has nonzero (real) eigenvalues in this eigensubspace $m=0$, it is natural to call $\hat r_{l,m=0}^{c(a)(0)}$ the $T_1$-decay modes of the cavity (qubit) photons since they describe the photon number decay.
% and the cavity/qubit $T_1$-decay rates from the (real) eigenvalue $\Gamma_{l,m=0}^{c/a(0)}=-\lambda_{l,m=0}^{c/a(0)}$. 
% On the other hand the eigenmodes in $m\ne 0$ eigensubspace involving coherences are referred to as the $T_2$-decay modes. 
% The spectrum of ${\mathcal L}_0$ is therefore fully characterized by the direct product of $T_1$ or $T_2$-decay modes of cavity and qubit, $\hat r_{\alpha_c=(k,m),\alpha_a=(k,m)}^{(0)}$. 

We now look into the specific form of the eigenstates. 
We first consider the $m=0$ sector of each respective species, i.e., the steady states and the $T_1$-decay modes for the cavity and qubit. 
Substituting Eq.~\eqref{app:right eigenstates 0} at $m=0$ into Eqs.~\eqref{app:L0c} and \eqref{app:L0a} (and similarly for the left eigenstates), one finds that the two equations can be collectively described as ($\mu=c,a$),
\begin{eqnarray}
    &&\lambda_{k,m=0}^{\mu(0)}r_{k,m=0,n}^{\mu(0)}
    = \tilde\kappa_\mu \big[
	(1+\tilde n_\mu)(n+1)r_{k,m=0,n+1}^{\mu(0)}
	\nonumber\\
	&&-(n+2n\tilde n_\mu + \tilde n_\mu)r_{k,m=0,n}^{\mu(0)}
	+n \tilde n_\mu  r_{k,m=0,n-1}^{\mu(0)}
	\big],
\end{eqnarray}
or
\begin{widetext}
\begin{eqnarray}
	\bm M \bm r_{k,m=0}^{\mu(0)}  
	= \lambda_{k,m=0}^{\mu(0)}
	\bm  r_{k,m=0}^{\mu(0)} , 
	\qquad
	\bm  r_{k,m=0}^{\mu(0)} = 
	\begin{pmatrix}
		r_{k,m=0,n=0}^{\mu(0)} \\
		r_{k,m=0,n=1}^{\mu(0)} \\
		r_{k,m=0,n=2}^{\mu(0)} \\
		r_{k,m=0,n=3}^{\mu(0)} \\
		\vdots
	\end{pmatrix},
	\\
	(\bm l_{k,m=0}^{\mu(0)})^{\mathsf T} 
	\bm M  
	= \lambda_{k,m=0}^{\mu(0)}
	(\bm  l_{k,m=0}^{\mu(0)})^{\mathsf T}  , 
	\qquad
	\bm  l_{k,m=0}^{\mu(0)} = 
	\begin{pmatrix}
		l_{k,m=0,n=0}^{\mu(0)} \\
		l_{k,m=0,n=1}^{\mu(0)} \\
		l_{k,m=0,n=2}^{\mu(0)} \\
		l_{k,m=0,n=3}^{\mu(0)} \\
		\vdots
	\end{pmatrix},
\end{eqnarray} 
with
\begin{equation}
	\bm M=\tilde\kappa_\mu
	\begin{pmatrix}
		-\tilde n_\mu & 1 + \tilde n_\mu  & 0 & 0 & 0 & \cdots \\
		\tilde n_\mu & -1-3\tilde n_\mu & 2(1+\tilde n_\mu) & 0 & 0 & \cdots \\
		0 & 2\tilde n_\mu & -(2+5\tilde n_\mu) & 3(1 + \tilde n_\mu) & 0 & \cdots \\  
		0 & 0 & 3\tilde n_\mu & -(3+7\tilde n_\mu) & 4(1+\tilde n_\mu) & \cdots \\
		\vdots & \vdots & \vdots & \vdots & \ddots & \vdots
	\end{pmatrix}.
\end{equation}
\end{widetext}
Note that by definition, the $m = 0$ modes consist of a linear combination of Fock state projectors. Since the coherent Hamiltonian is diagonal in the Fock basis, it follows that it does not affect these modes at all.
%Note how the coherent dynamics play no role (therefore the nonlinearities are absent), which is due to the property that they do not contain any processes that change the number of photon excitations.  
% \end{widetext}

This eigenvalue problem is known to be exactly solvable \cite{Honda2010,Prosen_Bosons_2010}, where the eigenvalues are given by,
\begin{eqnarray}
    \label{app:T1 decay higher}
    \lambda_{k,m=0}^{\mu(0)}
    = - k \tilde\kappa_\mu. 
    \qquad (k=0,1,2,...)
\end{eqnarray}
$k=0$ corresponds to the steady state solution, while $k\ge 1$ are the $T_1$-decay modes. 
Remarkably, the eigenvalues are \emph{independent} of thermal occupancy $\tilde n_\mu$. 

While there are many $T_1$-decay modes, we are especially interested in the \emph{slowest} mode that describes qubit population decay ($k=1$ for qubit) \emph{without} cavity decay ($k=0$ for cavity).  This eigenvectors of this mode have the form
\begin{eqnarray}
    \label{app:qubit T1 decay mode 0}
    \hat r_{\rm rel}^{(0)}
	&\equiv &
	\hat r_{\alpha_c=(k=0,m=0),\alpha_a=(k=1,m=0)}^{(0)}
	\nonumber\\
    &=&\hat r_{k=0,m=0}^{c(0)}\otimes\hat r_{k=1,m=0}^{a(0)}
    \equiv\hat\rho_{\rm ss}^{c(0)}\otimes\hat r_{\rm rel}^{a(0)},
    \\
    \hat l_{\rm rel}^{(0)}
	&\equiv &
	\hat l_{\alpha_c=(k=0,m=0),\alpha_a=(k=1,m=0)}^{(0)}
	\nonumber\\
    &=&\hat l_{k=0,m=0}^{c(0)}
    \otimes
    \hat l_{k=1,m=0}^{a(0)}
    \equiv\hat l_{\rm ss}^{c(0)}\otimes\hat l_{\rm rel}^{a(0)},
\end{eqnarray}
where $\hat l_{\rm ss}^{c(0)}=\hat 1$ is the left eigenstate of the steady state. 
% Here, we have ordered the $T_1$-decay modes in the descending order of the eigenvalues $\{\lambda_{k,m=0}^{a(0)}\}$.
% Among the multiple number of qubit $T_1$-decay modes, we are  interested in the slowest mode that changes the photon population, which  
We will refer to this mode as the `qubit $T_1$-decay mode' and its eigenvalue 
\begin{equation}
    \label{app:unperturbed T1 decay rate}
    \Gamma_{\rm rel}^{(0)}=-\lambda_{k=1,m=0}^{a(0)} = \tilde\kappa_a
\end{equation}
as the `qubit $T_1$-decay rate' (Eq.~(13) in the main text). 
The other $T_1$-modes (labelled by $k\ge 2$) will be referred to as `\emph{higher-order} qubit $T_1$-modes'. 
% We will call the qubit $T_1$-decay rate as $\Gamma_{\rm rel}^{(0)}=-\lambda_{k=1,m=0}^{a(0)}$.

% We list here the steady states and the slowest $T_1$-decay modes that would be used later.
The explicit form of the qubit $T_1$-decay mode (Eq.~\eqref{app:qubit T1 decay mode 0}) is listed below for the latter use. 
For the cavity part, the steady state
$\hat \rho_{\rm ss}^{c(0)}\equiv\sum_{n=0}^\infty p_{{\rm ss},n}^{c(0)}\ket{n_c}\bra{n_c}$ is given by
\begin{equation}
	p_{{\rm ss},n}^{c(0)} = \frac{1}{1+\tilde n_c}\Big(\frac{\tilde n_c}{1+\tilde n_c}\Big)^{n},
\end{equation}
with the corresponding left eigenstate $\hat l_{\rm ss}^{c(0)}=\hat 1$,
and the qubit part is given by,
\begin{eqnarray}
	r_{{\rm rel},n}^{a(0)}
	\equiv
	r^{a(0)}_{k=1,m=0,n}
	= -\frac{n-\tilde n_a}
	{1+\tilde n_a}
	\Big(\frac{\tilde n_a}{1+\tilde  n_a}\Big)^{n-1}
\end{eqnarray}
and
\begin{equation}
	l_{{\rm rel},n}^{a(0)}
	\equiv  l^{a(0)}_{k=1,m=0,n}= \frac{-n+\tilde n_a}{(1+\tilde n_a)^2}.
\end{equation}

In contrast to the $T_1$-decay modes,  the $T_2$-decay modes are affected by the coherent dynamics.
Therefore, the Kerr nonlinearity $\chi_{aa}\sim-U/2$ of the qubit does play a role, and describing the eigenstates and eigenvalues of these modes requires some care. 
% Below, we restrict ourselves to $m=\pm 1$, which is enough for our purpose. 

Let us start with the cavity part where such nonlinearities are absent. 
These can be computed exactly using the formalism of third-quantization \cite{Prosen_Fermions_2008, Prosen_Bosons_2010}, where the $T_2$-decay rates are given by \cite{Chaturvedi1991,Honda2010},
\begin{equation}
    \lambda_{k,m}^{c(0)}
    =-im\tilde\omega_c-\frac{\tilde\kappa_c}{2}(|m|+2k).
\end{equation}
The corresponding right and left eigenstate for $k=0$ and $m=\pm 1$ (which will be used in later sections) has the form \cite{Honda2010}
\begin{eqnarray}
    \hat r_{\uparrow}^{c(0)}
    &\equiv&
    \hat r_{k=0,m=1}^{c(0)}
    \nonumber\\
    &=&\sum_{n=1}^\infty 
	\Big( \frac{\tilde n_c}{1+\tilde n_c}\Big)^{n-1}
	\sqrt{n} \ket{n_c}\bra{(n-1)_c}, \\
    \hat r_{\downarrow}^{c(0)}
    &\equiv&
    \hat r_{k=0,m=-1}^{c(0)}
    \nonumber\\
    &=&	\sum_{n=1}^\infty 
	\Big( \frac{\tilde n_c}{1+\tilde n_c}\Big)^{n-1}
	\sqrt{n} \ket{(n-1)_c}\bra{n_c},
\end{eqnarray}
and
\begin{eqnarray}
    \hat l_{\uparrow}^{c(0)}
    &\equiv&
    \hat l_{k=0,m=1}^{c(0)}
    =\sum_{n=1}^\infty 
	\frac{\sqrt{n}}{1+\tilde n_c} \ket{n_c}\bra{(n-1)_c}, \\
	\hat l_{\downarrow}^{c(0)}
	&\equiv &
    \hat l_{k,m=-1}^{c(0)}
    =\sum_{n=1}^\infty 
	\frac{\sqrt{n}}{1+\tilde n_c}  \ket{(n-1)_c}\bra{n_c},
\end{eqnarray}
respectively. 

We now turn to the qubit part. As stressed earlier, the nonlinearity plays a role for the $T_2$-decay modes and rates, but surprisingly, this problem is known to be exactly solvable \cite{Dykman1984,Chaturvedi1991}. 
We will however not make use of the known exact solution  
and instead take advantage of the fact that most of the experiments are done in the regime $U\gg\tilde\kappa_a$; this leads to a massive simplification.  In this regime, the dissipation can be treated perturbatively: the right and left eigenvectors are simply outer products of Fock states \cite{Scarlatella2019}. Note crucially that this is only true of the $T_2$-decay modes. The $T_1$-decay modes are completely insensitive to the coherent dynamics and thus dissipation completely determines the structure of the eigenvectors, as seen above. Keeping this in mind, after a straightforward calculation, we arrive at the perturbative eigenvalue of the $T_2$ $m = \pm 1$ modes \cite{Scarlatella2019}
%\begin{equation}
%    \hat r_{(k+m,k)}^{a(0)}
%    \equiv\hat r_{k,m}^{a(0)}
%    =\ket{(k+m)_a}\bra{k_a},
%\end{equation}
%and $\hat l_{(k+m,k)}^{a(0)}\equiv\hat l_{k,m=\pm 1}^{a(0)}=\hat r_{(k+m,k)}^{a(0)}$
%with corresponding eigenvalues for $m=\pm 1$,
\begin{eqnarray}
    &&
    \lambda_{(k+m,m)}^{a(0)}
    \equiv\lambda_{k,m=\pm 1}^{a(0)}
    =-im (\tilde\omega_a - U k) 
	\nonumber\\
	&&\ \ \ \ \ 
	-\frac{\tilde\kappa_a}{2}\big[\tilde n_a(2k+3) + (1+\tilde n_a)(2k +1)\big], 
\end{eqnarray}
where we have used the relation $\chi_{aa}\simeq - U/2$.

\subsection{Derivation of Eq.~\eqref{eq:Gamma_rel thermal}}

We are now in the position to derive the photon dependence to the qubit $T_1$-decay rate $\Gamma_{\rm rel}$ (Eq.~\eqref{eq:Gamma_rel thermal} in the main text) in the full problem characterized by the Lindladian ${\mathcal L}$ (Eq.~(5)).  
% This is defined as the sum of the unperturbed $T_1$-decay rate $\Gamma_{\rm rel}^{(0)}$ (defined in Eq.~\eqref{app:unperturbed T1 decay rate}) plus the correction from the perturbation ${\mathcal L}_1$ to this mode.
This is defined, within the second-order perturbation, as the sum of the contribution from the unperturbed  $\Gamma_{\rm rel}^{(0)}$ (defined in Eq.~\eqref{app:unperturbed T1 decay rate}) and the perturbative correction to this mode: 
\begin{eqnarray}
    \Gamma_{\rm rel} = \Gamma_{\rm rel}^{(0)}
    +\Gamma_{\rm rel}^{(1)}
    +\Gamma_{\rm rel}^{(2)},
\end{eqnarray}
where $\Gamma_{\rm rel}^{(1)}=-\lambda_{\rm rel}^{(1)}=-\langle\hat l_{\rm rel}^{(0)},{\mathcal L_1}\hat r_{\rm rel}^{(0)}\rangle$ and
\begin{eqnarray}
    \Gamma_{\rm rel}^{(2)}=-\lambda_{\rm rel}^{(2)}
    =\sum_{\beta\ne{\rm rel}} 
		\frac{\avg{\hat l_{\rm rel}^{(0)},{\mathcal L}_1\hat r_\beta^{(0)}}
\avg{\hat l_\beta^{(0)},{\mathcal L}_1 \hat r_{\rm rel}^{(0)} }}{\lambda_\beta^{(0)} - \lambda_{\rm rel}^{(0)}}.
\nonumber\\
\end{eqnarray}
% Here, we have used the property that the first order correction vanishes $\lambda_{\rm rel}^{(1)}=0$.

The perturbative part ${\mathcal L}_1=\epsilon_{\rm crs}{\mathcal L}_{\rm crs}+\epsilon_{\rm nc}{\mathcal L}_{\rm nc}+\epsilon_{\rm cd}{\mathcal L}_{\rm cd}$ (Eq.~\eqref{app:L1}) is composed of three parts (where $\epsilon_{\rm crn}=\epsilon_{\rm nc}=\epsilon_{\rm cd}=1$ are book-keeping constants): cross-Kerr nonlinearity 
\begin{equation}
{\mathcal L}_{\rm crs}\bullet=-i[\hat H_{\rm crs},\bullet]
=-i\chi_{ca}[\hat c^\dagger\hat c\hat a^\dagger\hat a,\bullet],    
\end{equation}
nonlinear conversion 
\begin{equation}
    \label{app:L nonlinear conversion}
    {\mathcal L}_{\rm nc}\bullet=-i[\hat H_{\rm nc},\bullet]
    =-i\tilde\chi[\hat a^\dagger\hat a^\dagger\hat a\hat c+\hat c^\dagger\hat a^\dagger\hat a\hat a,\bullet],   
\end{equation}
and correlated dissipation 
\begin{eqnarray}
    \label{app:L correlated disspation}
    {\mathcal L}_{\rm cd}\hat\rho
    &=&-\frac{1}{2}(\tilde\gamma_{\uparrow}
    +\tilde\gamma_{\downarrow})
    \{\hat a^\dagger \hat c+\hat c^\dagger\hat a,\hat\rho\}
    \nonumber\\
    &+&\tilde\gamma_\downarrow
    (\hat a\hat\rho\hat c^\dagger + \hat c\hat\rho\hat a^\dagger)
    +\tilde\gamma_\uparrow
    (\hat a^\dagger\hat\rho\hat c + \hat c^\dagger\hat\rho\hat a).
\end{eqnarray}

Let us start by pointing out that the cross-Kerr nonlinearity gives no correction to the qubit $T_1$-decay rate $\Gamma_{\rm rel}$ to the order of our interest. 
This is due to the relation 
\begin{equation}
    {\mathcal L}_{\rm crs}\hat r_{\rm rel}^{(0)}
    ={\mathcal L}_{\rm crs}^\dagger\hat l_{\rm rel}^{(0)}=0,
\end{equation}
which follows from the property that the cross-Kerr nonlinearity does not change the number of excitation of each respective species. 
Therefore, in what follows, we only consider the correction from the nonlinear conversion ${\mathcal L}_{\rm nc}$ (Eq.~\eqref{app:L nonlinear conversion}) and correlated dissipation ${\mathcal L}_{\rm cd}$ (Eq.~\eqref{app:L correlated disspation}). 

Both of these perturbations ${\mathcal L}_{\rm nc},{\mathcal L}_{\rm cd}$ involve changes in the number of cavity/qubit excitations, and thus necessarily causes transitions between different eigensubspaces of ${\mathcal M}_{c}$ and ${\mathcal M}_a$.
More prosaically, they couple $T_1$ modes to $T_2$ modes and vice versa.
From this property, we can immediately conclude that the first order correction is absent,
\begin{eqnarray}
    \Gamma_{\rm rel}^{(1)}=-\langle\hat l_{\rm rel}^{(0)},{\mathcal L_1}\hat r_{\rm rel}^{(0)}\rangle
    = 0,
\end{eqnarray}
because $\hat l_{\rm rel}^{(0)}$ and ${\mathcal L_1}\hat r_{\rm rel}^{(0)}$ are in  different eigensubspaces.

\begin{widetext}
Therefore, the leading contribution is from the second order correction, which is composed of three terms,
\begin{eqnarray}
    \label{app:Gamma_rel second order}
    \Gamma_{\rm rel}^{(2)}
    =\sum_{\beta\ne{\rm rel}} 
	\Bigg[
	&&\epsilon_{\rm nc}^2
		\frac{\avg{\hat l_{\rm rel}^{(0)},{\mathcal L}_{\rm nc}\hat r_\beta^{(0)}}
        \avg{\hat l_\beta^{(0)},{\mathcal L}_{\rm nc} \hat r_{\rm rel}^{(0)} }}{\lambda_\beta^{(0)} - \lambda_{\rm rel}^{(0)}}
    +\epsilon_{\rm cd}^2
		\frac{\avg{\hat l_{\rm rel}^{(0)},{\mathcal L}_{\rm cd}\hat r_\beta^{(0)}}
        \avg{\hat l_\beta^{(0)},{\mathcal L}_{\rm cd} \hat r_{\rm rel}^{(0)} }}{\lambda_\beta^{(0)} - \lambda_{\rm rel}^{(0)}}
    \nonumber\\
    &&+\epsilon_{\rm nc}\epsilon_{\rm cd}
    \bigg[
		\frac{\avg{\hat l_{\rm rel}^{(0)},{\mathcal L}_{\rm nc}\hat r_\beta^{(0)}}
        \avg{\hat l_\beta^{(0)},{\mathcal L}_{\rm cd} \hat r_{\rm rel}^{(0)} }}{\lambda_\beta^{(0)} - \lambda_{\rm rel}^{(0)}}
    +		\frac{\avg{\hat l_{\rm rel}^{(0)},{\mathcal L}_{\rm cd}\hat r_\beta^{(0)}}
        \avg{\hat l_\beta^{(0)},{\mathcal L}_{\rm nc} \hat r_{\rm rel}^{(0)} }}{\lambda_\beta^{(0)} - \lambda_{\rm rel}^{(0)}}
    \bigg]
    \Bigg]
    \nonumber\\
    &&\equiv
    \epsilon_{\rm nc}^2\Gamma_{\rm rel}^{{\rm nc-nc(2)}}
    +\epsilon_{\rm cd}^2\Gamma_{\rm rel}^{{\rm cd-cd}(2)}
    +\epsilon_{\rm nc}\epsilon_{\rm cd}\Gamma_{\rm rel}^{{\rm nc-cd}(2)}.
\end{eqnarray}
% where the first, second, and the third term are the contribution from ${\mathcal O}({\mathcal L}_{\rm nc}^2)$, ${\mathcal O}({\mathcal L}_{\rm cd}^2)$, and ${\mathcal O}({\mathcal L}_{\rm nc}{\mathcal L}_{\rm cd})$, respectively.
The second term $\Gamma_{\rm rel}^{{\rm cd-cd}(2)}$ can be safely neglected in the regime of our interest $\kappa_c,\kappa_a\ll U,|\Delta|$ since they would only give contributions $\propto\kappa_\mu^2$.

We first consider the first term $\propto \epsilon_{\rm nc}^2$, that arises from the second-order process involving nonlinear conversion $\hat H^{\rm nc}_{\rm int}$. 
This is composed of two processes $\hat H^{\rm nc}_{\rm int}=\hat H^{\rm nc}_{c\rightarrow a}+\hat H^{\rm nc}_{a\rightarrow c}$;
$\hat H^{\rm nc}_{c\rightarrow a}=\chi_{ca}\hat a^\dagger\hat a^\dagger\hat a\hat c$ that converts the cavity excitation to the qubit excitation and $\hat H^{\rm nc}_{a\rightarrow c}=\chi_{ca}\hat c^\dagger\hat a^\dagger\hat a\hat a$ is its inverse process.
When these two processes act on the qubit $T_1$-decay mode $\hat r_{\rm rel}^{(0)}$, the resulting states
$\hat H_{c\rightarrow a}^{\rm nc}\hat r_{\rm rel}^{(0)}$ (and $\hat r_{\rm rel}^{(0)}H_{c\rightarrow a}^{\rm nc}$), 
$\hat H_{a\rightarrow c}^{\rm nc}\hat r_{\rm rel}^{(0)}$
(and $\hat \hat r_{\rm rel}^{(0)}H_{a\rightarrow c}^{\rm nc}$)
will overlap with the $T_2$-decay modes,
\begin{eqnarray}
    \label{app:intermediate states thermal}
    \hat r_{\downarrow_c,(n_a+1,n_a)}^{(0)}
    \equiv\hat r_{\downarrow}^{c(0)}
    \otimes \hat r_{(n_a+1,n_a)}^{a(0)}
    \qquad
    {\rm and}
    \qquad
    \hat r_{\uparrow_c,(n_a-1,n_a)}^{(0)}
    \equiv
    \hat r_{\uparrow}^{c(0)}
    \otimes 
    \hat r_{(n_a-1,n_a)}^{a(0)},
\end{eqnarray}
respectively.
These have the corresponding eigenvalues
\begin{eqnarray}
	%=============================
	\label{app:lambda_down_c,up_a}
	\lambda_{\downarrow_c,(n_a+1,n_a)}^{(0)} 
	&&=-\Big[
	\frac{\tilde\kappa_c}{2}
	+\frac{\tilde\kappa_a}{2}\big[\tilde n_a(2n_a+3) + (1+\tilde n_a)(2n_a+1)\big]
	\Big] 
	+i \big[\tilde\omega_c - (\tilde\omega_a - U n_a)\big],
	\\
% 	\nonumber\\
% 	&&=-\Big[
% 	\frac{\tilde\kappa_c}{2} 
% 	+\frac{\tilde\kappa_a}{2}
% 	\big[1+ 2n_a + 4\tilde n_a 
% 	+ 4\tilde n_a n_a \big]
% 	\Big] 
% 	+i \big[\omega_c -(\omega_a - U^{aa}_{\rm sc} n_a)\big].	%=============================
	\lambda_{\uparrow_c,(n_a,n_a+1)}^{(0)} 
	&&=-\Big[
	\frac{\tilde\kappa_c}{2} 
	+\frac{\tilde\kappa_a}{2}\big[\tilde n_a(2n_a+3) + (1+\tilde n_a)(2n_a+1)\big]
	\Big] 
	-i \big[\tilde\omega_c -(\tilde\omega_a - U n_a)\big],
% 	\nonumber\\
% 	&&=-\Big[
% 	\frac{\tilde\kappa_c}{2} 
% 	+\frac{\tilde\kappa_a}{2}
% 	\big[1+ 2n_a + 4\tilde n_a 
% 	+ 4\tilde n_a n_a \big]
% 	\Big] 
% 	-i \big[\omega_c -(\omega_a - U^{aa}_{\rm sc} n_a)\big]
	%=============================
\end{eqnarray}
As a result, $\Gamma_{\rm rel}^{\rm nc-nc(2)}$ is composed of two parts (where we introduce a short-hand notation $\ket{n_c,n_a}\equiv \ket{n_c}\otimes\ket{n_a}$.
),
%=========================
\begin{eqnarray}
%=========================
    \label{app:Gamma_rel nc-nc lengty}
	&&\Gamma_{\rm rel}^{{\rm nc-nc}(2)}
%=========================
	=-\sum_{n_c,n_a=0}^\infty 
	p_{{\rm ss},n_c}^{c(0)}
	r^{a(0)}_{{\rm rel},n_a}
	\nonumber\\
	&&\times
	\Bigg[
	2{\rm Re}\bigg[
	\frac{1}
	{\tilde\kappa_a +\lambda_{\downarrow_c,(n_a+1,n_a)}^{(0)}}
	\bigg]
	{\rm tr}\big[
	\hat l_{\rm rel}^{(0)\dagger}
	\hat H_{a\rightarrow c}^{\rm ns}
	\hat r_{\downarrow_c,(n_a+1,n_a)}^{(0)}
	-\hat l_{\rm rel}^{(0)\dagger}
	\hat r_{\downarrow_c,(n_a+1,n_a)}^{(0)}
	\hat H_{a\rightarrow c}^{{\rm ns}}
	\big]
	{\rm tr}\big[
	\hat l_{\downarrow_c,(n_a+1,n_a)}^{(0)\dagger}
	\hat H_{c\rightarrow a}^{\rm ns}
	\ket{n_c,n_a}\bra{n_c,n_a}
	\big]
	\nonumber\\
	&&+2{\rm Re}\bigg[
	\frac{1}{\tilde\kappa_a +\lambda_{\uparrow_c,(n_a-1,n_a)}^{(0)}}
	\bigg]
	{\rm tr}\big[
	\hat l_{\rm rel}^{(0)\dagger}
	\hat H_{c\rightarrow a}^{\rm ns}
	\hat r_{\uparrow_c,(n_a-1,n_a)}^{(0)}
	-\hat l_{\rm rel}^{(0)\dagger}
	\hat r_{\uparrow_c,(n_a-1,n_a)}^{(0)}
	\hat H_{c\rightarrow a}^{\rm ns} 
	\big]
	{\rm tr}\big[
	\hat l_{\uparrow_c,(n_a-1,n_a)}^{(0)\dagger}
	\hat H_{a\rightarrow c}^{\rm ns}
	\ket{n_c,n_a}\bra{n_c,n_a}
	\big]
	\Bigg],
	\nonumber\\
	%-----------------
	&&
	=-2\tilde\chi^2
	\sum_{n_c,n_a=0}^\infty 
	p_{{\rm ss},n_c}^{c(0)}
	r^{a(0)}_{{\rm rel},n_a}
	\Bigg[
	{\rm Re}\bigg[
	\frac{1}
	{\tilde\kappa_a +\lambda_{\downarrow_c,(n_a+1,n_a)}^{(0)}}
	\bigg]
	n_a
	C_{n_c}^{c,{\rm ss}\downarrow}
	C_{n_a}^{a,{\rm rel}\uparrow}
	+{\rm Re}\bigg[
	\frac{1}{\tilde\kappa_a +\lambda_{\uparrow_c,(n_a-1,n_a)}^{(0)}}
	\bigg]
	(n_a-1)
	C_{n_c}^{c,{\rm ss}\uparrow}
	C_{n_a}^{a,{\rm rel}\downarrow}	
	\Bigg],
	\nonumber\\
%=========================
\end{eqnarray}
%=========================
where
\begin{eqnarray}
	C_{n_c}^{c,{\rm ss}\downarrow}
	&=&{\rm tr}\big[
	\hat l_{\rm ss}^{c(0)\dagger}
	\hat c^\dagger 
	\hat r_{\downarrow}^{c(0)}
	-\hat l_{\rm ss}^{c(0)\dagger}
	\hat r_{\downarrow}^{c(0)}
	\hat c^\dagger
	\big]
	{\rm tr}\Big[
	\hat l_{\downarrow}^{c(0)\dagger}
	\hat c
	\ket{n_c}\bra{n_c}
	\Big]
	=
	\frac{n_c}{1+\tilde n_c}
	\sum_{n_c'=1}^\infty
	\bigg[
	\Big(\frac{\tilde n_c}{1+\tilde n_c}\Big)^{n_c'-1}
	\frac{n_c'}{1+\tilde n_c}
	\bigg]
% 	\nonumber\\
% 	&\approx &
% 	\frac{n_c}{(1+\tilde n_c)^2}
	\\
	%========================================
	C_{n_c}^{c,{\rm ss}\uparrow}
	&=&{\rm tr}\big[
	\hat l_{\rm ss}^{c(0)\dagger}
	\hat c
	\hat r_{\uparrow_c}^{(0)}
	-\hat l_{\rm ss}^{c(0)\dagger}
	\hat r_{\uparrow}^{c(0)}
	\hat c
	\big]
	{\rm tr}\Big[
	\hat l_{\uparrow}^{c(0)\dagger}
	\hat c^\dagger
	\ket{n_c}\bra{n_c}
	\Big]
	=
	\frac{1+n_c}{1+\tilde n_c}
	\sum_{n_c'=0}^\infty
	\bigg[
	\Big(\frac{\tilde n_c}{1+\tilde n_c}\Big)^{n_c'}
	\frac{1+n_c'}{1+\tilde n_c}
	\bigg]
	\\
%========================================
    C_{n_a}^{a,{\rm rel}\downarrow}
    &=&
    {\rm tr}\big[
	\hat l_{\rm rel}^{a(0)\dagger}
	\hat a^\dagger 
	\hat r_{(n_a-1,n_a)}^{a(0)}
	-\hat l_{\rm rel}^{a(0)\dagger}
	\hat r_{(n_a-1,n_a)}^{a(0)}
	\hat a^\dagger
	\big]
	{\rm tr}\Big[
	\hat l_{(n_a-1,n_a)}^{a(0)\dagger}
	\hat a
	\ket{n_a}\bra{n_a}
	\Big]
	={\rm tr}\big[
    \hat l_{\rm rel}^{a(0)}
    {\mathcal D}[\hat a]
    (\ket{n_a}\bra{n_a})
	\big]
    \\
%========================================
    C_{n_a}^{a,{\rm rel}\uparrow}
    &=&{\rm tr}\big[
	\hat l_{\rm rel}^{a(0)\dagger}
	\hat a
	\hat r_{(n_a+1,n_a)}^{a(0)}
	-\hat l_{\rm rel}^{a(0)\dagger}
	\hat r_{(n_a+1,n_a)}^{a(0)}
	\hat a
	\big]
	{\rm tr}\Big[
	\hat l_{(n_a+1,n_a)}^{a(0)\dagger}
	\hat a^\dagger
	\ket{n_a}\bra{n_a}
	\Big]
	={\rm tr}\big[
    \hat l_{\rm rel}^{a(0)}
    {\mathcal D}[\hat a^\dagger]
    (\ket{n_a}\bra{n_a})
	\big].
\end{eqnarray}

% After a lengthy but a straightforward computation, we arrive at,

At the low temperature regime $\bar n_c^0, \bar n_a^0\ll 1$, it is sufficient to sum up the first several Fock states, 
% \alexcom{Can you be more precise quantitatively?}
\begin{eqnarray}
    \label{app:Gamma_rel nc-nc}
    \Gamma_{\rm rel}^{{\rm nc-nc}(2)}
    &\approx&    
    -\tilde\chi^2
    \frac{\tilde\kappa_a+\tilde\kappa_c}{(\Delta-U)^2}
    \bigg[
    p_{{\rm ss},n_c=1}^{c(0)}r_{{\rm rel},n_a=1}^{a(0)}
	C_{n_c=1}^{c,{\rm ss}\downarrow}
	C_{n_a=1}^{a,{\rm rel}\uparrow}	
    + p_{{\rm ss},n_c=0}^{c(0)}
    r_{{\rm rel},n_a=2}^{a(0)}
	C_{n_c=0}^{c,{\rm ss}\uparrow}
	C_{n_a=2}^{a,{\rm rel}\downarrow}	
	\bigg]
    \nonumber\\
    &\approx&
    -\frac{g^2U^2}{\Delta^2}
    \frac{\tilde\kappa_a+\tilde\kappa_c}{(\Delta-U)^2}
    (
    2\tilde n_c
    - 4\tilde n_a
    )
\end{eqnarray}
giving the third term of Eq.~\eqref{eq:Gamma_rel thermal} in the main text. 
\end{widetext}

% An interesting feature of Eq.~\eqref{app:Gamma_rel nc-nc} is that 
Equation~\eqref{app:Gamma_rel nc-nc} tells us that, 
for the contribution solely from nonlinear conversion, 
increasing the qubit thermal population $\tilde n_a$ {\it increases} the qubit $T_1$-decay rate.  In contrast, increasing cavity thermal population $\tilde n_c$ {\it decreases} this rate. 
Note that the former can be important even when the \emph{bare} qubit population is absent $(\bar n_a^0=0, \bar n_c^0>0$) in the regime $\kappa_a\ll\kappa_{\rm P}$,
as the qubit population can be comparable to the cavity population $\tilde n_a\simeq \tilde n_c$ , see Eq.~(8) in the main text.  
This intriguing property can be understood by using Fermi's Golden rule: there is an effective incoherent pumping or decay process between the qubit $n=1$ and $n=2$ Fock state mediated by the cavity.
%This intriguing property can be understood as the Fermi's Golden rule rates of incoherent pumping or decay process between the qubit $n=1$ and $n=2$ Fock state. 
(Note crucially that the excitations between $n=0$ and $n=1$ qubit Fock state is absent because the nonlinear conversion can only take place when at least one qubit photon present.)
To see this, it is instructive to rewrite Eq.~\eqref{app:Gamma_rel nc-nc} as 
\begin{eqnarray}
    \label{app:Gamma_rel nc-nc effective dissipation}
    \Gamma_{\rm rel}^{{\rm nc-nc}(2)}
    \approx
    -\langle\hat l_{\rm rel}^{a(0)},
    {\mathcal L}_{\rm eff}^{\rm nc}
    \hat r_{\rm rel}^{a(0)}
    \rangle
\end{eqnarray}
with 
\begin{eqnarray}
    \label{app:Leff nonlinear conversion}
    {\mathcal L}_{\rm eff}^{\rm nc}
    &=&\kappa_{\rm eff}^{\rm nc}
    (\tilde n_c{\mathcal D}[\hat a_{n\ge 1}^\dagger]
    +{\mathcal D}[\hat a_{n\ge 2}]),
    \\
    \label{app:kappa_eff^nc}
    \kappa_{\rm eff}^{\rm nc}
    &=&\tilde\chi^2
    % \frac{g^2U^2}{\Delta^2}
    \frac{\tilde\kappa_a+\tilde\kappa_c}{(\Delta-U)^2},
\end{eqnarray}
where we have introduced annihilation/creation operators $\hat a_{n\ge 2}$ and $\hat a_{n\ge 1}^\dagger$ that \emph{only} acts on higher-number Fock states, i.e. 
\begin{eqnarray}
    \hat a_{n\ge 2}\ket{n}=\sqrt{n}\ket{n-1},
    \qquad
    \hat a_{n\ge 1}^\dagger\ket{n}=\sqrt{n+1}\ket{n+1}
    \nonumber\\
\end{eqnarray}
but $\hat a_{n\ge 2}\ket{n}=0(n\le 1)$, $\hat a_{n\ge 1}^\dagger\ket{n}=0(n=0)$, reflecting the absence of the excitations between $n=0$ and $n=1$ states from the nonlinear conversion process. 

The fact that $\Gamma_{\rm rel}^{{\rm rel}(2)}$ can be expressed as Eqs.~\eqref{app:Gamma_rel nc-nc effective dissipation} and \eqref{app:Leff nonlinear conversion} shows that $\Gamma_{\rm rel}^{{\rm nc-nc}(2)}$ can be interpreted as a first-order correction from the effective dissipator ${\mathcal L}_{\rm eff}^{\rm nc}$ (\textit{cf.} Eq.~\eqref{app:eigenvalue first order correction}).
The effective dissipation rate $\kappa_{\rm eff}^{\rm nc}$ (Eq.~\eqref{app:kappa_eff^nc}) can be understood as the Fermi's Golden rule rate of the transition from the $(n_c,n_a)=(1,1)$ state to $(n_c,n_a)=(0,2)$ state (and its inverse), where it is given as the product of the transition rate $\tilde\chi^2$ and the density of states.
% $\rho(\omega=\tilde\omega_a)$, where $\rho(\omega)=(\tilde\kappa_a+\tilde\kappa_c)/[(\omega-\omega_c)^2+]$
The first term of ${\mathcal L}_{\rm eff}^{\rm nc}$ that describes the effective incoherent qubit pumping process from qubit $n=1$ to $n=2$ state is proportional to $\tilde n_c$ because the nonlinear conversion that excites the qubit can only activate in the presence of the cavity photon population.
This incoherent pumping process to the higher qubit states contributes as the \emph{decrease} of the qubit $T_1$-decay rate, which follows from the relation
$\langle\hat l_{\rm rel}^{a(0)},
    {\mathcal D}[\hat a_{n\ge 1}^\dagger]
    \hat r_{\rm rel}^{a(0)}
    \rangle\approx 2$.
Note that the fact that this involves the $n=2$ qubit Fock state is essential in obtaining the negative contribution to $\Gamma_{\rm rel}$, since the  the transition between $n=0$ and $n=1$ would affect the $T_1$-decay rate in the opposite way, following from $\langle\hat l_{\rm rel}^{a(0)},
{\mathcal D}[\hat a^\dagger]
\hat r_{\rm rel}^{a(0)}
\rangle\approx -1$ at low temperature.
    
On the other hand, its inverse process from the qubit $n=2$ to $n=1$ state contribute as the increase of the qubit $T_1$-decay rate. 
Since the $n=2$ state can only be populated when the qubit is populated ($\tilde n_a>0$), this term is proportional to $\tilde n_a$, which follows from the relation
$\langle\hat l_{\rm rel}^{a(0)},
    {\mathcal D}[\hat a_{n\ge 2}]
    \hat r_{\rm rel}^{a(0)}
    \rangle\approx -4\tilde n_a$.

% \begin{widetext}
We can similarly compute the third term $\Gamma_{\rm rel}^{{\rm nc-cd}(2)}$ of Eq.~\eqref{app:Gamma_rel second order}, which is the ``cross term" contribution of the nonlinear conversion and correlated dissipation. 
As in the first term $\Gamma_{\rm rel}^{{\rm nc-nc}(2)}$, the intermediate states are given by Eq.~\eqref{app:intermediate states thermal}.
After a lengthy but straightforward computation, we arrive at,
\begin{widetext}
\begin{eqnarray}	
%================================================
%================================================
	&&\Gamma_{\rm rel}^{{\rm nc-cd}(2)}
	=\sum_{n_c,n_a=0}^\infty 
	p_{{\rm ss},n_c}^{c(0)}
	r_{{\rm rel},n_a}^{a(0)}
	\nonumber\\
	&&
	\times\Bigg[
	%---------------------------------------------
	2\tilde\gamma_\downarrow 
	\tilde \chi
	{\rm Im}
	\bigg[
	\frac{1}
	{\tilde\kappa_a +\lambda_{\downarrow_c,(n_a,n_a-1)}^{(0)}}
	\bigg]
	(n_a-1)
	C_{n_c}^{c,{\rm ss}\downarrow}
	C_{n_a}^{a,{\rm rel}\downarrow}
	%---------------------------------------------
	+2\tilde\gamma_\downarrow
	\tilde\chi
	{\rm Im}
	\bigg[
	\frac{1}
	{\tilde\kappa_a +\lambda_{\uparrow_c,(n_a-1,n_a)}^{(0)}}
	\bigg]
	(n_a-1)
	C_{n_c}^{c,{\rm ss}\uparrow}
	C_{n_a}^{c,{\rm rel}\downarrow}
	\nonumber\\
	%---------------------------------------------
	&&	\ \ 
	+2\tilde\gamma_\uparrow
	\tilde\chi
	{\rm Im}
	\bigg[
	\frac{1}
	{\tilde\kappa_a +\lambda_{\uparrow_c,(n_a,n_a+1)}^{(0)}}
	\bigg]
    n_a
	C_{n_c}^{c,{\rm ss}\uparrow}
	C_{n_a}^{a,{\rm rel}\uparrow}
	%---------------------------------------------
	+2\tilde\gamma_\uparrow
	\tilde\chi
	{\rm Im}
	\bigg[
	\frac{1}
	{\tilde\kappa_a +\lambda_{\downarrow_c,(n_a+1,n_a)}^{(0)}}
	\bigg]
	n_a
	C_{n_c}^{c,{\rm ss}\downarrow}
	C_{n_a}^{a,{\rm rel}\uparrow}
	\Bigg].
	%==========
\end{eqnarray}
A notable difference from Eq.~\eqref{app:Gamma_rel nc-nc lengty} is that the \emph{imaginary} part of the ``propagator" $G_\beta=1/(-\tilde\kappa_a-\lambda_\beta^{(0)})$ (where $\beta$ labels the intermediate state) enters the expression, while the \emph{real} part of $G_\beta$ shows up in Eq.~\eqref{app:Gamma_rel nc-nc lengty}. 
The physical meaning of the latter is the density of states of the system, while the former is related to that by the Kramers-Kronig relation.
This difference reflects the property that this term originates from the combination of the coherent and dissipative perturbation. 
Comparing the relation 
${\rm Im}G_\beta\sim 1/(\Delta-U)$
and
${\rm Re}G_\beta\sim \tilde\kappa_{c,a}/(\Delta-U)^2$, 
one finds that this gives rise to one factor of $\Delta-U$ larger  compared to $\Gamma_{\rm rel}^{{\rm nc-nc}(2)}$
and the peculiar sign dependence to the sign of $\Delta-U$.
Indeed, at low temperature $n_c^0,n_a^0\ll 1$, we find
\begin{eqnarray}	
%================================================
%================================================
\label{app:Gamma_rel^nc-cd}
	\Gamma_{\rm rel}^{{\rm nc-cd}(2)}
	&\approx &
	%---------------------------------------------
    2\tilde\gamma_\downarrow
	\tilde\chi
	\frac{-1}{\Delta-U}
	p_{{\rm ss},n_c=0}^{c(0)}
	r_{{\rm rel},n_a=2}^{a(0)}
	C_{n_c=0}^{c,{\rm ss}\uparrow}
	C_{n_a=2}^{c,{\rm rel}\downarrow}
	%---------------------------------------------
	+2\tilde\gamma_\uparrow
	\tilde\chi
	\frac{-1}{\Delta-U}
	p_{{\rm ss},n_c=0}^{c(0)}
	r_{{\rm rel},n_a=1}^{a(0)}
	C_{n_c=0}^{c,{\rm ss}\uparrow}
	C_{n_a=1}^{a,{\rm rel}\uparrow}
	\nonumber\\
	%==========
	&\approx &
	%---------------------------------------------
    2\tilde\gamma_\downarrow
	\tilde\chi
	\frac{-1}{\Delta-U}
	\langle\hat l_{\rm rel}^{a(0)}
	{\mathcal D}[\hat a_{n\ge 2}]
	\hat r_{\rm rel}^{a(0)}\rangle
	%---------------------------------------------
	+2\tilde\gamma_\uparrow
	\tilde\chi
	\frac{-1}{\Delta-U}
	\langle\hat l_{\rm rel}^{a(0)}
	{\mathcal D}[\hat a^\dagger_{n\ge 1}]
	\hat r_{\rm rel}^{a(0)}\rangle
	\nonumber\\
	%==========
	&\approx &
	2(\kappa_c-\kappa_a)
	\frac{g^2U}{\Delta^2}
	\frac{-1}{\Delta-U}
	(-4\tilde n_a)
% 	\nonumber\\
	%---------------------------------------------
	+2(\kappa_c\bar n_c^0-\kappa_a\bar n_a^0)
	\frac{g^2U}{\Delta^2}
	\frac{-1}{\Delta-U}
	\cdot 2
	\nonumber\\
	&=&
	\frac{g^2}{\Delta^2}
	\frac{U}{\Delta-U}
	[8(\kappa_c-\kappa_a)\tilde n_a
	-4(\kappa_c\bar n_c^0-\kappa_a\bar n_a^0)],
\end{eqnarray}
\end{widetext}
giving the second term of Eq.~\eqref{eq:Gamma_rel thermal}. 
Note crucially that, again, the transition between $(n_c,n_a)=(1,1)$ Fock state to $(n_c,n_a)=(0,2)$ state and its inverse is playing the dominant role to this term as well, as one can see from the second line of Eq.~\eqref{app:Gamma_rel^nc-cd}. 

This completes the derivation of Eq.~\eqref{eq:Gamma_rel thermal}. 
% \alexcom{I think this is a nice way of understanding why the larger factor appears, but is this the best way to explain it to SC/quantum optics people? I don't think they are used to thinking about response functions...}
% \AC{I think the above text/explanation is fine, though I don't know how much intuition it provides besides explaing the different power of $\Delta$ in the denominator.  Formally this difference corresponds to an energy shift, though I don't know how to directly use this to provide a simple picture.  Could perhaps stress that the dissipation show up here directly as a matrix element driving a transition, and not via a DOS.  Fine though to leave things here as is I think.}
% \end{widetext}

\subsection{Limitation of Eq.~\eqref{eq:Gamma_rel thermal}}
\label{SM:Limitation of Eq.(14)}

So far, we have analytically derived the second order corrections to the qubit $T_1$-decay rate $\Gamma_{\rm rel}$ in terms of the nonlinear conversion and correlated dissipation ${\mathcal L}_1$.
As seen in Figs.~2 and 3 in the main text, the obtained formula (Eq.~\eqref{eq:Gamma_rel thermal} in the main text) gives an excellent agreement with our numerical simulation (which we provide details in Appendix~\ref{sec:numerical simulation}) in most regimes.
However, we see a slight deviation when the nonlinearity is relatively large $U=0.1|\Delta|$ and is in the regime where the qubit decay $\tilde\kappa_a$ is dominated by Purcell decay contribution  $\kappa_a\ll\kappa_{\rm P}$.
% (but we stress that a qualitative agreement is still obtained and is off only by some numerical factor, see Fig.~3).
Notably, the formula recovers its predictive power in the opposite regime $\kappa_a>\kappa_{\rm P}$, even with large nonlinearity $U=0.1|\Delta|$, see Fig.~3.

We argue below that this deviation is due to the missing higher order correction in terms of ${\mathcal L}_1$, that can become important when $\kappa_a\ll\kappa_{\rm P}$.
We show that there exists correction to $\Gamma_{\rm rel}$ of ${\mathcal O}((g^2 U^2/\Delta^4)\kappa_c\tilde n_c)$ from the higher-order perturbation \emph{only} in the regime $\kappa_a\ll\kappa_{\rm P}$, which is comparable to $\Gamma_{\rm rel}^{\rm nc-nc(2)}$ (the third term of Eq.~\eqref{eq:Gamma_rel thermal} in the main text).
This is due to the appearance of ``resonant'' processes that involves higher order qubit $T_1$-decay modes as its intermediate state.
These results are in agreement with what is seen in the numerics. 
We stress, however, that the qualitative features and the order of magnitude of the correction are well captured already in Eq.~\eqref{eq:Gamma_rel thermal}, since the most dominant correction of ${\mathcal O}((g^2 U/\Delta^3)\kappa_\mu\tilde n_\mu)$ from the second term of Eq.~\eqref{eq:Gamma_rel thermal} is already appropriately included in our second order perturbation theory.

% \begin{widetext}
The next order correction would be from the fourth order perturbation in terms of ${\mathcal L}_1$.
From the recursion relation \eqref{app:general order perturbation}, we have
\begin{widetext}
\begin{eqnarray}
    &&
    \Gamma_{\rm rel}^{(4)}
    =-\langle\hat l_{\rm rel}^{(0)},
    {\mathcal L}_1\hat r_{\rm rel}^{(3)}\rangle
    % -\Gamma_{\rm rel}^{(2)}
    % \langle\hat l_{\rm rel}^{(0)},
    % \hat r_{\rm rel}^{(2)}\rangle
    % \nonumber\\
    % &&
    =-\langle \hat l_{\rm rel}^{(0)},
    {\mathcal L}_1
    ({\mathcal L}_0+\tilde\kappa_a)^{-1}
    {\mathcal L}_1
    ({\mathcal L}_0+\tilde\kappa_a)^{-1}
    {\mathcal L}_1
    ({\mathcal L}_0+\tilde\kappa_a)^{-1}
    {\mathcal L}_1
    \hat r_{\rm rel}^{(0)}\rangle
    % -\Gamma_{\rm rel}^{(2)}
    % \langle\hat l_{\rm rel}^{(0)},
    % ({\mathcal L}_0+\tilde\kappa_a)^{-1}
    % {\mathcal L}_1
    % ({\mathcal L}_0+\tilde\kappa_a)^{-1}
    % {\mathcal L}_1
    % \hat r_{\rm rel}^{(2)}\rangle
    \nonumber\\
    &&=
    -\sum_{\mu={\rm crs,nc,cd}}
    \sum_{\beta,\gamma,\delta\ne{\rm rel}}
    \langle \hat l_{\rm rel}^{(0)},
    {\mathcal L}_1^{\mu}
    \hat r_{\beta}^{(0)}
    \rangle
    \frac{1}
    {\lambda_{\beta}^{(0)}+\tilde\kappa_a}
    \langle \hat l_{\beta},
    {\mathcal L}_1^{\mu}
    \hat r_{\gamma}^{(0)}
    \rangle
    \frac{1}
    {\lambda_{\gamma}^{(0)}+\tilde\kappa_a}
    \langle \hat l_{\gamma}^{(0)},
    {\mathcal L}_1^{\mu}
    \hat r_{\delta}^{(0)}
    \rangle
    \frac{1}
    {\lambda_{\delta}^{(0)}+\tilde\kappa_a}
    \langle \hat l_{\delta}^{(0)},
    {\mathcal L}_1^{\mu}
    \hat r_{\rm rel}^{(0)}
    \rangle
    % \nonumber\\
    % &&-\Gamma_{\rm rel}^{(2)}
\end{eqnarray}
where we have introduced a compact notation for the cross-Kerr nonlinearity ${\mathcal L}_1^{\mu ={\rm crs}}={\mathcal L}_{\rm crs}$, nonlinear conversion ${\mathcal L}_1^{\mu ={\rm nc}}={\mathcal L}_{\rm nc}$,
and correlated dissipation
${\mathcal L}_1^{\mu={\rm cd}}={\mathcal L}_{\rm cd}$.
% The physical picture of this is that the correction occurs from the virtual excitation process that perturbs the unperturbed qubit $T_1$-decay mode $\hat r_{\rm rel}^{(0)}$ (Eq.~\eqref{app:qubit T1 decay mode 0}) four times to a state that has overlap with $\hat l_{\rm rel}^{(0)}$, i.e., to the $m=0$ subspace.
% In what follows, we will concentrate on the contribution from the first term, as it is enough to convey our message.
As in Rayleigh-Schr\"odinger perturbation theory for quantum mechanics, this can be understood as a result of the summation over all possible processes involving four steps of virtual excitations from the unperturbed initial state.

Let us consider in particular the process where two nonlinear conversion (${\mathcal L}_{\rm nc}$) and two correlated dissipation (${\mathcal L}_{\rm cd}$) are involved,
% In particular, let us consider a fourth-order process 
which evolves the qubit $T_1$-decay mode as follows: 
\begin{eqnarray}
    \hat r_{\rm rel}^{(0)}
    \xrightarrow{{\mathcal L}_{\rm nc}}
    \hat r_\delta
    =\hat r_{\downarrow}^{c(0)}
    \otimes
    \hat r_{(n_a+1,n_a)}^{a(0)}
    \xrightarrow{{\mathcal L}_{\rm cd}}
    \hat r_\gamma
    =\hat \rho_{\rm ss}^{c(0)}
    \otimes
    \hat r_{k=2,m=0}^{a(0)}
    \xrightarrow{{\mathcal L}_{\rm nc}}
    \hat r_\beta
    =\hat r_{\downarrow}^{c(0)}
    \otimes
    \hat r_{(n_a+1,n_a)}^{a(0)}
    \xrightarrow{{\mathcal L}_{\rm cd}}
    \hat r_{\rm rel}^{(0)}.
\end{eqnarray}
Here, $n_a$ only takes $n_a\ge 1$ because the nonlinear conversion only activates when qubit excitation is present.
This is a process where the qubit $T_1$-decay mode is excited to a $T_2$ mode by the nonlinear conversion ${\mathcal L}_{\rm nc}$, then converted to a \emph{higher order} qubit $T_1$-decay mode by the correlated dissipation ${\mathcal L}_{\rm cd}$, and then ultimately transfering back to the qubit $T_1$-decay mode by the further perturbation from ${\mathcal L}_{\rm nc}$ and ${\mathcal L}_{\rm cd}$.  
The contribution from this process can be estimated as
\begin{eqnarray}
    \label{app:Gamma rel 4}
    \Gamma_{\rm rel}^{(4)}
    &\sim &
    \tilde\chi^2 \tilde\gamma_{\uparrow}^2
    p_{{\rm ss},n_c=1}^{c(0)}
    r_{{\rm rel},n_a=1}^{a(0)}
    {\rm Re}
    \Bigg[
    \frac{1}{\lambda_{\downarrow_c,(n_a=2,n_a=1)}^{(0)}+\tilde\kappa_a}
    \frac{1}{\lambda_{k=2,m=0}^{a(0)}+\tilde\kappa_a}
    \frac{1}{\lambda_{\downarrow_c,(n_a=2,n_a=1)}^{(0)}+\tilde\kappa_a}
    \Bigg]
    \nonumber\\
    &\sim &
    \tilde\chi^2 \tilde\gamma_{\uparrow}^2
    \tilde n_c
    \frac{1}{(\Delta-U)^2}
    \frac{1}{\tilde\kappa_a}    
    \simeq
    \frac{g^4}{\Delta^4}
    \frac{U^2}{(\Delta-U)^2}
    \frac{(\kappa_c-\kappa_a)^2}{\tilde\kappa_a}
    \tilde n_c
\end{eqnarray}
\end{widetext}
at low temperature.
Here, 
it is proportional to $\tilde\chi^2\tilde\gamma_\uparrow^2$ because two nonlinear conversion and correlated dissipation are involved, and have used  Eqs.~\eqref{app:T1 decay higher} and \eqref{app:lambda_down_c,up_a} to estimate the contribution from the propagators of the intermediate states.

% \begin{eqnarray}
%     \ket{n_c=1,n_a=1}\bra{n_c=1,n_a=1}
%     \xrightarrow{}
%     \ket{n_c=0,n_a=2}\bra{n_c=1,n_a=}
%     \xrightarrow{}
%     \hat \rho_{\rm ss}^{c(0)}
%     \otimes
%     \hat r_{k=2,m=0}^{a(0)}
%     \xrightarrow{}
%     \hat r_{\downarrow}^{c(0)}
%     \otimes
%     \hat r_{(n_a=2,n_a=1}^{a(0)}
%     \xrightarrow{}
%     \hat r_{\rm rel}^{(0)}
% \end{eqnarray}

% \begin{eqnarray}
%     \ket{n_c=1}\bra{n_c=1}
%     \otimes
%     \ket{n_a=1}\bra{n_a=1}
%     \xrightarrow{{\mathcal L}_{}}
%     \hat r_{\downarrow}^{c(0)}
%     \otimes
%     \hat r_{(n_a=2,n_a=1)}^{a(0)}
%     \xrightarrow{}
%     \hat \rho_{\rm ss}^{c(0)}
%     \otimes
%     \hat r_{k=2,m=0}^{a(0)}
%     \xrightarrow{}
%     \hat r_{\downarrow}^{c(0)}
%     \otimes
%     \hat r_{(n_a=2,n_a=1}^{a(0)}
%     \xrightarrow{}
%     \hat r_{\rm rel}^{(0)}
% \end{eqnarray}

% \begin{eqnarray}
%     \alpha_c={\rm ss},\alpha_a=(k=1,m=0)
%     \xrightarrow{}
%     \alpha_c=(k=0,m=1),\alpha_a=(k=0,m=-1)    \xrightarrow{}
%     \xrightarrow{}
%     \hat r_{\downarrow}^{c(0)}
%     \xrightarrow{}
%     \alpha_c={\rm ss},\alpha_a=(k=1,m=0)
% \end{eqnarray}

% \end{widetext}

Due to the fact that this is a contribution from the fourth-order correction, $\Gamma_{\rm rel}^{(4)}$ is proportional to $g^4/\Delta^4$. This, at a glance, seems to always give only subleading order correction to $\Gamma_{\rm rel}$ compared to the corrections given in Eq.~\eqref{eq:Gamma_rel thermal} in the main text that are $\propto g^2/\Delta^2$. 
However, note the appearance of the qubit decay rate $\tilde\kappa_a$ in the \emph{denominator} in its expression, which is due to the property that this process involves a (higher order) qubit $T_1$-decay mode as its intermediate state. 
Because of this ``resonant'' structure, in the regime $\kappa_a\ll \kappa_{\rm P}=(g^2/\Delta^2)(\kappa_c-\kappa_a)$,
we can further estimate the correction as (recall that $\tilde\kappa_a=\kappa_a+\kappa_{\rm P}\simeq\kappa_{\rm P}$ in this regime)
\begin{eqnarray}
    \Gamma_{\rm rel}^{(4)}
    & \sim &
    \frac{g^4}{\Delta^4}
    \frac{U^2}{(\Delta-U)^2}
    \frac{\kappa_c^2}{\frac{g^2}{\Delta^2}(\kappa_c-\kappa_a)}
    \tilde n_c
    \nonumber\\
    &\sim&
    \frac{g^2}{\Delta^2}
    \frac{U^2}{(\Delta-U)^2}
    \kappa_c \tilde n_c
\end{eqnarray}
which is comparable to the third term in Eq.~\eqref{eq:Gamma_rel thermal} in the main text. 
Note how one of the $g^2/\Delta^2$ factor in the numerator is canceled out with that in the denominator to yield this large magnitude. 
The obtained order of magnitude matches with the magnitude of the deviation between Eq.~\eqref{eq:Gamma_rel thermal} and  the numerics. 
This is also consistent with the observation made in Fig.~2 in the main text that Eq.~\eqref{eq:Gamma_rel thermal} matches with the numerics at very small $U\ll|\Delta|$, where the second term of Eq.~\eqref{eq:Gamma_rel thermal} dominates over $\Gamma_{\rm rel}^{(4)}$.

On the other hand, when $\kappa_a\gg\kappa_{\rm P}$,  $\tilde\kappa_a$ in the denominator of Eq.~\eqref{app:Gamma rel 4} does not get as small and the cancellization of $g^2/\Delta^2$ seen above does not occur. 
As a result, $\Gamma_{\rm rel}^{(4)}\propto g^4/\Delta^4$ 
only gives subleading correction compared to the terms obtained in Eq.~\eqref{eq:Gamma_rel thermal} in the main text.
This is consistent with the results obtained in Fig.~3, where an excellent agreement is still obtained even with a relatively large nonlinearity $U(=0.1|\Delta|)$.

% Recalling that $\tilde\kappa_a = \kappa_a + \kappa_{\rm P} = \kappa_a+\frac{g^2}{\Delta^2}(\kappa_c-\kappa_a)$,
%  since $(\kappa_c-\kappa_a)^2/\tilde\kappa_a\sim \kappa_c$, the fourth order correction $\Gamma_{\rm rel}^{(4)}$ is ${\mathcal O}(g^4/\Delta^4)$. Therefore, 

% From the above arguments, we attribute the quantitative deviation between Eq.~\eqref{eq:Gamma_rel thermal} and the numerics seen at $U=0.1|\Delta|$ and $\kappa_a=0<\kappa_{\rm P}$ to these processes involving higher order qubit $T_1$-decay modes.

% The dissipative harmonic oscillator (Eq.~\eqref{eq si: dissipative harmonic oscillator}) has been solved exactly [CITE] 

\section{Bare cavity coherent drive}\label{SM:Section_Coherent_Drive}

This section deals with the bare cavity coherent drive $\hat H_{\rm D}=f_c e^{-i\omega_{\rm D}t}\hat c_0+{\rm h.c.}$ and will derive Eq.~(20) in the main text. 
As sketched in the main text, we first move to the rotating frame that eliminates the time dependence from the Hamiltonian and further make a displacement transformation that eliminates linear terms $\propto \hat c_0,\hat a_0$ from the Hamiltonian.
The resulting equation of motion at weak drive regime $|\alpha_a|^2\ll 1$ and zero temperature $\bar n_c^0=\bar n_a^0=0$ is given by, 
\begin{eqnarray}
    &&\partial_t\hat\rho
    =-i[\hat H_{\rm s}',\hat\rho]
    +\kappa_a 
    {\mathcal D}[\hat a_0]\hat\rho
    +    \kappa_c
    {\mathcal D}[\hat c_0]\hat\rho
    \equiv {\mathcal L}'\hat\rho,
\end{eqnarray}
with
\begin{eqnarray}
    \hat H_{\rm s}'
    \simeq \hat H_0'
    +\hat H_{\rm int}', 
\end{eqnarray}
where
\begin{eqnarray}
    \hat H_0'
    &=&\hat H_0+\hat H_{\rm quad}
    \nonumber\\
    &\simeq&
    (\omega_a - \omega_{\rm D} -2U|\alpha_a|^2)\hat a_0^\dagger\hat a_0
    + g(\hat a_0^\dagger \hat c_0 + {\rm h.c.})
    \nonumber\\
    &+& (\omega_c-\omega_{\rm D})\hat c_0^\dagger \hat c_0,
\end{eqnarray}
is a \emph{modified} quadratic Hamiltonian from the energy shift that arises due to the drive
(where the squeezing terms $\sim \ha_0 \ha_0$ and $\ha_0^\dagger \ha_0^\dagger$ are omitted as they play no role to the order of our interest),
and
\begin{eqnarray}
    \label{app:Hint'}
    \hat H_{\rm int}'
    =\hat H_{\rm int}
    +\hat V[\alpha_a],
    \qquad
    \hat V[\alpha_a]=-U(\alpha_a\hat a_0^\dagger\hat a_0^\dagger\hat a_0
    +{\rm h.c.}).
    \nonumber\\
\end{eqnarray}
The displacement fields $\alpha_c,\alpha_a$ satisfy the relation,
\begin{eqnarray}
    \begin{pmatrix}
        \omega_c - \omega_{\rm D}- i\kappa_c/2 & g \\
        g & \omega_a - \omega_{\rm D} - i \kappa_a/2
    \end{pmatrix}
    \begin{pmatrix}
        \alpha_c \\
        \alpha_a
    \end{pmatrix}
    = -
    \begin{pmatrix}
        f_c \\
        0
    \end{pmatrix}.
    \nonumber
\end{eqnarray}

In what follows, we diagonalize the modified quadratic Hamiltonian as $\hat H_0'=(\tilde\omega_a'-\omega_{\rm D})\hat a'^\dagger\hat a'
+(\tilde\omega_c'-\omega_{\rm D})\hat c'^\dagger\hat c'$
by introducing the modified polariton operators
$\hat c'\simeq [1-g^2/(2\Delta'^2)]\hat c_0 - (g/\Delta')\hat a_0$
and
$\hat a'\simeq [1-g^2/(2\Delta'^2)]\hat a_0 + (g/\Delta')\hat c_0$,
where 
$\tilde\omega_a\simeq \omega_a+g^2/\Delta'$ and
$\tilde\omega_c\simeq \omega_c+g^2/\Delta'$.
These expressions has the same form as the the usual blackbox basis without coherent drive (see main text), where the qubit-cavity detuning $\Delta=\omega_a-\omega_c$ is replaced by
\begin{eqnarray}
    \Delta'[\alpha_a]=\Delta-2U|\alpha_a|^2,
\end{eqnarray}
since the only difference between $\hat H_0$ and $\hat H_0'$ is the energy shift from the drive.

For the interaction term $\hat H_{\rm int}'\approx \hat H_{\rm int}^{\rm slf}{}'+\hat H_{\rm int}^{\rm crs}{}'+\hat H_{\rm int}^{\rm nc}{}'$, they are given as the sum of the self- and cross-Kerr nonlinearity and the nonlinear conversion as before (\textit{cf.} Eqs.~(2) and (3) in the main text), 
\begin{eqnarray}
    \hH_{\rm int}^{\rm slf}{}'
    &=& 
    \chi_{aa}'
    \hat a'^\dagger \hat a'^\dagger \hat a' \hat a',
    \\
    \hH_{\rm int}^{\rm crs}{}'
    &=&
    \chi_{ca}'
    \hat c'^\dagger\hat c'
    \hat a'^\dagger\hat a',
    \\
    \hH_{\rm int}^{\rm nc}{}'
    &=& \tilde \chi'
    (\hat a'^\dagger\hat a'^\dagger\hat a'\hat c'
    +\hat c'^\dagger\hat a'^\dagger\hat a'\hat a'),
\end{eqnarray}
with $\chi_{aa}'=-(U/2)
    [1-g^2/(2\Delta'^2)],\chi_{ca}'=-2g^2U/\Delta'^2$, and $\tilde \chi' = g U / \Delta'$.
The drive term $\hat V[\alpha_a]$ transforms as
\begin{eqnarray}
    \hat V[\alpha_a]\approx
    -U(\alpha_a\hat a'^\dagger\hat a'^\dagger\hat a'
    +{\rm h.c.}),
\end{eqnarray}
where we have dropped the higher-order corrections ${\mathcal O}((g^2/\Delta^2)U\alpha_a,\alpha_a^3)$.

The resulting Lindblad master equation in this basis takes the form 
\begin{eqnarray}
    \label{app:Lindblad coherent}
    \partial_t\hat\rho
    ={\mathcal L}'\hat\rho
    ={\mathcal L}_{\rm ind}'\hat\rho
    +{\mathcal L}_{\rm cd}'\hat\rho.
\end{eqnarray}
Here, ${\mathcal L}_{\rm ind}'$ and ${\mathcal L}_{\rm cd}'$ take the similar form to ${\mathcal L}_{\rm ind}$ (Eq.~(6) in the main text) and ${\mathcal L}_{\rm cd}$ (Eq.~(9) in the main text), respectively, except for the replacement of parameters and some additional terms arising from the drive $\hat V[\alpha_a]$:
\begin{eqnarray}
    {\mathcal L}_{\rm ind}'\hat\rho
    &=&-i[\hH_{\rm s}',\hat\rho]
    \nonumber\\
    &+&\kappa_a'[\alpha_a]
    {\mathcal D}[\hat a']\hat\rho
    +    \kappa_c'[\alpha_a]
    {\mathcal D}[\hat c']\hat\rho
    \\
    {\mathcal L}_{\rm cd}'\hat\rho
    &=&-\frac{\tilde\gamma_{\downarrow}'[\alpha_a]}{2}
    \{\hat a'^\dagger \hat c'
    +\hat c'^\dagger\hat a',\hat\rho\}
    \nonumber\\
    && +\tilde\gamma_\downarrow'[\alpha_a]
    (\hat a'\hat\rho\hat c'^\dagger + \hat c'\hat\rho\hat a'^\dagger).
\end{eqnarray}
As stressed in the main text, the intrinsic qubit polariton damping rate is modified to 
\begin{eqnarray}
    \tilde\kappa'_a[\alpha_a]
    & = &
    \kappa_a +\frac{g^2}{(\Delta'[\alpha_a])^2}
    (\kappa_c - \kappa_a) 
    \nonumber\\
    &\simeq &
    \kappa_a + \kappa_{\rm P}
    +\frac{g^2}{\Delta^2}\frac{4U}{\Delta}(\kappa_c-\kappa_a)
    |\alpha_a|^2
\end{eqnarray}
and similarly for the cavity polariton dissipation rate and correlated dissipation rate,
\begin{eqnarray}
    \tilde\kappa'_c[\alpha_a]
    & = &
    \kappa_c +\frac{g^2}{(\Delta'[\alpha_a])^2}
    (\kappa_a - \kappa_c) 
    \nonumber\\
    &\simeq &
    \kappa_c - \kappa_{\rm P}
    -\frac{g^2}{\Delta^2}\frac{4U}{\Delta}(\kappa_c-\kappa_a)
    |\alpha_a|^2,
    \\
    \tilde\gamma_\downarrow'[\alpha_a]
    &=&
    \frac{g}{\Delta'[\alpha_a]}
    (\kappa_c  - \kappa_a)
    \\
    &\simeq&
    \tilde\gamma_\downarrow
    +\frac{2gU}{\Delta}(\kappa_c-\kappa_a)|\alpha_a|^2.
\end{eqnarray}

We now compute the qubit $T_1$-decay rate $\Gamma_{\rm rel}[\alpha_a]$ of this system. 
Since the form of the master equation~\eqref{app:Lindblad coherent} is very much similar to that of the thermal case (Eq.~(5) in the main text), most of the analysis below would be parallel to the previous section.
Similarly to the thermal case, we regard the qubit-cavity coupling terms and $\hat V[\alpha_a]$ that gives rise to non-secular nonlinearity, given by,
\begin{eqnarray}
    {\mathcal L}_1'\hat\rho
    =-i[
    \epsilon_{\rm crs}\hH_{\rm int}^{\rm crs}{}'
    +\epsilon_{\rm ns}\hH_{\rm int}^{\rm nc}{}'
    +\epsilon_{\rm V}\hat V[\alpha_a],\hat\rho]
    +\epsilon_{\rm cd}{\mathcal L}_{\rm cd}'\hat\rho
    \nonumber\\
\end{eqnarray}
as the perturbation on top of the unperturbed part ${\mathcal L}_0'={\mathcal L}'-{\mathcal L}_1'$, or
\begin{eqnarray}
    \label{app:L0 coherent}
    {\mathcal L}_0'\hat\rho
    =-i[\hH_0'+\hH_{\rm int}^{\rm slf}{}',\hat\rho]
    +\kappa_a'[\alpha_a]
    {\mathcal D}[\hat a']\hat\rho
    +    \kappa_c'[\alpha_a]
    {\mathcal D}[\hat c']\hat\rho
    \nonumber\\
\end{eqnarray}
Here, again, $\epsilon_{\rm crs}=\epsilon_{\rm ns}=\epsilon_{\rm V}=\epsilon_{\rm cd}=1$ is the book-keeping constant. 
We will compute up to the second-order in ${\mathcal L}_1'$, where the qubit $T_1$-decay rate is given by $\Gamma_{\rm rel}[\alpha_a]=\Gamma_{\rm rel}^{(0)}+\Gamma_{\rm rel}^{(2)}$. (The first order correction vanishes as before.)

The unperturbed part Eq.~\eqref{app:L0 coherent} has the same form as that of the thermal case Eq.~\eqref{app:L0}. Therefore, those results can be directly applied by the appropriate replacement of the parameters such as $\tilde\kappa_\mu\rightarrow\tilde\kappa_\mu'[\alpha_a]$. This gives the unperturbed qubit $T_1$-decay rate (\textit{cf.} Eq.~\eqref{app:unperturbed T1 decay rate})
\begin{eqnarray}
    \Gamma_{\rm rel}^{(0)}[\alpha_a]
    =\tilde\kappa_a'[\alpha_a]
    =\kappa_a+\kappa_{\rm P}
    +\frac{g^2}{\Delta^2}\frac{4U}{\Delta}
    (\kappa_c-\kappa_a)|\alpha_a|^2.
    \nonumber\\
\end{eqnarray}
Note that the coherent drive strength dependence is already included in this unperturbed part.

We move on to consider the perturbative correction to the qubit $T_1$-decay rate from ${\mathcal L}_1'$. 
Actually, it turns out that the only contribution at zero temperature is that from $\hat V[\alpha_a]$.
As before, the cross-Kerr nonlinearity does not contribute to the qubit $T_1$-decay rate since they do not change the photon excitation number.
For the nonlinear conversion and the correlated dissipation to be activated, photon excitations should be present in the steady state, which, however, are absent at zero temperature $\bar n_c^0=\bar n_a^0=0$.

Since $\hat V[\alpha_a]$ only excites the qubit, the intermediate state that participates to the second-order correction to the qubit $T_1$-decay rate would be of the form that solely involves the coherence of the qubit,
\begin{eqnarray}
    \hat r_{{\rm ss},(n_a+1,n_a)}^{(0)}
    \equiv\hat \rho_{\rm ss}^{c(0)}
    \otimes
    \hat r_{(n_a+1,n_a)}^{a(0)}
    % =\ket{0,n_a+1}\bra{0,n_a}
    % \nonumber\\
\end{eqnarray}
% and
% \begin{eqnarray}
%     \hat r_{{\rm ss},(n_a,n_a+1)}^{(0)}
%     \equiv\hat \rho_{\rm ss}^{c(0)}
%     \otimes
%     \hat r_{(n_a,n_a+1)}^{a(0)}
%     =\ket{0,n_a}\bra{0,n_a+1},
%     \nonumber\\
% \end{eqnarray}
with corresponding eigenvalue
\begin{eqnarray}
	\lambda_{{\rm ss},(n_a+1,n_a)} ^{(0)}
	&=& -\frac{\tilde\kappa_a}{2}(1+2n_a)
	+i(\tilde\omega_a'-\omega_D-Un_a).
	\nonumber\\
% 	\\
% 	\lambda_{{\rm ss},(n_a,n_a+1)} ^{(0)}
% 	&=& -\frac{\tilde\kappa_a}{2}(1+2n_a)
% 	-i(\tilde\omega_a'-\omega_D-Un_a).
% 	\nonumber\\
\end{eqnarray}
As a result, the second order correction is computed as
\begin{widetext}
\begin{eqnarray}
    \Gamma_{\rm rel}^{(2)}[\alpha_a]
    &\approx&
    -\epsilon_{\rm V}^2
    U^2 |\alpha_a|^2 2{\rm Re}\bigg[\frac{1}{\tilde\kappa_a+\lambda_{{\rm ss},(n_a=2,n_a=1)}^{(0)}}\bigg]
    % \nonumber\\
    % &\times&
    \langle\hat l_{\rm rel}^{a(0)},
    {\mathcal D}[\hat a_{\rm n\ge 1}'^\dagger]
    \hat r_{\rm rel}^{a(0)}\rangle
    \nonumber\\
    &\approx&
    -\epsilon_{\rm V}^2
    \frac{2U^2}{(\tilde\omega_a-\omega_{\rm D}-U)^2}\tilde\kappa_a|\alpha_a|^2,
\end{eqnarray}
\end{widetext}
where the creation operator $\hat a_{n\ge1}'^\dagger$ only acts on the Fock states with $n\ge 1$.
It is proportional to $\tilde\kappa_a$ and the denominator has the form $(\tilde\omega_a-\omega_{\rm D}-U)^2$  (in contrast to Eq.~\eqref{app:Gamma_rel nc-nc} where it is proportional to $\tilde\kappa_a+\tilde\kappa_c$ and $1/(\Delta-U)^2$),
reflecting the eigenvalue of the intermediate state. 
This is the second term of Eq.~(20) in the main text and hence completes the derivation.

\section{Numerical simulation}
\label{sec:numerical simulation}

    %%%%%%%%%%%
	% FIG S1 diagonalization vs simulation
	%%%%%%%%%%%
	\begin{figure}[t]
	\centering
    \includegraphics[width=0.7\linewidth,keepaspectratio]{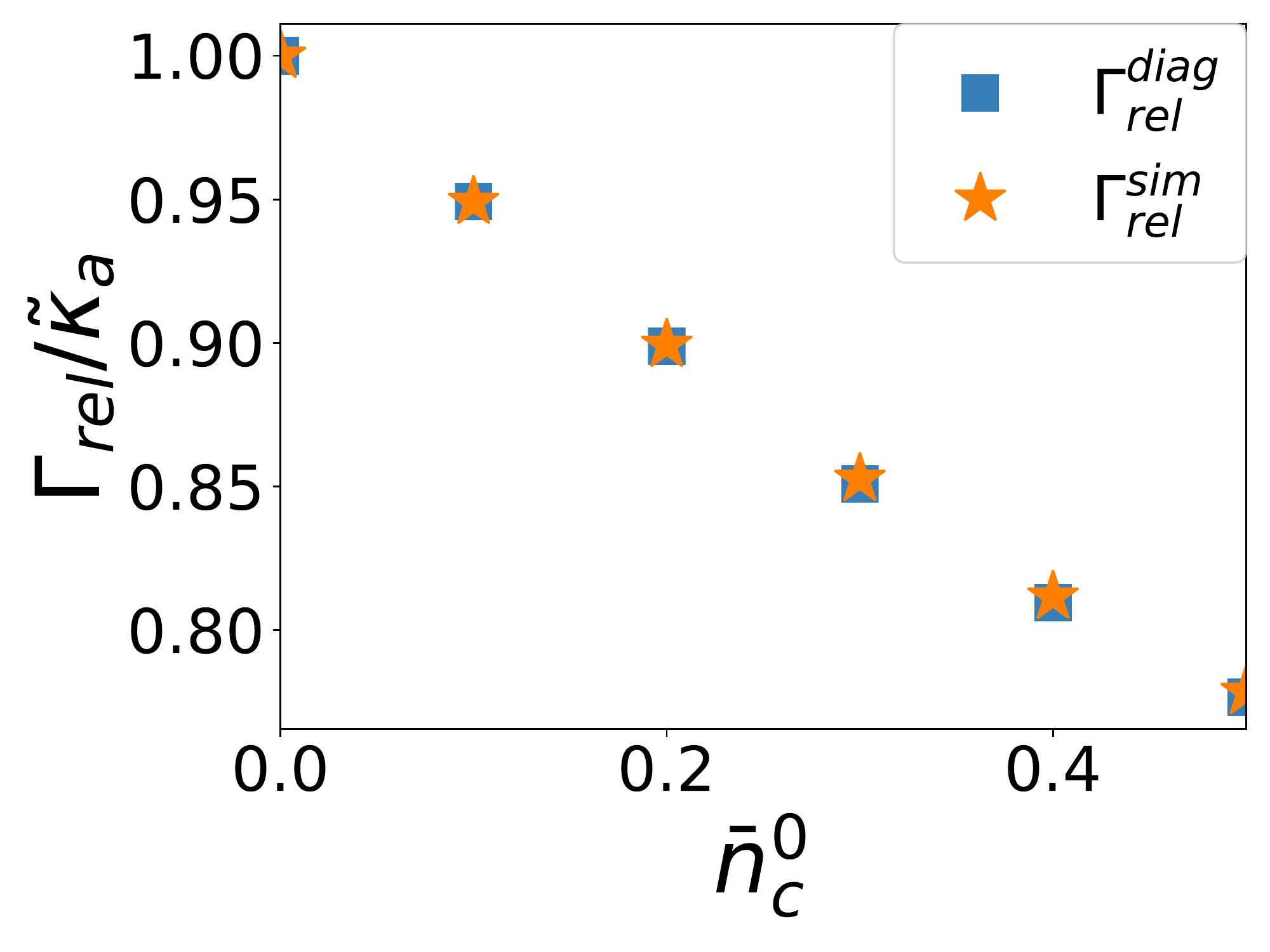}
	\caption{ 
	Comparison between the two different definitions of qubit $T_1$-decay rate $\Gamma_{\rm rel}^{\rm diag}$ and $\Gamma_{\rm rel}^{\rm sim}$. Here, $\Gamma_{\rm rel}^{\rm diag}$ is determined from numerical diagonalization of the Lindbladian ${\mathcal L}$ (see main text).
	$\Gamma_{\rm rel}^{\rm sim}$ is determined by numerically computing $\avg{\hat a^\dagger\hat a}(t)$ and fitting to the relation $\avg{\hat a^\dagger\hat a}(t)\sim A e^{-\Gamma_{\rm rel}^{\rm sim}t}$ at late times. This is computed by simulating Eq.~(5) in the main text, with the initial state set to the form of Eq.~\eqref{app:initial condition}. For the fitting, we have used the data set at $t=[0.95 T,T]$, where $T=8\times10^4|\Delta|$.
	The two different definitions of qubit $T_1$-decay rates $\Gamma_{\rm rel}^{\rm diag}$ and $\Gamma_{\rm rel}^{\rm sim}$ are in excellent agreement.
	We have taken the same parameters as that of Fig.~3(a) (with $\kappa_a=0$) in the main text.
    % 	Thermal cavity drive dependence on the qubit $T_1$-decay rate $\Gamma_{\rm rel}$ for a larger nonlinearity $U=0.1|\Delta|$.
    % 	Lines (points) correspond to analytical (numerical) results.
    % 	Higher order corrections are more important here, but the analytic results still give good qualitative agreement.
    % 	(b) The ratio of the numerically ($s_{\rm num}$) to analytically ($s_{\rm th}$) evaluated slope parameter $s = d \Gamma_{\rm rel} / d \bar{n}_c^0$ (evaluated at zero temperature). 
    % 	For both the panels, we set $g=0.1|\Delta|,\kappa_c = 0.01|\Delta|$ and $\Delta<0$.
    %     For large $U$ and $\kappa_{\rm P} > \kappa_a$, higher order corrections become important, but our analytical expression still provides a reasonable qualitative agreement (see SM). 
	}
	\label{fig:diagonalization vs simulation}
	\end{figure}
    %%%%%%%%%%%

We outline here the procedure we took to determine the qubit $T_1$-decay rate $\Gamma_{\rm rel}^{\rm diag}$ from numerical diagonalization of the Linbladian both for the thermal case ${\mathcal L}$  (Eq.~(4) in the main text) and ${\mathcal L}'$ for the coherent drive case (Eq.~\eqref{app:Lindblad coherent}).
% Below, we outline the procedure we took. 
% We first numerically diagonalize the Lindbladian ${\mathcal L}$ (${\mathcal L}'$) to obtain the right and left eigenstates $\hat r_\alpha$ and $\hat l_\alpha$.
% We then sort the modes into the steady state $\hat\rho_{ss}$, $T_1$-decay and $T_2$-decay modes, which can practically be distinguished from the eigenvalue $\lambda_\alpha$ being zero, real or complex, respectively.
To identify the qubit $T_1$-decay rate, recall that the density matrix evolves as 
% \cite{Scarlatella2019}
% \ryocom{Xander, could you check whether this equation is there in the Gardiner-Zoller textbook? If not, add a citation that does have this written?}
% \alexcom{Couldn't find it, so I put Aash's NJP as a reference.}
% \ryocom{OK...}
\begin{eqnarray}\label{app:Decomp_Eigenmodes}
    \hat\rho(t)
    =\hat\rho_{\rm ss}
    +\sum_\alpha w_\alpha e^{\lambda_\alpha t}\hat r_\alpha
\end{eqnarray}
with $w_\alpha= \langle \hat l_\alpha, \hat\rho(0)\rangle$. 
Here, we took the normalization of $\hat r_{\alpha}$ and $\hat l_{\alpha}$ to satisfy
${\rm Tr}[\hat l_\alpha^\dagger \hat l_\alpha]=1$ together with the bi-orthogonality relation ${\rm Tr}[\hat l_\alpha^\dagger \hat r_\beta]=\delta_{\alpha,\beta}$.
Equation~ \eqref{app:Decomp_Eigenmodes} tells us that if one starts in an initial state where the qubit is excited, e.g.~ by adding a single photon to the steady state
% However, in the long-time limit, 
%The natural question is whether the $T_1$ rate as defined in the main text is physically relevant: does an initially excited qubit really decay to the ground state at this rate or not?  Given the relatively weak nonlinearity, we take the initial qubit-excite state to be one where we add a single qubit photon to the steady state  
%Since we are interested in the mode that involves the decay of the qubit photon number excitation, we consider the situation where the initial state is given by the ``qubit-excited" steady state,
\begin{eqnarray}
    \label{app:initial condition}
    \hat\rho(0) \propto \hat a^\dagger\hat\rho_{\rm ss}\hat a.
\end{eqnarray}
then one will excite a set of Liouvillian eigenmodes $\hat{r}_\alpha$ each with its own exponential decay rate $- \textrm{Re }\lambda_{\alpha}$.  We can define operationally the $T_1$ decay rate of the qubit by the slowest decay rate associated with such an initial state (as this will define the long-time relaxation).  Alternatively, we could define the $T_1$ decay rate as the rate associated with the excited mode with the largest weight $|w_\alpha|$ (i.e.~the mode that most closely represents the deviation between the initial state and the steady state).  Either of these definitions can be used from numerical simulations to extract the qubit $T_1$ rate; for all parameters shown in our numerical plots, both definitions yield the same result.   
%Since the decay modes that has the largest overlap with the initial state  (with eigenvalues) dominates the long time behavior of the qubit photon number $\langle\hat a^\dagger\hat a\rangle(t)$, we identify the mode with the largest $|w_\alpha|$ as the qubit $T_1$-decay or relaxation mode $\alpha={\rm rel}$ .
The resulting numerically-obtained qubit $T_1$-decay rate defined as $\Gamma_{\rm rel}^{\rm diag}=-\lambda_{\rm rel}$ is computed using QuTiP code \cite{QUTIP,QUTIP2} and are plotted in Figs.~1-3 in the main text.

A more direct and experimentally relevant way to compute the $T_1$-decay rate is to fit the photon number to an exponentially decaying curve after transient dynamics have damped out. 
To gain further confidence in our approach, we have also computed the qubit $T_1$-decay rate $\Gamma_{\rm rel}^{\rm sim}$ in this manner from the direct simulation of the Lindblad master equation~(5) with  initial condition given by Eq.~\eqref{app:initial condition}. We have fitted the time evolution of the qubit photon number to
$\langle\hat a^\dagger\hat a\rangle(t)\sim e^{-\Gamma_{\rm rel}^{\rm sim}t}$ at long times,
%which gives another way to determine the qubit $T_1$-decay rate $\Gamma_{\rm rel}^{\rm sim}$.
and compared $\Gamma_{\rm rel}^{\rm diag}$ and $\Gamma_{\rm rel}^{\rm sim}$ in Fig. \ref{fig:diagonalization vs simulation}.
% for both thermal and coherent drive case.
As one sees, we find an excellent agreement between the two qubit $T_1$-decay rates, supporting our approach of determining $\Gamma_{\rm rel}$ numerically.

% To gain confidence to our approach, we have compared $\Gamma_{\rm rel}^{\rm diag}$ obtained above and that obtained directly simulating Eq.~(5) in the main text and fitting $\langle\hat a^\dagger\hat a\rangle(t)\sim e^{-\Gamma_{\rm rel}^{\rm sim}t}$ with the initial condition given by Eq.~\eqref{app:initial condition}.

% We then look for the relevant $T_1$-decay mode (that has real eigenvalues) by searching for the mode that has the largest overlap $|w_\alpha|$ with the \emph{perturbed} steady state $\tilde\rho_{\rm ss}$, where $\tilde\rho_{\rm ss}$ has one additional qubit excitation to the steady state $\hat\rho_{\rm ss}$,

\section{Extensions to other models - Qubit-Mediated Cross-Kerr Interaction}
\label{SM:extensions}

In this section, we briefly explain how our general formalism could be applied to other relevant driven-dissipative circuit QED models studied in the literature. We will not explicitly perform the calculation, but rather setup the problem and sketch the solution so that any interested reader could apply the same technique to their model. We focus on a ubiquitous system studied in the context of bosonic error correction:
two (bare) linear cavity modes $\hb_0$ and $\hc_0$ are both off-resonantly coupled to a common transmon qubit $\ha_0$ \cite{Zhang2021}.  The goal here is to use the transmon to mediate nonlinear mode-mode interactions.  Of course, an issue is the transmon will also generate unwanted dissipative interactions which are drive dependent.  Our approach provides a powerful menas to treat this system.  

The starting Hamiltonian describing the isolated system is
\begin{eqnarray}
    &&\hat{H}_0
    =
    \omega_b \hb_0^\dagger \hb_0
    +
    \omega_c \hc_0^\dagger \hc_0
    +
    \omega_a \ha_0^\dagger \ha_0
    -
    \frac{U}{2}
    \ha_0^\dagger \ha_0^\dagger \ha_0 \ha_0
    \nonumber\\
    &&+
    \left(
    g_b\ha_0^\dagger \hb_0
    +
    g_c\ha_0^\dagger \hc_0
    +
    f_b e^{-i \omega_D t}  \hb_0
    +
    f_c e^{-i \omega_D t} \hc_0
    +
    \text{h.c.}
    \right)
    \nonumber\\
\end{eqnarray}
where $\omega_k$ is the bare resonant frequency, $g_k$ is the coupling between the linear mode and the qubit and $f_k$ the strength of the coherent drive on each linear mode, which we take to be at the same frequency $\omega_D$. As we have done throughout, we shall assume that each mode is subject to it's own independent Markoivian environment with decay rate $\kappa_{k_0}$ and thermal occupation $\bar{n}_{k_0}$ for $k \in \{a,b,c\}$ . Following the procedure outlined in Appendix~\ref{SM:Keldysh_Section}, integrating out the Markovian baths leads to the master equation
\begin{eqnarray}
    &&\partial_t \hrho
    =
    -i[\hat{H}_0, \hrho]
    \nonumber\\
    &&+
    \sum_{k\in a,b,c}
    \Big(
    \kappa_{k_0}(\bar{n}_{k_0}+1)\mathcal{D}[\hat{k}_0]\hrho
    +\kappa_{k} \bar{n}_{k_0} \mathcal{D}[\hat{k}^\dagger_0]\hrho
    \Big)
\end{eqnarray}
which, note, is written in terms of the bare qubit and cavity modes. We can now succinctly outline how to apply our method to this problem, which can be readily generalized to more complicated setups.
\begin{itemize}
    \item First diagonalize the linear and quadratic parts of $\hat{H}_0$. This involves moving to a rotating frame at frequency $\omega_D$,  diagonalizing the quadratic $3 \times 3$ matrix via a simple unitary transformation and then performing a simple displacement transformation to eliminate the linear drives. With this procedure, one obtains a set of polariton modes $\hat{k}$ which, by construction, can be written as a linear combination of the bare modes $\hat{k}_0$ and the drives $f_k$. Given that the detuning between the bare qubit and caivty modes are much larger than their coupling $|g_{b/c}/(\omega_a-\omega_{b/c})|^2 \ll 1$, these polaritons will serve as the starting point for Lindblad perturbation theory.
    
    \item   One then writes the coherent Hamiltonian $H_0$ in this new polariton basis. This gives rises to the usual self and cross-Kerr interaction. In addition, to lowest order in $|g_{b/c}/(\omega_a-\omega_k)|$, there will be a set of non-linear conversion processes of the form $\tilde{\chi}_{b/c a} \ha^\dagger \ha^\dagger \ha \hb/\hc$ and
    $\alpha_{a}^* \ha^\dagger \ha \ha$, where $\alpha_a$ is proportional to the drive strength. Although they are non-resonant, one must keep these terms since, as explained in the main text, the interplay between the dissipative conversion and this coherent non-linear conversion will lead to the leading-order correction in the qubit $T_1$ decay rate.
    
    \item The jump terms are also written in terms of the polariton creation and annihilation operators. Writing these out explicitly, one makes a distinction between the terms which describe damping of the polariton modes and those which describe a dissipative conversion process. Retaining only the former amounts to making the standard secular approximation; our method relies on keeping all such terms.
    
    \item With all relevant terms in hand, one splits the Lindbladian in two parts $\hat{\mathcal{L}} = \hat{\mathcal{L}}_0+\hat{\mathcal{L}}_1$. The first term $\hat{\mathcal{L}}_0$ consists solely of terms which do not couple the various polariton modes.
    We stress that $\hat{\mathcal{L}}_0$ includes self-Kerr interactions which (as discussed in the main text) can be treated exactly.
    The remaining part describes the dissipative and coherent polariton-polariton coupling terms. All such terms are necessarily are of the order $|g_{b/c}/(\omega_a-\omega_{b/c})|$ (assuming $|g_{b}/(\omega_a-\omega_{b})| \sim |g_{b}/(\omega_a-\omega_{c})| $) and thus serve as the correct starting point for Lindblad perturbation theory.
    
    \item The eigenvalues and eigenvectors of $\hat{\mathcal{L}}_0$ must then be found, usually within some approximate scheme (using e.g. the smallness of the bare non-linearity $U$). This only involves solving three single-mode problems, since $\hat{\mathcal{L}}_0$ does not couple the different polaritons. This was done explicitly in Secs.~\ref{SM:Section_Thermal_Occupation}-\ref{SM:Section_Coherent_Drive}. 
    
    \item One can now perform Lindblad perturbation in $\hat{\mathcal{L}_1}$ using $\hat{\mathcal{L}}_0$ as the unperturbed Lindbladian. With this information, one can systematically investigate e.g. how qubit-induced dissipation depends on drive amplitudes and the strength of the various thermal noise terms.  

\end{itemize}

\begin{acknowledgments}
This work was supported by the Department of Energy BES quantum information science program under award DE-SC0020152.  
\end{acknowledgments}

\bibliography{main_cqed}

\end{document}